\documentclass{aastex63}

\newcommand\aastex{AAS\TeX}

\usepackage{threeparttable}
\usepackage{comment}
\usepackage{bm}
\turnoffeditone

\submitjournal{APJ}
\accepted{2021 August 11}

\shorttitle{Triggered star formation by shocks}
\shortauthors{Kinoshita et al.}
\graphicspath{{./}{figures/}}

\begin{document}

\title{Triggered star formation by shocks }

\correspondingauthor{Shinichi.W.Kinoshita}
\email{kinoshita.shinichi@grad.nao.ac.jp}

\author{Shinichi.W.Kinoshita}
\affiliation{Department of Astronomy, the University of Tokyo, 7-3-1 Hongo Bunkyo, 113-0033 Tokyo, Japan}
\affiliation{National Astronomical Observatory of Japan, NINS, 2-21-1 Osawa, Mitaka, Tokyo 181-8588, Japan}

\author{Fumitaka Nakamura}
\affiliation{Department of Astronomy, the University of Tokyo, 7-3-1 Hongo Bunkyo, 113-0033 Tokyo, Japan}
\affiliation{National Astronomical Observatory of Japan, NINS, 2-21-1 Osawa, Mitaka, Tokyo 181-8588, Japan}
\affiliation{The Graduate University for Advanced Studies (SOKENDAI), 2-21-1 Osawa, Mitaka, Tokyo 181-0015, Japan}

\author{Benjamin Wu}
\affiliation{National Astronomical Observatory of Japan, NINS, 2-21-1 Osawa, Mitaka, Tokyo 181-8588, Japan}
\affiliation{NVIDIA Research, 2788 San Tomas Expressway, Santa Clara, CA 95050, USA}



\begin{abstract}
Star formation can be triggered by compression from shock waves. In this study, we investigated the interaction of hydrodynamic shocks with Bonnor-Ebert spheres using 3D hydrodynamical simulations with self-gravity. Our simulations \edit1{indicated} that the cloud evolution \edit1{primarily} depends on two parameters: the shock speed and initial cloud radius. The stronger shock can compress the cloud more efficiently, and \edit1{when} the central region becomes gravitationally unstable, a shock triggers the cloud contraction. However, if it is \edit1{excessively} strong, it shreds the cloud more violently and the cloud is destroyed. From simple theoretical considerations, we derived the condition of triggered gravitational collapse\edit1{,} which \edit1{agreed} with the simulation results.
Introducing sink particles, we followed the further evolution after star formation.
Since stronger shocks tend to shred the cloud material more efficiently, the stronger the shock is, the smaller the final (asymptotic) masses of \edit1{the} stars formed (i.e., sink particles) become.
In addition, the shock accelerates the cloud, promoting mixing of shock-accelerated \edit1{interstellar medium} gas. As a result, the separation between the sink particles and the shocked cloud center and their relative speed increase over time. We also investigated the effect of cloud turbulence on shock-cloud interaction.
We \edit1{observed} that the cloud turbulence prevents rapid cloud contraction \edit1{;} thus\edit1{,} the turbulent cloud is destroyed more rapidly than the thermally-supported cloud. Therefore, the masses of stars formed become smaller. Our simulations can provide a general guide to the evolutionary process of dense cores and Bok globules impacted by shocks.

\end{abstract}



\section{Introduction}

\subsection{Triggered star formation}
Galactic star formation is \edit1{occasionally} classified \edit1{into} two main processes. One is “spontaneous star formation," in which the contraction of molecular clouds and subsequent star formation proceeds without any significant external disturbances. The other is “triggered star formation," in which external factors promote the compression of molecular clouds, which induces star formation. The external compression is driven by shock waves generated by stellar winds, supernova explosions, cloud-cloud collision\edit1{, etc}. For example, \citet{Elmegreen_1977} suggested that the compressed dense layer between \edit1{a} shock and an ionization front can be gravitationally unstable. Cloud-cloud collisions have been proposed as an important mechanism in the formation of high-mass stars (e.g., \citealp{Stolte_2008}; \citealp{Furukawa_2009}; \citealp{Wu_2017ApJ...835..137W}).
There is also \edit1{increasing} observational evidence for star formation triggered by supernovae, ionization fronts, cloud-cloud \edit1{collisions,} and other shocks in the \edit1{interstellar medium (ISM)} (e.g., \citealp{Yokogawa_2003}; \citealp{Ortega_2004}; \citealp{Hester_2005}; \citealp{Furukawa_2009}; \citealp{Kinoshita_2021}).

\edit1{Thus}, triggered star formation can occur when supersonic shocks sweep over clouds. The shock-cloud interaction is a fundamental and important physical phenomenon for star formation.
\edit1{Generally}, shock-cloud interactions are highly nonlinear hydrodynamic processes thus numerical simulation is \edit1{an effective} tool to understand \edit1{these} details.
Over the past decades, the shock-cloud interaction has been explored by many groups (e.g., \citealp{Klein_1994}; \citealp{Xu_1995}; \citealp{Boss_1995}; \citealp{Nakamura_2006}; \citealp{Pittard_2009}; \citealp{Banda_2018}). Most models assumed the two-dimensional (2D) axial-symmetric geometry.
The first three-dimensional (3D) simulations were \edit1{conducted} by \citet{Stone_1992}. These simulations demonstrated that cloud destruction occurs faster in 3D \edit1{because of} the rapid growth of hydrodynamic instabilities in 3D.  
Later, the effects of various physical factors were explored by including turbulence (e.g.,\citealp{Pittard_2009}), magnetic felds (e.g., \citealp{Fragile_2005}), and radiative cooling (e.g., \citealp{Mellema_2002}).
\edit1{Some numerical simulations of these studies} incorporated self-gravity. These studies revealed the details of triggered star formation by shocks\edit1{, particularly} in the context of the formation of our \edit1{solar} System triggered by a supernova shock (e.g., \citealp{Boss_1995}; \citealp{Foster_1996}; \citealp{Vanhala_1998}; \citealp{Vanhala_2002} \citealp{Boss_2008}; \citealp{Leao_2009};  \citet{Boss_2013}; \citealp{Li_2014}; \citealp{Falle_2017}). 
These previous studies demonstrated that isothermal shocks can trigger \edit1{the} gravitational collapse of stable clouds.
\citet{Boss_2013} \edit1{observed} that faster shocks destroy and disperse the cloud material before its collapse.
\edit1{In contrast}, \citet{Falle_2017} \edit1{demonstrated} that slower shocks cannot induce collapse. 
These previous results indicate that only intermediate-speed shocks can trigger gravitational collapse. 

Additionally, \edit1{actual} clouds and cores contain turbulent motions. 
Recently, \citet{Banda_2018} performed numerical experiments to investigate how cloud turbulence influences the shock-cloud evolution in the absence of self-gravity. They suggested that cloud turbulence \edit1{results in faster} cloud destruction and influences several ISM properties such as cloud porosity. Supersonic turbulence also enhances the acceleration of clouds \edit1{owing} to shocks.

In this \edit1{study}, we \edit1{used} 3D hydrodynamic numerical simulations to study shock-cloud \edit1{interactions}. We \edit1{considerd} an isothermal shock interacting with a Bonnor-Ebert sphere.
In these simulations, self-gravity and sink particles \edit1{were} included.
This initial setup \edit1{was} similar to those of \citet{Li_2014} and \citet{Falle_2017}. \citet{Li_2014} considered an application to the formation process of our Sun \edit1{through} a shock-cloud interaction.  Thus, their initial conditions \edit1{were} very specific. \citet{Falle_2017} investigated the early stages of shock-cloud \edit1{interactions} and derived the condition for \edit1{the} triggered \edit1{gravitational} collapse of stable Bonner-Ebert spheres. Here, we \edit1{considered} more general cases in a wider parameter range (from stable to unstable clouds). The inclusion of sink particles \edit1{enabled} us to follow \edit1{a} much longer evolution of the shock-cloud evolution. We also \edit1{examined} the effects of cloud turbulence on the shock-cloud interaction by including transonic cloud turbulence, which is often observed in the cloud cores in star-forming regions.

In the context of astronomical objects, we \edit1{envisioned} these simulations to represent the interaction of ISM shocks with dense cores and Bok globules in star-forming regions. Although the observed dense cores and Bok globules \edit1{were} not in perfect equilibrium, some \edit1{observations indicate} that the density structures of the dense cores and Bok globules \edit1{were} in reasonable agreement with those of Bonner-Ebert spheres (e.g.,  \citealp{Bacmann_2000}; \citealp{Kandori_2005}; \citealp{Alves_2001}). In some star-forming regions such as rho Oph and Orion A, the majority of the cores is likely to be pressure-confined (\citealp{2010ApJ...714..680M}; \citealp{2017ApJ...846..144K}).
Therefore, our initial setup of the simulations is expected to represent reasonable conditions \edit1{that} are \edit1{occurring} in the ISM.

This paper is organized as follows. In Section \ref{sec:Method}, we describe the numerical methods, initial conditions, and simulation models which we \edit1{employed} in this \edit1{study}. In Section \ref{sec:result}, we present the numerical results. 
\edit1{Thereafter}, we discuss interpretations of simulation results in Section \ref{sec:Discussion}. We derived a simple analytic condition for triggered gravitational collapse. Finally, we summarize our results and conclusion in Section \ref{sec:summary_1}.

\section{Numerical Methods}
\label{sec:Method}
\subsection{Basic Equations and Numerical Code}
In this \edit1{study}, the simulations \edit1{were} conducted using {\tt Enzo}\footnote{http://enzo-project.org}, \edit1{a magnetohydrodynamics} adaptive mesh refinement (AMR) code \citep{Bryan_2014}.
We used Version 2.6 of the Enzo code in a 3D Cartesian coordinate system $(x, y, z)$. 
We numerically \edit1{solved} the following hydrodynamic equations for mass, momentum\edit1{,} and energy conservation:

\begin{eqnarray}
\label{eq:mass_con}
 \frac{\partial \rho}{\partial t}+\nabla\cdot(\rho \bm{v})&=&0 , \\
\label{eq:momentum_con}
 \frac{\partial (\rho \bm{v})}{\partial t}+\nabla\cdot(\rho \bm{v}\otimes\bm{v})+\nabla{p}&=&-\rho\nabla\phi , \\
\end{eqnarray}
and 
\begin{eqnarray}
 \label{eq:energy_con}
 \frac{\partial}{\partial t} \left\{\rho(\frac{1}{2}v^{2}+\epsilon)\right\}+  \nabla\cdot \left\{ \rho \bm{v} \right(\frac{1}{2}v^{2}+h\left)\right\}&=&-\rho\bm{v}\cdot\nabla\phi,
\end{eqnarray}
where $\rho$ is the density,  $\bm{v}$ is the velocity, $p$ is the pressure, $\epsilon$ is the internal energy, $h=\epsilon+p/\rho$ is the enthalpy, \edit1{and} $\phi$ is the gravitational potential.
We \edit1{used} the ideal gas law:
\begin{eqnarray}
\label{eq:ideal_gas}
 p=(\gamma-1)\rho\epsilon.
\end{eqnarray}

$\phi$ \edit1{can be} determined by solving the following Poisson's equation\edit1{:}
\begin{eqnarray}
\label{eq:poisson}
 \nabla^{2}\phi=4\pi G(\rho+\rho_{\rm particle}),
\end{eqnarray}
where $G$ is the gravitational constant \edit1{and} $\rho_{\rm particle}$ is the density of sink particles assigned onto the finest grids by using a second-order cloud-in-cell interpolation technique \citep{1988csup.book.....H}. See Section \ref{subsec:AMR and sink particle condition} for the details of the sink particles.

 We \edit1{assumed} a mean molecular weight $\mu$= 2.3, and $\gamma$ \edit1{was} set to 1.00001 for an approximate isothermal assumption. In this \edit1{study}, we \edit1{considered} the purely hydrodynamic problem\edit1{,} ignoring radiative cooling, heating, magnetic fields, and thermal conduction.

The hydrodynamic equations \edit1{were} solved \edit1{using} a Runge Kutta second-order based \edit1{monotone upstream-centered scheme for conservation laws (MUSCL)} \citep{van_Leer_1977}. The Riemann problem \edit1{was} solved \edit1{using} the \edit1{Harten-Lax-van Leer (HLL) solver}, a two-wave, three-state solver with no resolution of contact waves, while 
the reconstruction method for the MUSCL solver \edit1{was a piecewise linear model (PLM)}.
\edit1{Refer to} \citet{Bryan_2014} for more details.

\subsection{Initial conditions}
\subsubsection{Problem setup}
\begin{figure}[hbtp]
\begin{center}
\includegraphics[width=60mm]{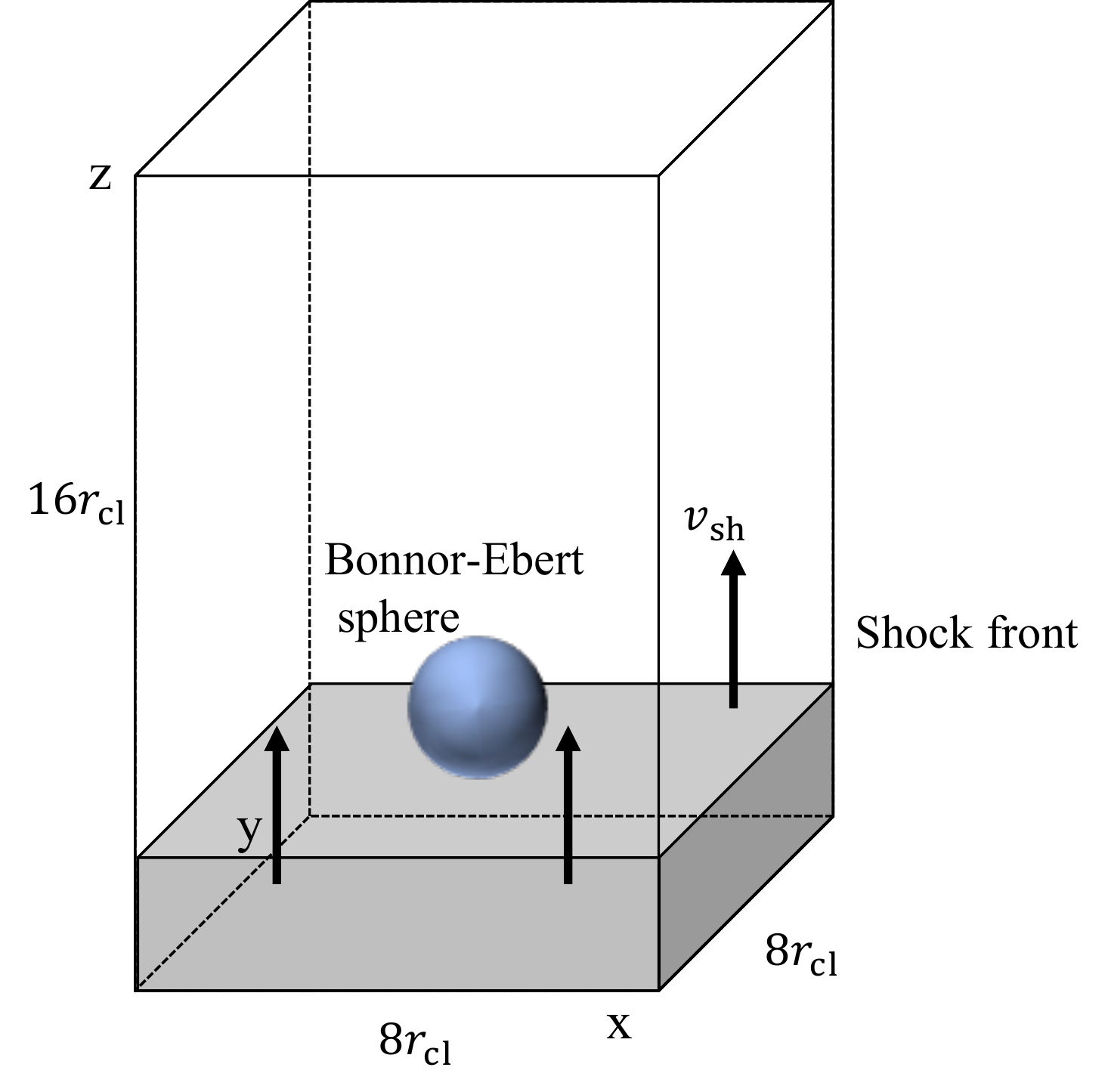}
\begin{flushleft}
\caption{\edit1{Schematic} of the simulation setup. The cloud is centered at \edit1{(0, 0, 0)}.
The shock is propagating upwards through the ISM with the velocity $v_{\rm sh}$. The shocked ISM is \edit1{indicated} in gray.}
\label{fig:setup}
\end{flushleft}
\end{center}
\end{figure}
In the simulations, we \edit1{considered} the interaction between a planar shock and a Bonnor-Ebert sphere initially at equilibrium. Observations have indicated that the density profiles of dense cores and Bok Globules such as B68 can be approximated by that of a Bonner-Ebert sphere (e.g., \citealp{Alves_2001}).
Figure \ref{fig:setup} shows the schematic of our simulation setup. The simulation domain \edit1{was} a rectangular prism. The domain length of each side \edit1{was} 8$r_{\rm cl}$, 8$r_{\rm cl}$, and 16$r_{\rm cl}$, for \edit1{the} $x$, $y$, and $z$ directions, respectively, where $r_{\rm cl}$ is the cloud radius. To follow the shock-cloud evolution, we set the side length in the z direction\edit1{, in which} the shock propagates\edit1{, to be} longer than the other two \edit1{sides} ($x$ and $y$). The computational domain \edit1{was} set to $-4r_{\rm cl}\leq x \leq4r_{\rm cl}$, $-4r_{\rm cl}\leq y \leq4r_{\rm cl}$, and $-4r_{\rm cl}\leq z \leq12r_{\rm cl}$. The initial Bonnor-Ebert sphere with a central density $\rho_{\rm c}$ \edit1{was} placed at the coordinate origin \edit1{(0, 0, 0)}, and outside the cloud, we set the ISM gas with the constant density of $\rho_{\rm ism}$.
The density contrast of the cloud to the ambient gas at the cloud surface \edit1{was} specified by $\chi_{\rm s}$\edit1{:}
\begin{equation}
\chi_{s}=\rho_{\rm s}/\rho_{\rm ism}
\end{equation}
where $\rho_{\rm s}$ is the
cloud surface density.
Initially, the cloud \edit1{had} a temperature of $T_{\rm cl}$ and \edit1{was} set in pressure equilibrium with the ambient gas, that is, 

\begin{eqnarray}
\label{eq:ambient}
  P_{0}=\rho_{\rm s}c_{\rm cl}^{2}=\rho_{\rm ism}c_{\rm ism}^{2},
\end{eqnarray}
where $P_{0}$ is the cloud surface pressure, $c_{\rm cl}$ is the sound speed in the cloud,  
$c_{\rm ism}$ is the sound speed of the ambient gas.
Therefore, the temperature of the external ISM \edit1{was} $T_{\rm ism}=\chi_{\rm s}T_{\rm cl}$. 

We set an inflow condition on the bottom plane ($x=-4r_{\rm cl}$) in Figure \ref{fig:setup} for a shock. For other boundaries, we \edit1{adopted} an outflow boundary condition. \edit1{When} the simulation \edit1{began}, a planar isothermal shock \edit1{moved} to the positive z-direction toward the cloud through the ISM with a Mach number \edit1{of}
\begin{eqnarray}
\label{eq:ismshock}
M_{\rm sh}=\frac{|v_{\rm sh}|}{c_{\rm ism}},
\end{eqnarray}
where $v_{\rm sh}$ is the shock velocity.

Table \ref{tab:fixed parameter} shows the initial simulation parameters.
We set the cloud central density and temperature as $\rho_{\rm c}=10^4 \rm ~cm^{-3}$ and $T=10~\rm K$, respectively\edit1{;} and $\chi_{\rm s}$ \edit1{was} fixed to $\chi_{\rm s}=100$, \edit1{thus} the ISM temperature \edit1{became} $1000$ K.

\begin{table}

\caption{Initial simulation parameters}
\normalsize
\centering
\begin{tabular}{llll}
\toprule
Parameter             & unit        & \edit1{Value}    & \edit1{Caption} \\ \hline
$T_{\rm cl}$              & (K)         & 10       & Cloud temperature.       \\
$T_{\rm ism}$             & (K)         & 10$^{3}$       & Preshocked amibient ISM temperature.       \\
$\rho_{\rm c}$            & (cm$^{-3}$) & 10$^{4}$ & Central number density of the cloud.        \\
$\chi_{\rm s}$ &             & 10$^{2}$ & Density contrast between the cloud surface and the ambient ISM gas.         \\
$c_{\rm cl}$ &   ($\rm km ~s^{-1}$)          & 0.19 & Sound speed in the cloud.      \\
$c_{\rm ism}$ &   ($\rm km ~s^{-1}$)          & 1.9 & Sound speed in the ambient gas.      \\
$\xi$                 &             & 2.04$-$14.22  & Dimensionless radius of the cloud.        \\
$M_{\rm sh}$              &             & 1.20$-$7.00        & Mach number of the propagating shock.         \\ \hline
$\xi_{\rm crit}$              &             & 6.45       & Critical dimensionless radius (see Appendix \ref{app:bonnor}).    \\ \hline

\end{tabular}
\vskip5pt
\begin{tablenotes}

\item Summary of the initial physical parameters. 
\end{tablenotes}

\label{tab:fixed parameter}

\end{table}

\subsubsection{Model Parameters}
\label{sec:Simulation models}
we \edit1{selected an initially stable cloud of $\xi$ = 3.22 ($P_{0}/P_{\rm crit}$ = 0.5) to be our fiducial model}. In this case, we \edit1{applied} shocks of $M_{\rm sh}$ = 1.20, 1.41, 3.15, 4.00, 4.46, 5.00, and 5.64. These Mach numbers correspond to postshock ambient gas pressure being $\sqrt{2}$, 2, 10, 16, 20, 25, and 32 times greater than those of preshocked gas.
In addition to fiducial models, we considered a number of different simulation models with different dimensionless radii and shock Mach numbers, as summarized in Figure \ref{fig:parameter}. Appendix \ref{tab:simulation model} \edit1{specifies} initial parameters of each model numerically. We \edit1{considered} both stable ($\xi<6.45$) and unstable initial clouds ($\xi\geq6.45$). In stable clouds, we set $\xi$ = 2.04, 2.48, 3.22, 4.05\edit1{,} and 6.45.  In our setting, $\xi=1$ \edit1{corresponded} to \edit1{approximately} 0.034 pc (cf. Equation (\ref{eq:bonnor_set})). These dimensionless radii \edit1{corresponded} to $P_{0}$ /$P_{\rm crit}$ = 0.125, 0.25, 0.5, 0.75, and 1.0, respectively. In $\xi>6.45$ cases, we \edit1{considered} $\xi$ = 8.52 and 14.22. These \edit1{corresponded} to $\rho_{\rm c}/\rho_{\rm s}$ = 28.08 (twice the density ratio of $\xi$ = 6.45) and 100.0, respectively.
Although the initial cloud \edit1{was} unstable in the $\xi > 6.45 $ cases, the free-fall time \edit1{was} still \edit1{significantly} longer than the shock arrival time and the cloud crushing time $t_{\rm cc}$(see Appendix \ref{Cloud crushing time}). We \edit1{also} discuss the cloud evolution triggered by shocks for $\xi>6.45$ cases.

\subsubsection{Turbulent cloud models}
\begin{figure}[hbtp]
\begin{center}
\includegraphics[width=100mm]{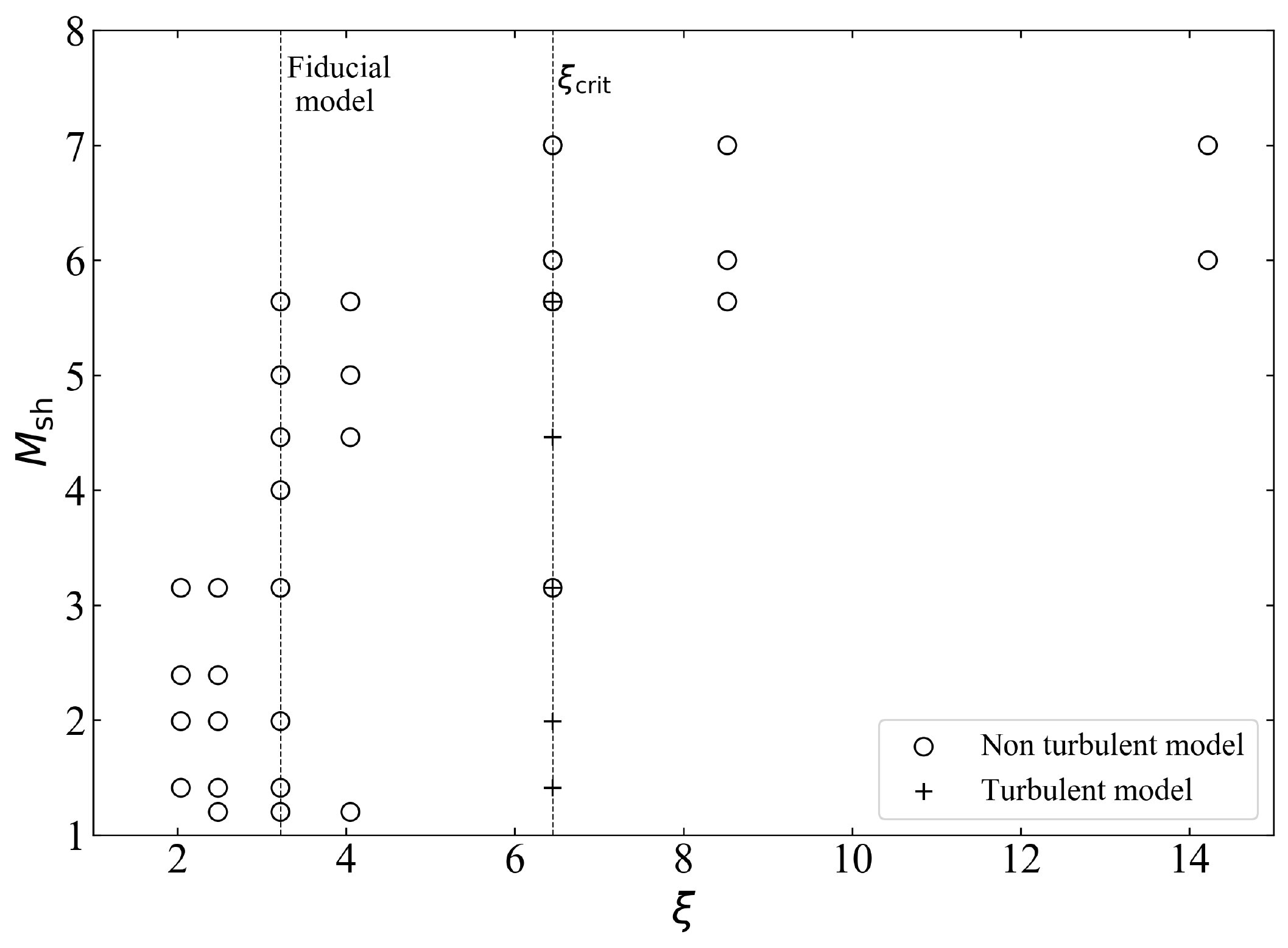}
\begin{flushleft}
\caption{\edit1{Values employed} for each model. The position of each circle indicates the initial condition, dimensionless radius $\xi$ of the initial cloud\edit1{,} and Mach number $M_{\rm sh}$ of the propagating shock. The position of each cross point \edit1{indicates} that of turbulent model. Appendix \ref{app:Employed Values for each model} also specifies initial parameters of each model numerically. }
\label{fig:parameter}
\end{flushleft}
\end{center}
\end{figure}

In addition to the aforementioned models, we \edit1{included} pure solenoidal turbulence to the fiducial Bonner-Ebert models where $\xi = \xi_{\rm crit}$.  A velocity power spectrum of 
$v_{\rm k}^{2}\propto k^{-4}$ \edit1{was} added to the gas, where $k$ is the wavenumber. This power spectrum \edit1{corresponded} to the expected spectrum given by Larson's law \citep{Larson_1981}. The initial amplitude of the turbulence \edit1{was} prescribed by the sonic Mach number 

\begin{equation}
\label{eq:tur_cloud}
M_{\rm tur}\equiv \frac{\sigma}{c_{\rm cl}}=1.0,
\end{equation}
where $\sigma$ is the velocity dispersion. 
In simulations, initially, we \edit1{evolved} the clouds without shocks for 0.48 Myr to form the turbulent density structures in the clouds. \edit1{Thereafter}, clouds with turbulent \edit1{densities} and velocity structures \edit1{interacted} with shocks. 
The parameters of these turbulent clouds are indicated in Figure \ref{fig:parameter} (see also Table \ref{tab:simulation model} , No.8).

\subsubsection{Color variable}
\label{subsec:color}
Similarly to previous simulations (e.g., \citealp{Xu_1995}), to follow the evolution of shocked clouds quantitatively, we \edit1{introduced} a Lagrangian tracer variable $C$\edit1{,} represented by
\begin{equation}
\label{eq:color}
\frac{(\partial \rho C)}{\partial t} + \nabla \cdot (\rho C {\bm v}) = 0.
\end{equation}

Initially, we \edit1{defined} $C=1$ for the \edit1{entire} cloud and $C=0$ for the ambient gas everywhere else. During the shock-cloud evolution, the cloud material \edit1{mixed} with the ambient gas, resulting in regions with $0<C<1$. 
We \edit1{used} the variable $C$ to quantify cloud mixing rate as 
\begin{equation}
\label{eq:mixing rate}
\frac{m_{\rm mix}}{m_{\rm cl}}=\frac{\int_{0.1<C<0.9} \rho C dV}{m_{\rm cl}},
\end{equation}
where $m_{\rm mix}$ is the total mass in the zones with $0.1<C<0.9$\edit1{,} and $m_{\rm cl}$ is the cloud mass expressed as 
\begin{equation}
\label{eq:cloud mass}
m_{\rm cl}=\int_{V} \rho C dV.
\end{equation}
Equation (\ref{eq:color}) describes the conservation law of $m_{\rm cl}$.

We also \edit1{defined} the cloud living rate as 
\begin{equation}
\label{eq:living rate}
\frac{m_{\rm live}}{m_{\rm cl}}=\frac{\int_{C>0.9} \rho C dV}{m_{\rm cl}},
\end{equation}
where $m_{\rm live}$ is the total mass in the zones with $C>0.9$.

To investigate the cloud motion, we \edit1{used} the mass-weighted  averaged cloud position in the z-direction\edit1{,} defined by 
\begin{equation}
\label{eq:z}
\langle z\rangle=\frac{1}{m_{\rm cl}}\int_{V}z\rho C dV.
\end{equation}

\subsubsection{AMR and sink particle condition}
\label{subsec:AMR and sink particle condition}
The simulation domain \edit1{had} a top level root grid of $256\times256\times512$ with additional levels of AMR. We \edit1{used} the following two criteria as the AMR condition to follow \edit1{closely} cloud evolution and collapse. In all models, the refinement \edit1{was permitted} until the finest resolution \edit1{reached} $\Delta x_{\rm min} \sim 2.0\times10^{-4}\rm ~pc$. 

One AMR criterion \edit1{was} based on $C$. If the local region \edit1{had} $C>0.1$, one level AMR \edit1{was} applied. \edit1{Using} this refinement, in all simulation models, the initial cloud radius \edit1{was} divided into 64 cells. A resolution of 64 zones per cloud radius \edit1{was} sufficient \edit1{to quantitatively follow} the shock-cloud evolution (e.g., \citealp{Klein_1994}; \citealp{Pittard_2016}).

The other AMR criterion \edit1{was} the Jeans criterion to prevent spurious numerical fragmentation. \citet{Truelove_1997} \edit1{suggested} that four cells per Jeans length are the minimum cells \edit1{required} to prevent spurious numerical fragmentation. We \edit1{adopted}
the limit \edit1{in which} the Jeans length \edit1{does} not fall below eight cells: $\Delta x < \lambda_{\rm j}/8$, where $\lambda_{\rm j}=\pi^{1/2}c/(G\rho)^{1/2}$ is the Jeans length. 
In our simulations, the refinement \edit1{continued} until the density \edit1{reached} the threshold value $\rho_{\rm th}=1.5\times 10^{4} \rho_{\rm c}$, where $\rho_{\rm c}$ is the initial cloud central density. That is, $\Delta x < \lambda_{\rm j}/8$ is \edit1{satisfied} as long as $\rho<\rho_{\rm th}$ (see Equation (29) in  \citealp{Bryan_2014}). 
Applying the sound speed $c_{\rm cl}$ in the cloud and $\rho_{\rm th}$, $\lambda_{\rm j}/8$ \edit1{was approximately} $\sim2.0\times10^{-4}\rm ~pc$.  Therefore, in all models the refinement \edit1{was permitted} until $\Delta x_{\rm min} \sim 2.0\times10^{-4}\rm ~pc$. When the local density \edit1{increased to} more than $\rho_{\rm th}$, instead of creating another AMR level, we \edit1{used} the sink particle technique. \edit1{With} this method, any excess mass in the cell above the $\rho_{\rm th}$  \edit1{was} transferred to the newly created point particle to avoid artificial fragmentation when the Jeans length \edit1{decreased} further during the collapse. By that time\edit1{,} it \edit1{was} clear that the cloud collapse \edit1{became} unstoppable. We set $\rho_{\rm c}$ to $10^{4} \rm ~cm^{-3}$ (see Section \ref{sec:Simulation models}) and $\rho_{\rm th}(=1.5\times10^{8} \rm cm^{-3})$  is three or four orders of magnitude higher than the density of general molecular cloud cores. For $\rho<\rho_{\rm th}$, the isothermal approximation \edit1{was} valid (for $\rho\gtrsim10^{6}\rho_{\rm c}=10^{10} \rm cm^{-3}$, the dense cores become adiabatic). \edit1{When formed}, these particles \edit1{moved} through the grid \edit1{via} gravitational interactions with the surrounding gas and other particles.




\section{Numerical Results}
\label{sec:result}

\ifpdf
    \graphicspath{{Chapter3/Figs/Raster/}{Chapter3/Figs/PDF/}{Chapter3/Figs/}}
\else
    \graphicspath{{Chapter3/Figs/Vector/}{Chapter3/Figs/}}
\fi
\label{sec:results}
Here, we will provide some numerical results. The clouds with $\xi$=3.22 ($P_{0}/P_{\rm crit}=0.5 $) \edit1{were} stable initially. First, as a representative example, we will discuss cloud evolution using the results of $\xi$=3.22 at different Mach numbers. In Appendix \ref{app:model_results}, we \edit1{provide} other dimensionless radii cases. Finally, we will \edit1{discuss} the results of turbulent cloud models. 

In all cases, we tracked shock-cloud evolution until 10$\%$ of the initial cloud mass exited the simulation box.

\subsection{Evolution of maximum density}
\label{sec:result_max_density}

\begin{figure*}
  
    \begin{tabular}{c}

      \begin{minipage}{0.5\hsize}
        \begin{center}
          \includegraphics[clip, width=8.0cm]{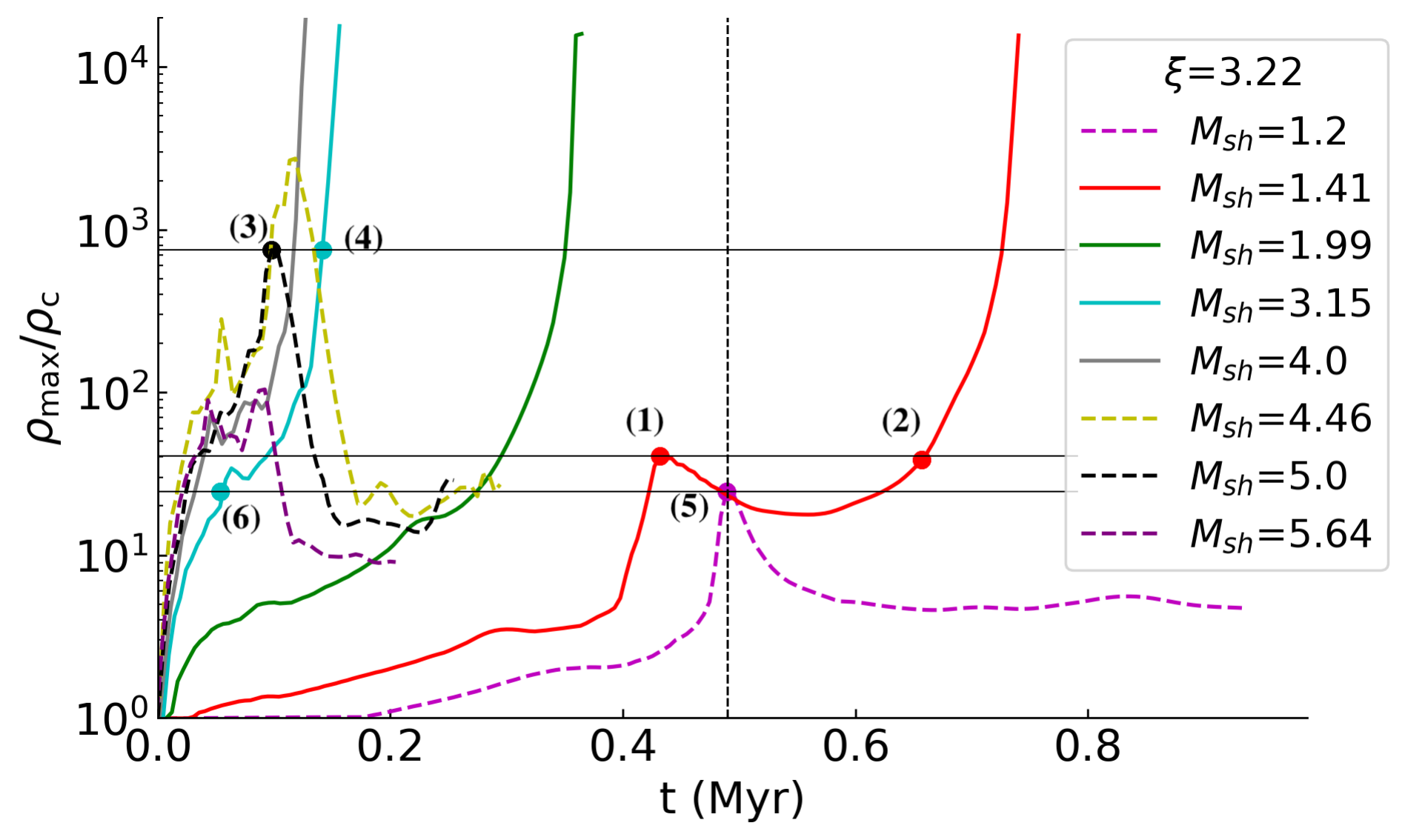}
        \end{center}
      \end{minipage}

      \begin{minipage}{0.5\hsize}
        \begin{center}
          \includegraphics[clip, width=8.0cm]{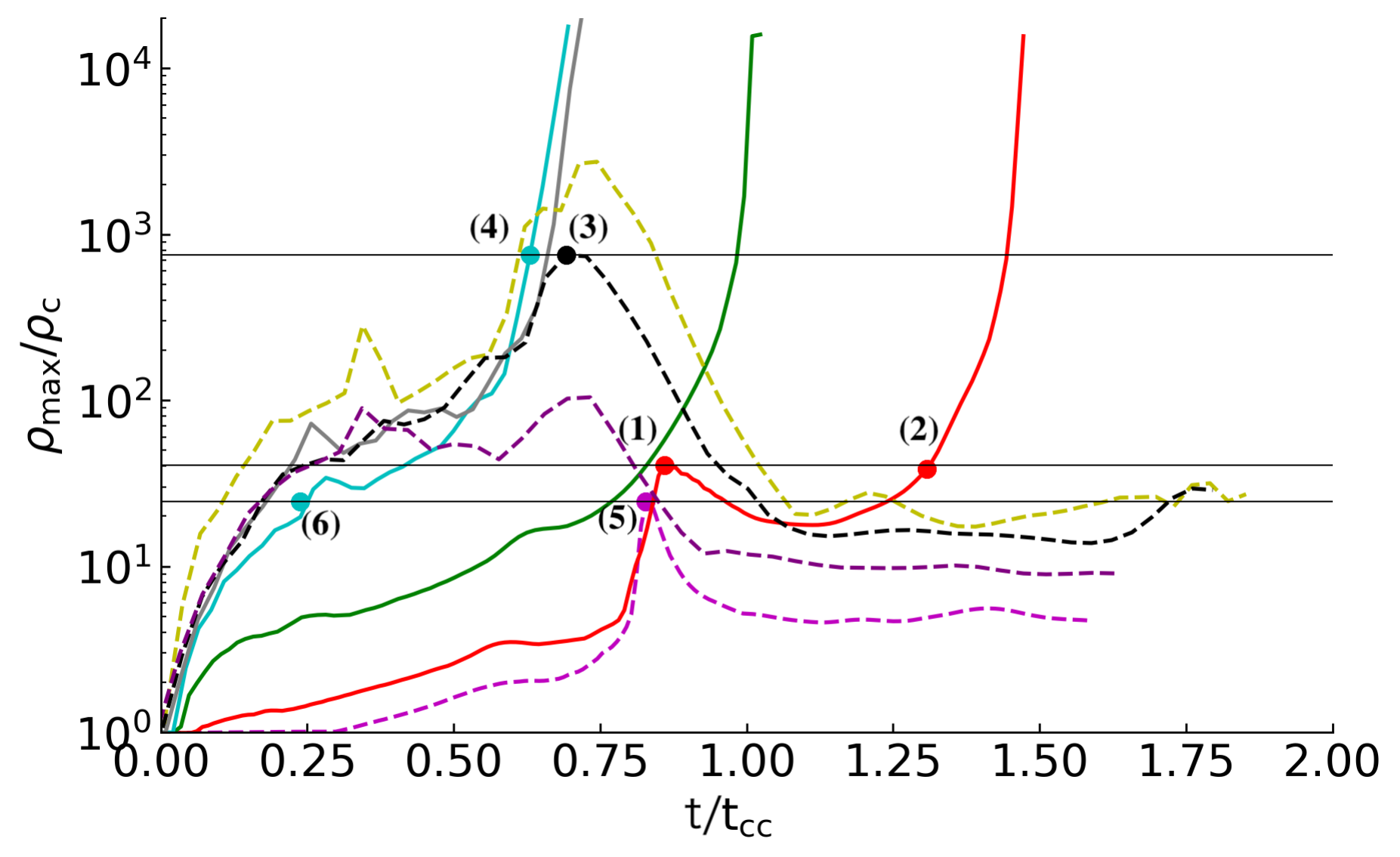}
        \end{center}
      \end{minipage}

       \end{tabular}

    \vskip5pt  
    \caption{\edit1{Left} panel: Time evolution of the maximum density normalized to the initial central cloud density $\rho_{\rm max}/\rho_{\rm c}$ for the \edit1{initially} \edit1{gravitational} stable sphere with $\xi$=3.22 in the range of the shock Mach numbers of $M_{\rm sh}=1.2\edit1{-}5.64$. \edit1{Right} panel: Same as the \edit1{left} panel, but as a function of the evolution time normalized to the cloud crushing time $t_{\rm cc}$. The solid lines \edit1{indicate} the models for which the gravitational collapses are triggered by the shocks. The cases \edit1{in which} the clouds do not collapse after shock passage \edit1{are indicated by} dashed lines. \edit1{The vertical} dashed line \edit1{indicates} the free fall time $t_{\rm ff}$ calculated using \edit1{the} mean density of initial \edit1{Bonnor-Ebert} sphere. \edit1{Point} (1) corresponds to the rebounding phase for $M_{\rm sh}$=1.41, while point (2) corresponds to the increasing density phase when the maximum density is equal to point (1) for $M_{\rm sh}$=1.41. Point (3) corresponds to the rebounding phase for $M_{\rm sh}$=5.00, while point (4) corresponds to the increasing density phase when the maximum density is equal to point (3) for $M_{\rm sh}$=3.15. Point (5) corresponds to the rebounding phase for $M_{\rm sh}$=1.20, while point (6) corresponds to the increasing density phase when the maximum density is equal to point (5) for $M_{\rm sh}$=3.15. }
    \label{fig:rho_max_3.22_3}

\end{figure*}

The maximum density is a good indicator of cloud stability.
Figure \ref{fig:rho_max_3.22_3} shows the maximum density normalized to the initial central cloud density as functions of time after a shock \edit1{arrived} at \edit1{a} cloud. Hereafter, in all figures, $t=0$ indicates the time when the shock front first \edit1{reached} the surface of the cloud. The vertical dashed line \edit1{indicates} one free fall time $t_{\rm ff}=(3\pi/32G\langle\rho_{\rm cl}\rangle)^{1/2}$, where $\langle\rho_{\rm cl}\rangle$ is the mean density of the initial cloud. The solid lines \edit1{indicates} the cases for which gravitational collapse is triggered by the shocks. The cases \edit1{in which} the clouds \edit1{did} not collapse after shock passage are \edit1{indicated by} dashed lines.
Below, we \edit1{term} these two cases “triggered-collapse case" and “no-collapse case" respectively.

One important feature is that only intermediate shocks of $M_{\rm sh}=1.41\edit1{-}4.00$ can trigger cloud contraction. For no-collapse cases, the maximum density \edit1{increased} at the beginning but \edit1{decreased} to lower values after rebounding. For example, 
for the model with a weak shock of $M_{\rm sh}=1.2$, the maximum density \edit1{reached} \edit1{$\sim 10 \rho_{\rm c}$} by one cloud free-fall time.  However, it gradually \edit1{decreased} with time. For the strong shock of $M_{\rm sh}=5.64$, the maximum density \edit1{reached} \edit1{$\sim 100 \rho_{\rm c}$} at \edit1{approximately} $t=0.1$ Myr. However, the maximum density \edit1{subsequently decreased} to \edit1{$\sim 10 \rho_{\rm c}$}.
Even for $M_{\rm sh} = 1.41$, the cloud \edit1{rebounded} once before collapsing at t = 0.43 Myr (point (1) in Figure \ref{fig:rho_max_3.22_3}).
In triggered-collapse cases, the rate of density-increase becomes \edit1{decreased} at the beginning. After some time, the rate of density-increase \edit1{increased} and the maximum density \edit1{reached} more than $\rho_{\rm th}$ and sink particles are formed.

\subsection{Density distribution}
As \edit1{discussed} in Section \ref{sec:result_max_density}, the density evolution depends on the Mach number $M_{\rm sh}$. Here, we present density distribution results for four cases: (1) a weak shock with $M_{\rm sh}$=1.41, in which the cloud slowly collapses, (2) an intermediate shock with $M_{\rm sh}$=3.15, in which the cloud rapidly collapses after shock passage, (3) a strong shock with $M_{\rm sh}$=5.0, in which the cloud does not collapse, and (4) a weakest shock with $M_{\rm sh}$=1.2, in which the cloud does not collapse. 

\subsubsection{Weak shock collapse ($M_{\rm sh}$=1.41)}
Figure \ref{fig:slice_05_1_41} shows the time evolution of the mass density distribution in the $(x,y)$ plane for $M_{\rm sh}=1.41$. In each panel, we \edit1{magnify} the maximum density point.
In addition to the times in Myr, the times in dimensionless units normalized to the cloud crushing time $t_{\rm cc}$ (see Appendix \ref{sec:timescales}) are also shown. We derived $t_{\rm cc}$ as $\chi \equiv \langle\rho_{\rm cl}\rangle/\rho_{\rm ism}$. The time evolution of the maximum density are also shown in these panels. 
As Figure \ref{fig:slice_05_1_41} (a) and (b) \edit1{show}, the cloud surface \edit1{was} compressed by the shock. The area surrounded by the white dotted lines indicates the compressed shocked layer formed by shocks propagating in the cloud. \edit1{At approximately} $t = 0.43$ Myr, the shock propagating in the cloud from downstream \edit1{collided} with that from upstream, and the density \edit1{increased dramatically} at this collision \edit1{point} (the area enclosed by the white circle in Figure \ref{fig:slice_05_1_41} (c)). This phase \edit1{corresponded} to the maximum density rebounding (point (1) in Figure \ref{fig:rho_max_3.22_3}). After this rebounding, the higher density region \edit1{was} not formed for a while. \edit1{At approximately} $t = 0.66$ Myr, the \edit1{entire} cloud \edit1{began} contracting again and the high density region \edit1{was} formed again. As a result, a sink particle \edit1{was} created at $t=0.74$ Myr. While the gas around the cloud center \edit1{accreted} on the sink particle, as shown in Figure \ref{fig:slice_05_1_41} (e), the gas at the cloud surface \edit1{was} shredded gradually by the hydrodynamic instability.

Figure \ref{fig:pdf-rho-v_1_41} shows the \edit1{mass per unit velocity and density interval}. As shown in Figure \ref{fig:pdf-rho-v_1_41} (a) and (b), \edit1{as} shocks \edit1{propagate} in the cloud, some shock-compressed gas \edit1{was} accelerated and \edit1{became} denser than the initial central density $\rho_{\rm c}$. Figure \ref{fig:pdf-rho-v_1_41} (c) corresponds to the rebounding phase (point (1) in Figure \ref{fig:rho_max_3.22_3}). Figure \ref{fig:pdf-rho-v_1_41} (d) corresponds to a density increasing phase when the maximum density \edit1{reached} the same value as Figure \ref{fig:pdf-rho-v_1_41} (c) again (point (2) in Figure \ref{fig:rho_max_3.22_3}). Comparing the \edit1{mass per unit velocity and density interval} of Figures \ref{fig:pdf-rho-v_1_41} (c) and (d), the mass fraction contained in the dense part \edit1{seemed} to be larger for Figure \ref{fig:pdf-rho-v_1_41} (d). By the time of Figure \ref{fig:pdf-rho-v_1_41} (c), the cloud \edit1{did} not contain sufficient mass to become gravitationally unstable. \edit1{However}, by the time of Figure \ref{fig:pdf-rho-v_1_41} (d), the central part of the shocked cloud \edit1{contained} more mass \edit1{such} that the gravitational collapse \edit1{was} initiated. After the rebound, the amount of dense gas \edit1{increased} as shown in Figure \ref{fig:pdf-rho-v_1_41} (e), \edit1{resulting in} collapse.

\subsubsection{Intermediate shock collapse ($M_{\rm sh}$=3.15) }
Figure \ref{fig:slice_05_3_15} is the same as Figure \ref{fig:slice_05_1_41} but for $M_{\rm sh}$=3.15. 
As in the $M_{\rm sh}$ = 1.41 case, the cloud surface \edit1{was} compressed by the shock and \edit1{the} shocked layer \edit1{progressed} to the cloud center. Unlike \edit1{for} $M_{\rm sh}$ = 1.41, there \edit1{was} no rebounding phase. A large high-density region \edit1{was} formed, \edit1{resulting in} direct gravitational contraction and the creation of a sink particle \edit1{at approximately} t = 0.16 Myr. After the sink particle creation, the cloud around the particle \edit1{was} stripped gradually. Eventually, it \edit1{had} a comet-like structure with the sink as the head and the stripped gas as the tail.
Figure \ref{fig:pdf-rho-v_3_15} shows the \edit1{mass per unit velocity and density interval} similar to Figure \ref{fig:pdf-rho-v_1_41}. \edit1{As} the shock \edit1{propagated} in the cloud, high density gas developed monotonically, \edit1{resulting in} gravitational collapse.

\subsubsection{Strong shock no-collapse ($M_{\rm sh}$=5.00) }
Figure \ref{fig:slice_05_5s} is the same as Figure \ref{fig:slice_05_1_41} but for $M_{\rm sh}$=5.00. As shown in Figure \ref{fig:slice_05_5s} (a) and (b), the cloud surface \edit1{was} compressed by the shock, and the \edit1{entire} cloud \edit1{contracted} but \edit1{did} not collapse. After density rebounding \edit1{at approximately} $t=0.08$ Myr, the cloud \edit1{was} destroyed and swept downstream of the shock mixing with the ambient gas.

Figure \ref{fig:pdf-rho-v_5} shows the \edit1{mass per unit velocity and density interval} for $M_{\rm sh}$=5.00. 
As shown in Figure \ref{fig:pdf-rho-v_5} (a) and (b), \edit1{the} high-density gas \edit1{initially} increased. Figure \ref{fig:pdf-rho-v_5} (b) corresponds to density rebounding point for the $M_{\rm sh}$=5.00 case (corresponding to the point (3) in Figure \ref{fig:rho_max_3.22_3}). After rebounding, \edit1{the} evolution towards the high-density side stopped. Gradually, gas accelerated and distributed on the low-density side. This \edit1{indicated} that the \edit1{entire} cloud \edit1{was} accelerated by the shock, and cloud \edit1{was} dispersed and \edit1{flowed} at a higher speed downstream.
In Figure \ref{fig:pdf-rho-v_5} (b), the central denser cloud material \edit1{had} velocity \edit1{of approximately} $2~\rm km/s$, and \edit1{the lower-density} outer regions \edit1{had} higher \edit1{velocities}. While Figure \ref{fig:pdf-rho-v_3_15} (c) (corresponding to (4) in Figure \ref{fig:rho_max_3.22_3}) has the same maximum density as Figure \ref{fig:pdf-rho-v_5} (b), the cloud material \edit1{had a} lower velocity ($\lesssim 2.0~ \rm km/s$). 
For $M_{\rm sh}$=5.00, a high-density region initially formed, but the cloud accelerated more and dispersed before the gravitational collapse.

\subsubsection{Weak shock no-collapse ($M_{\rm sh}$=1.20) }
\label{subsec:1.20 case}
\begin{figure}[hbtp]
\begin{center}
\includegraphics[width=150mm]{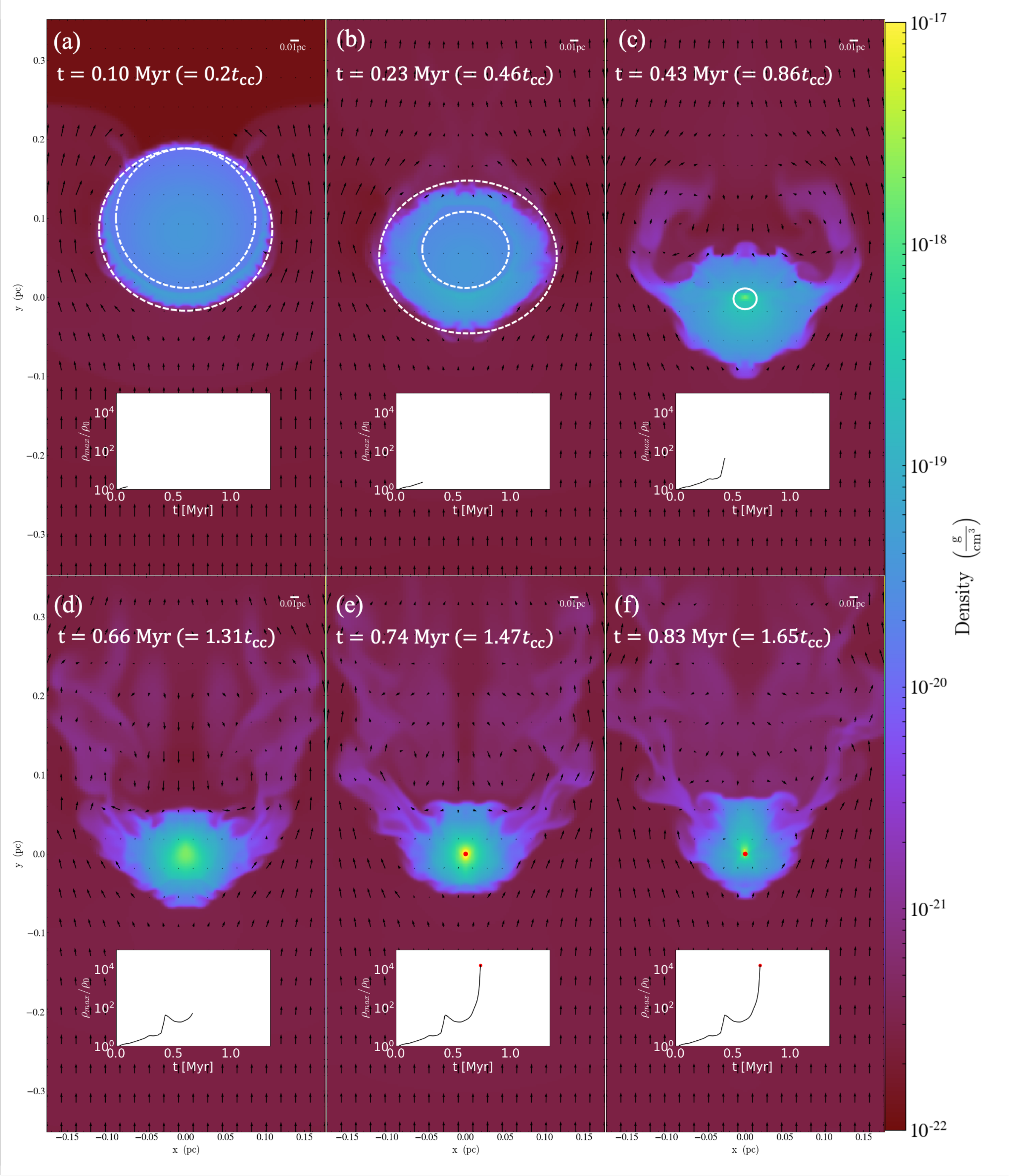}
\begin{flushleft}
\caption{Enlarged slice of the density distribution in the $(x,y)$ plane centered on the maximum density point, and velocity arrows for $\xi=3.22$ and $M_{\rm sh}=1.41$. The region \edit1{in which the maximum density occurred is shown} at the center of the figure. Red points \edit1{indicate} the sink particles. \edit1{The density} evolution as \edit1{in} Figure \ref{fig:rho_max_3.22_3} are \edit1{indicated} on \edit1{the} figures. The area surrounded by the white dotted line in panel (a) and (b) indicate the compressed shocked layer formed by shocks propagating in clouds.
    The area enclosed by the white circle in panel (c) indicates the part where upstream and downstream shocks \edit1{collided}.}
    \label{fig:slice_05_1_41}
\end{flushleft}
\end{center}
\end{figure}

\begin{figure}[hbtp]
\begin{center}
\includegraphics[width=150mm]{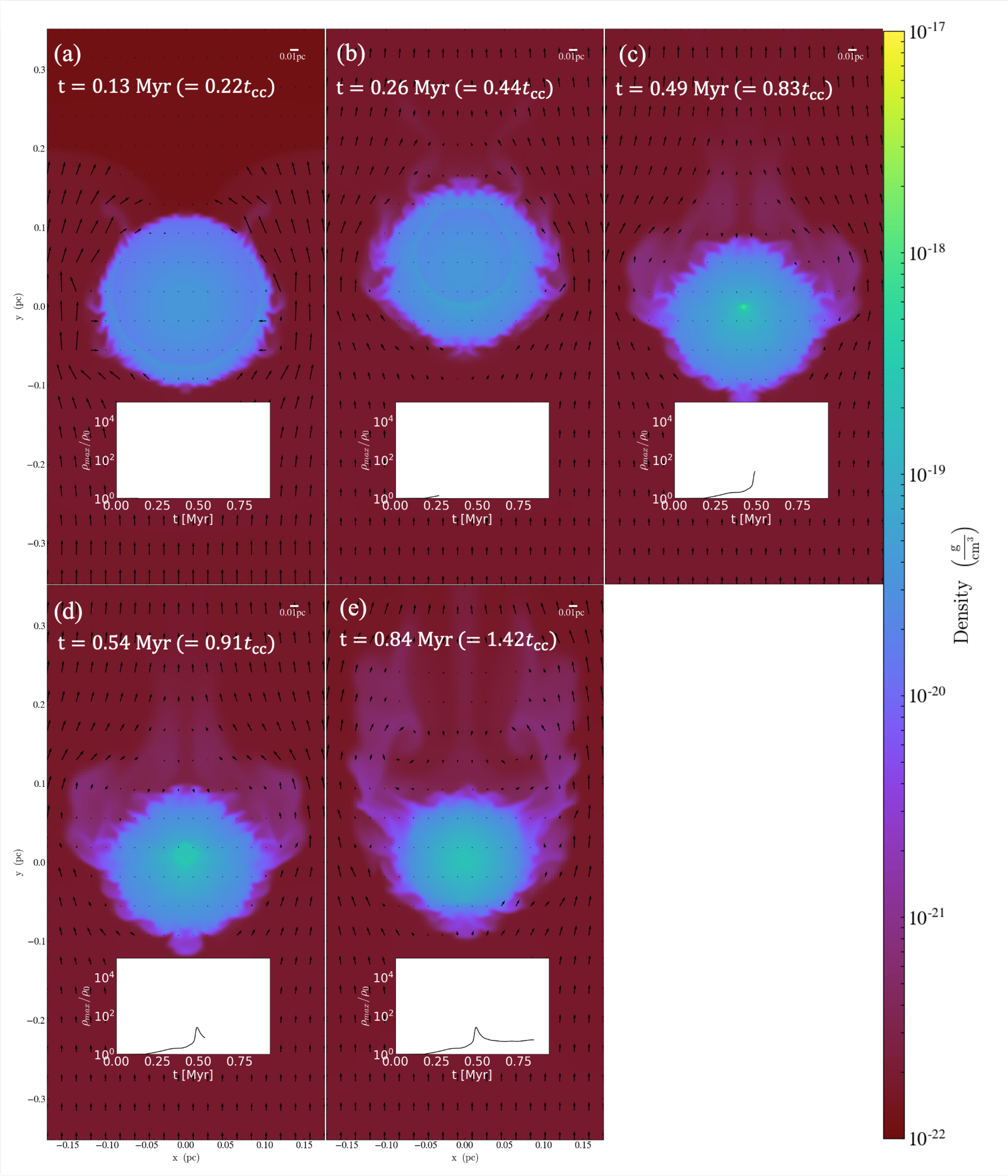}
\begin{flushleft}
\caption{As \edit1{Figure} \ref{fig:slice_05_1_41} for $\xi=3.22$ and $M_{sh}=1.20$ model.}
    \label{fig:slice_05_1_2s}
\end{flushleft}
\end{center}
\end{figure}

Figure \ref{fig:slice_05_1_2s} is the same as Figure \ref{fig:slice_05_1_41} but for $M_{\rm sh}$=1.20. 
\edit1{Similar to} other $M_{\rm sh}$ cases, as shown in Figure \ref{fig:slice_05_1_2s} (a) and (b), the cloud surface \edit1{was} compressed by the shock, and the compressed shocked layer \edit1{advanced} to the cloud center. The maximum density \edit1{rebounded at approximately} t=0.49 Myr and \edit1{decreased} gradually. 

Figure \ref{fig:pdf-rho-v_1_2} shows the \edit1{mass per unit velocity and density interval} for $M_{\rm sh}$=1.20.
Throughout \edit1{the} simulation times, the high-density regions \edit1{were} not as large as \edit1{those in} higher $M_{\rm sh}$ examples.
Comparing Figure \ref{fig:pdf-rho-v_1_2} (c) and Figure \ref{fig:pdf-rho-v_3_15} (b), which have a common maximum density (points (5) and (6) in Figure \ref{fig:rho_max_3.22_3}), for Figure \ref{fig:pdf-rho-v_1_2} (c), \edit1{the} cloud mass has \edit1{a lower} distribution on the high density side compared \edit1{with} Figure \ref{fig:pdf-rho-v_3_15} (b) \edit1{for} $M_{\rm sh}$=3.15. That is, \edit1{at} $ M_{\rm sh} $ = 1.20, a denser gas region \edit1{was} not sufficiently formed compared \edit1{with} the case of successful collapse. For $M_{\rm sh}$=1.20, after rebounding, the surface of the cloud \edit1{was gradually} stripped toward the downstream shock.

\begin{figure}[hbtp]
\begin{center}
\includegraphics[width=150mm]{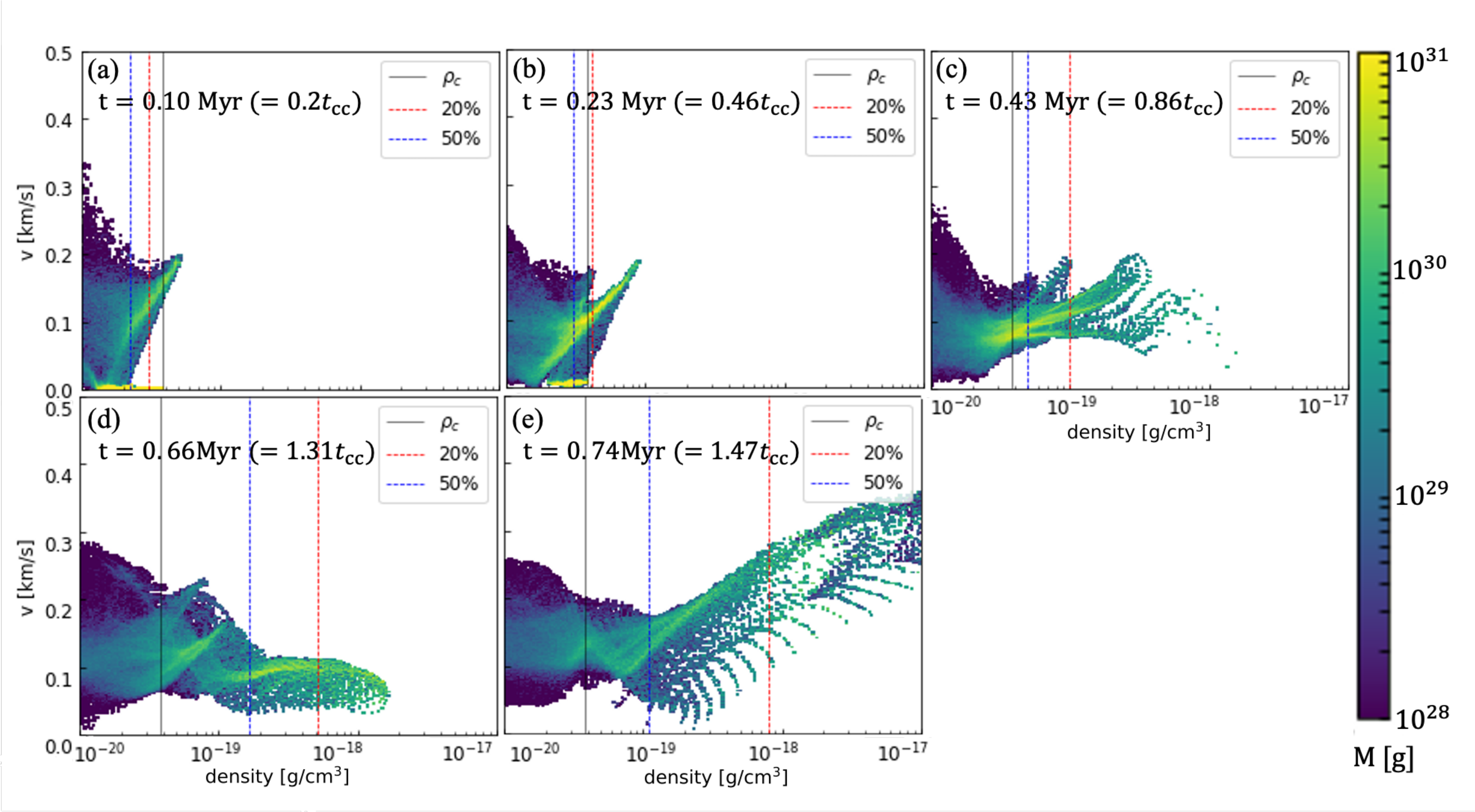}
\begin{flushleft}
\caption{\edit1{Mass per unit density and velocity(i.e. $\sqrt{v_{x}^{2}+v_{x}^{2}+v_{z}^{2}}$) interval} for the $\xi=3.22$ and $M_{\rm sh}$=1.41 model. Only gas of C$>$0.1 region \edit1{is} displayed on this figure. \edit1{The black} vertical line indicates the initial cloud central density $\rho_{\rm c}$. \edit1{The blue} and red two dashed lines show that the total mass of the gas located to the right of these lines \edit1{accounted} for 50$\%$ and 20$\%$ of the initial cloud mass, respectively.}
    \label{fig:pdf-rho-v_1_41}
\end{flushleft}
\end{center}
\end{figure}

\begin{figure}[hbtp]
\begin{center}
\includegraphics[width=150mm]{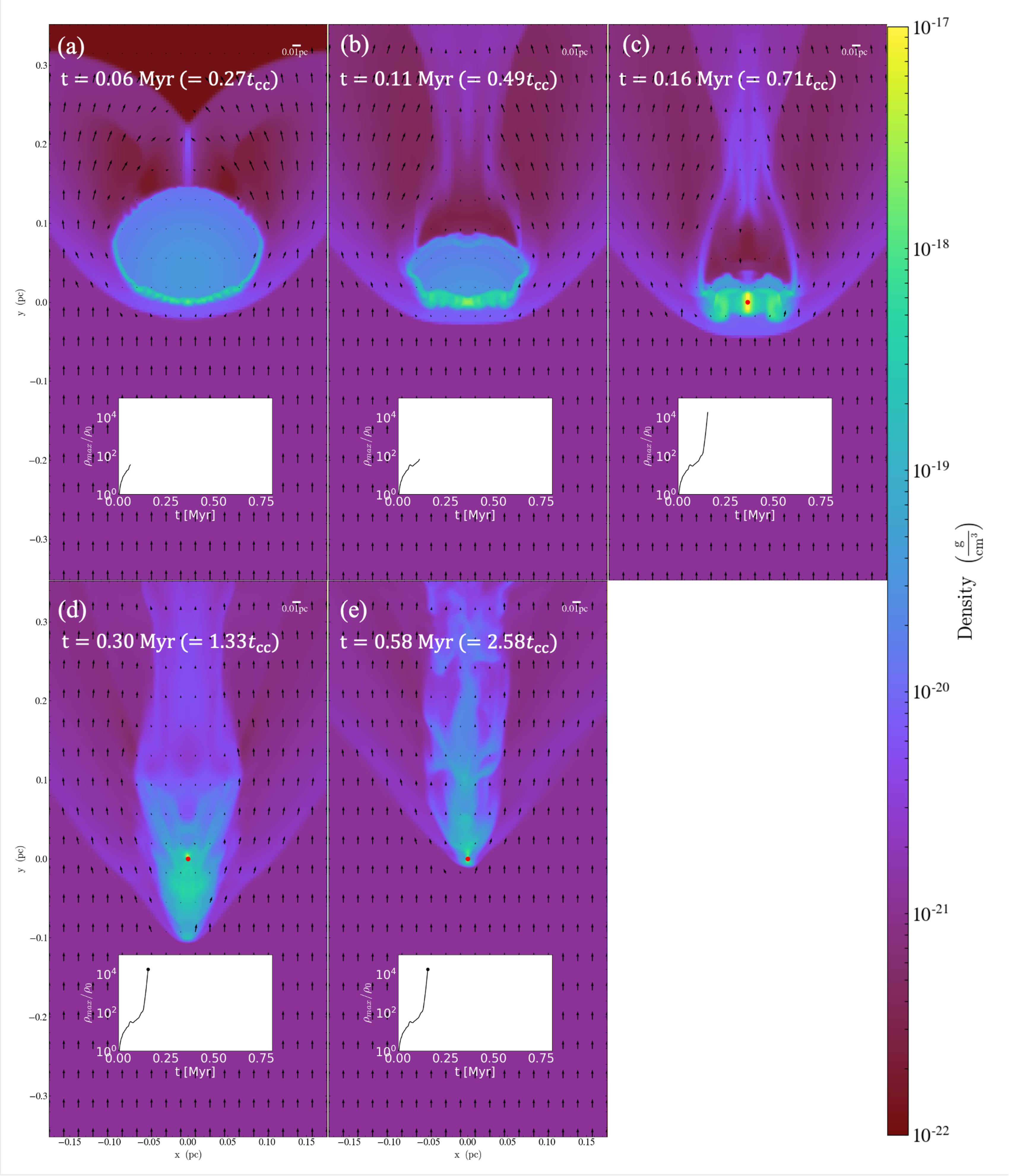}
\begin{flushleft}
\caption{As Figure \ref{fig:pdf-rho-v_1_41} for the $\xi=3.22$ and $M_{\rm sh}$=3.15 case.}
    \label{fig:slice_05_3_15}
\end{flushleft}
\end{center}
\end{figure}

\begin{figure}[hbtp]
\begin{center}
\includegraphics[width=150mm]{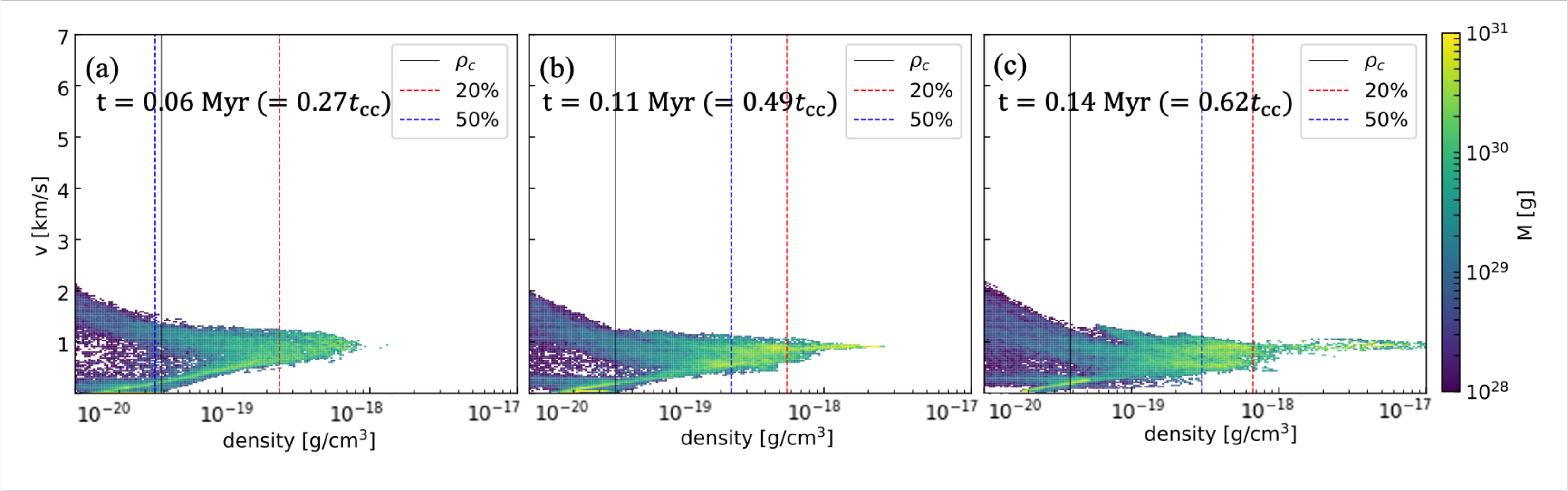}
\begin{flushleft}
\caption{As Figure \ref{fig:pdf-rho-v_1_41} for the $\xi=3.22$ and $M_{\rm sh}$=3.15 case.}
    \label{fig:pdf-rho-v_3_15}
\end{flushleft}
\end{center}
\end{figure}

\begin{figure}[hbtp]
\begin{center}
\includegraphics[width=150mm]{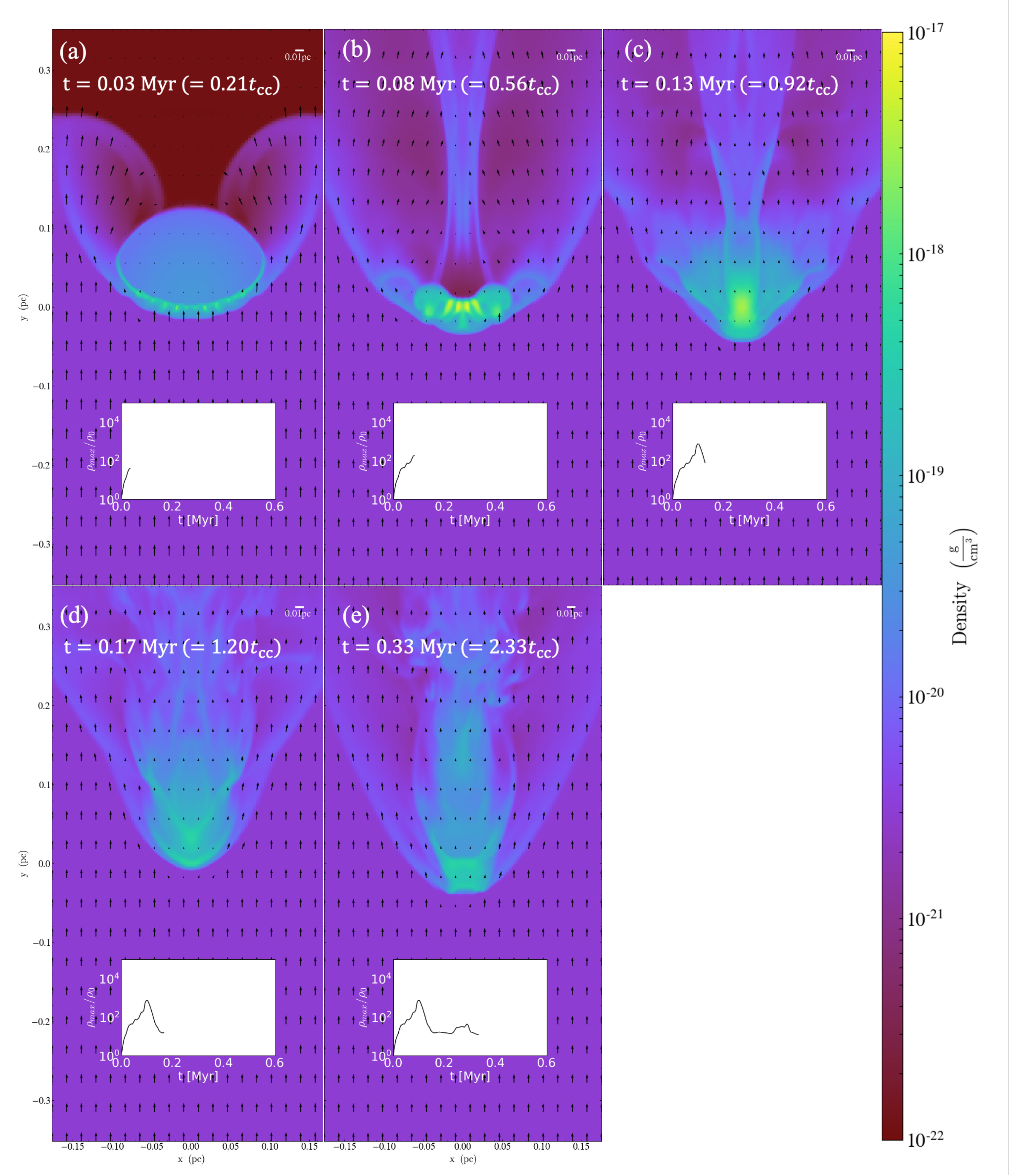}
\begin{flushleft}
\caption{As Figure \ref{fig:pdf-rho-v_1_41} for $\xi=3.22$ and $M_{\rm sh}$=5.0.}
    \label{fig:slice_05_5s}
\end{flushleft}
\end{center}
\end{figure}

\begin{figure}[hbtp]
\begin{center}
\includegraphics[width=150mm]{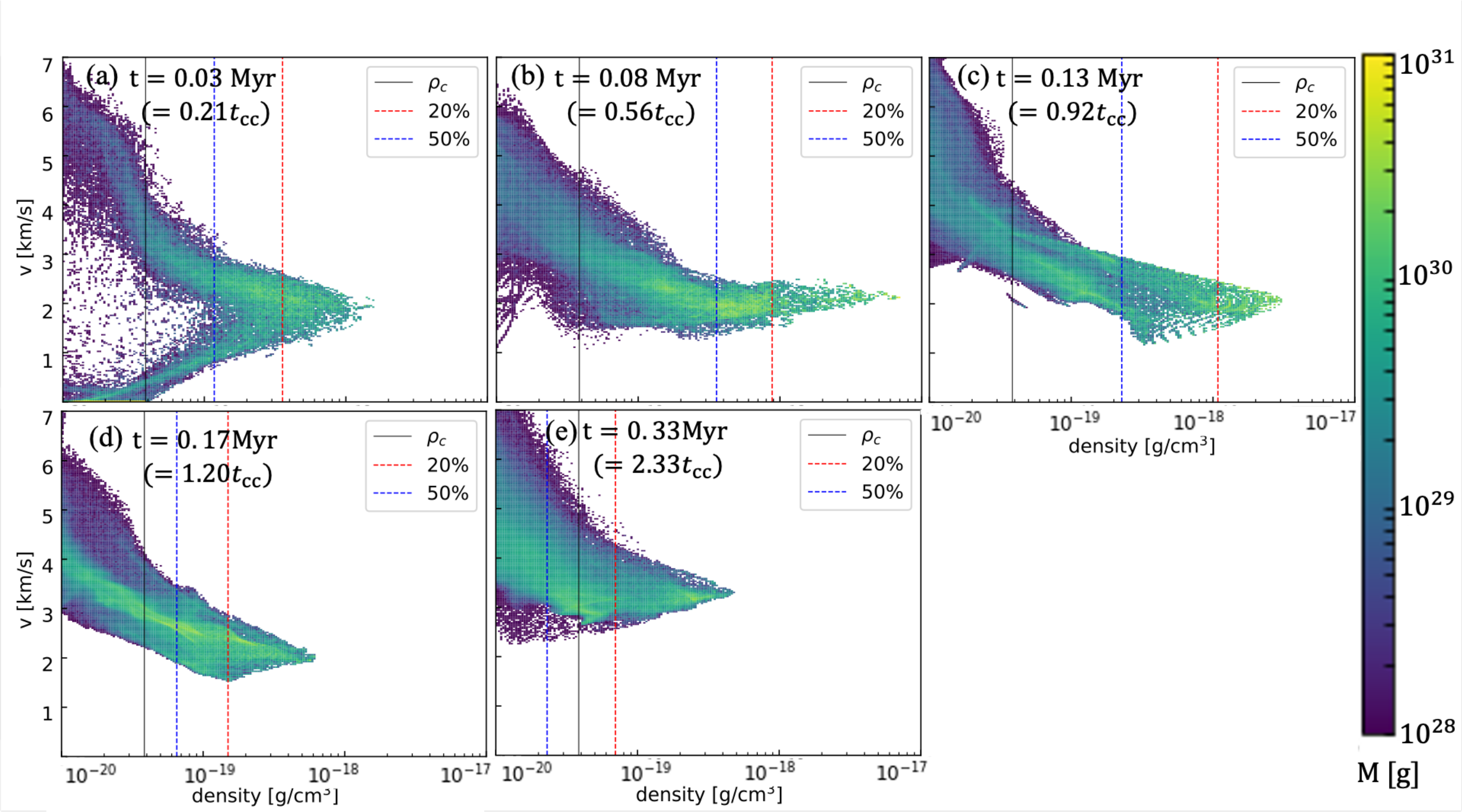}
\begin{flushleft}
\caption{Same as Figure \ref{fig:pdf-rho-v_1_41} for the $\xi=3.22$ and $M_{\rm sh}$=5.00 model.}
    \label{fig:pdf-rho-v_5}
\end{flushleft}
\end{center}
\end{figure}

\begin{figure}[hbtp]
\begin{center}
\includegraphics[width=150mm]{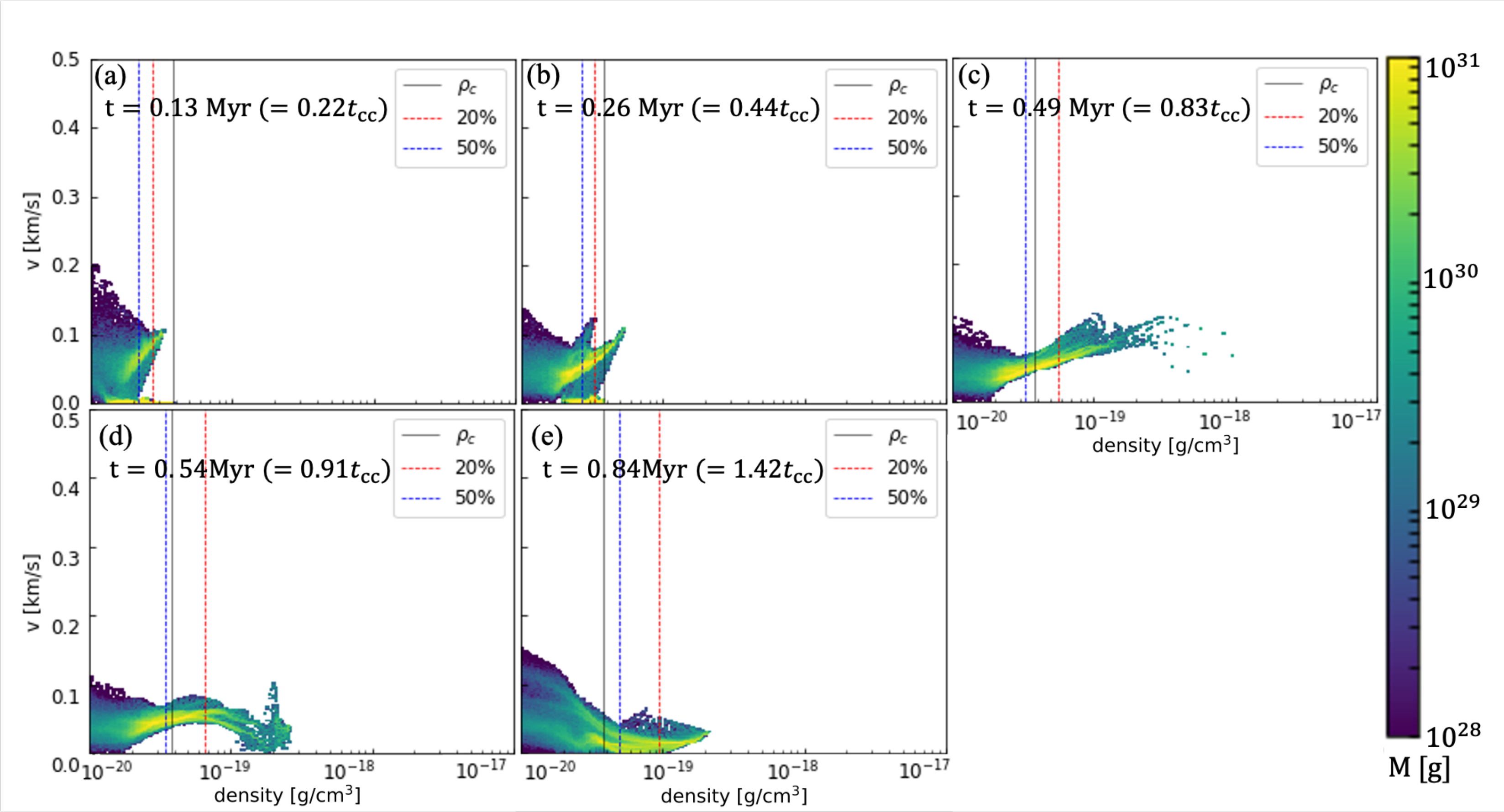}
\begin{flushleft}
\caption{As Figure \ref{fig:pdf-rho-v_1_41} for the $\xi=3.22$ and $M_{sh}$=1.20 model.}
    \label{fig:pdf-rho-v_1_2}
\end{flushleft}
\end{center}
\end{figure}

\subsection{Mixing and living rates}

\begin{figure}[hbtp]
\begin{center}
\includegraphics[width=100mm]{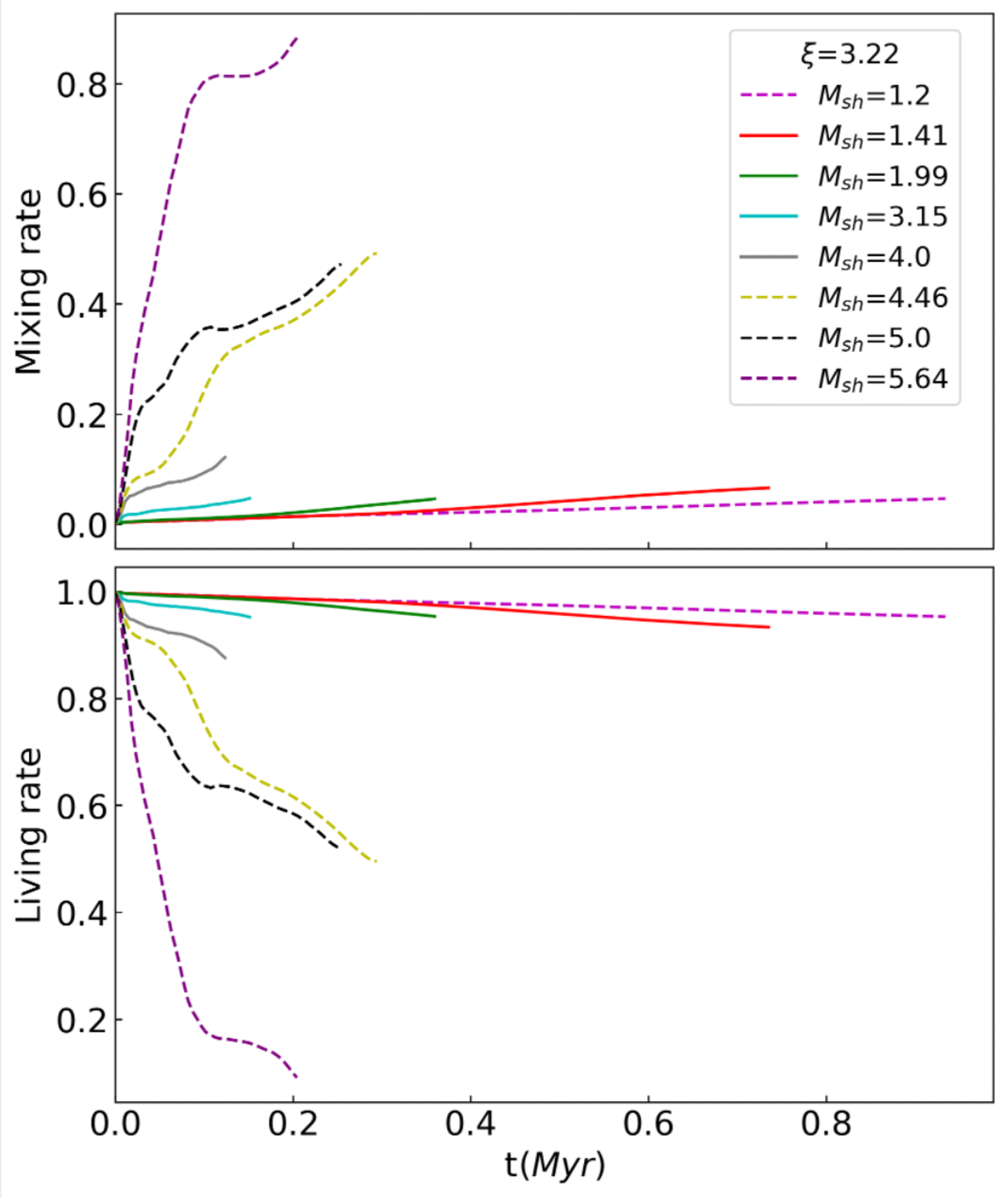}
\begin{flushleft}
\caption{Mixing rate (top) and living rate (bottom) defined in \edit1{Equations} (\ref{eq:mixing rate}) and (\ref{eq:living rate}) as functions of time after shocks \edit1{arrived at the} cloud for $\xi=3.22$. For triggered-collapse cases, the transition is \edit1{indicated} by a solid line. After the time at which \edit1{the} sink particle \edit1{was} introduced, transition is not shown. For no-collapse cases, the transition is represented by a dashed line.}
\label{fig:mix_live_r_3.22_3}
\end{flushleft}
\end{center}
\end{figure}

The top panel in Figure \ref{fig:mix_live_r_3.22_3} shows the time evolution of the mixing rate defined in Equation (\ref{eq:mixing rate}).
The larger the shock Mach number is, the faster the mixing rate \edit1{increases}. For $M_{\rm sh}$ = 4.46$-$5.64, at \edit1{approximately} $t=0.1$ Myr when the maximum density rebound \edit1{occurred}, the mixing rate \edit1{exceeded} $0.2-0.3$, i.e.\edit1{,}  $20\%-30\%$ cloud gas mixed with the ambient gas.
\edit1{In contrast}, for lower $M_{\rm sh}$  cases, the mixing rate \edit1{did} not become higher than $0.2-0.3$ even when the cloud \edit1{rebounded}.
A similar trend is indicated in the bottom panel in Figure \ref{fig:mix_live_r_3.22_3}, which shows the time evolution of living rate defined in Equation (\ref{eq:living rate}). For \edit1{4.46$-$5.64} cases, \edit1{during} rebounding, \edit1{the} living rates \edit1{were lower} $\sim$ 0.8, \edit1{whereas}, for lower $M_{\rm sh}$ cases, living rates \edit1{were higher} than $\sim$ 0.8 when maximum densities \edit1{increased dramatically}. During the cloud contraction, more gas \edit1{was} removed in the larger $M_{\rm sh}$ cases than in the lower $M_{\rm sh}$ \edit1{ones}.

\subsection{Evolution of sink particles}
\label{subsub:Evolution of sink particles}

\begin{figure}[hbtp]
\begin{center}
\includegraphics[width=100mm]{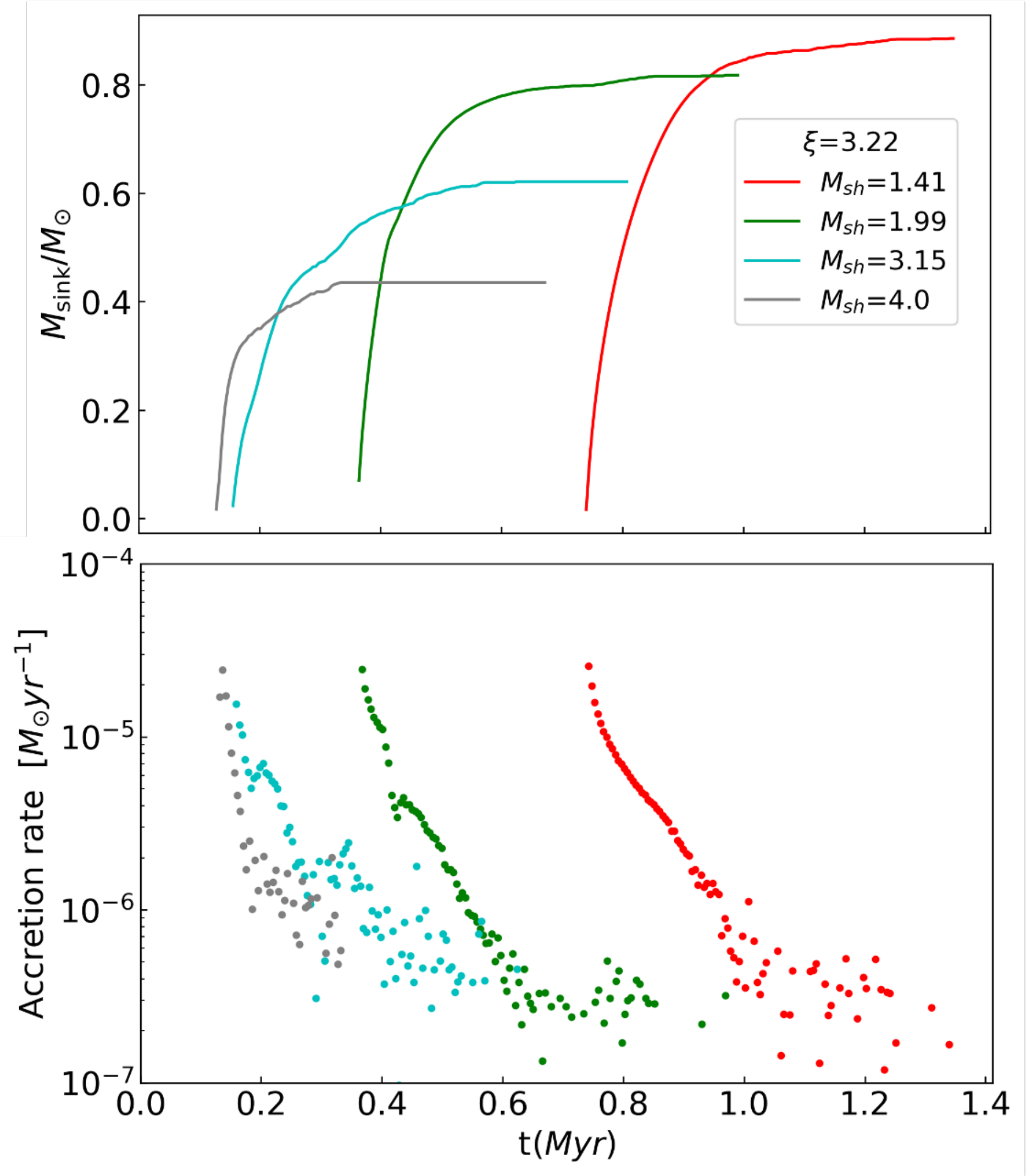}
\begin{flushleft}
\caption{Sink particle mass normalized by \edit1{the} initial cloud mass (top) and mass accretion rate of sink particles (bottom) for the $\xi=3.22$.}
\label{fig:particle_3.22_3}
\end{flushleft}
\end{center}
\end{figure}

For triggered-collapse cases, the evolutions of mass and mass accretion rates of sink particles are shown in Figure \ref{fig:particle_3.22_3}.
\edit1{The accretion} rates gradually \edit1{decreased} by a few orders of magnitude and the mass of sink particles \edit1{converged} asymptotically. The top panel in Figure \ref{fig:particle_3.22_3} indicates that the higher the Mach number \edit1{was}, the lower the asymptotic sink particle mass \edit1{became}.
For $M_{\rm sh}$=1.41, 1.99, 3.15, and 4.00, periods when mass accretion rates \edit1{were} above $10^{-6} M_{\odot} \rm yr^{-1}$, \edit1{were} 0.22, 0.18, 0.13, and 0.07 Myr, respectively. The higher the Mach number \edit1{was}, the shorter the time with \edit1{a} high accretion rate \edit1{was}. This trend of accretion timescale would affect the asymptotic sink particle mass.

\subsection{Initially turbulent cloud cases}
\label{sec:Initially turbulent cloud cases}
\begin{figure}[hbtp]
\begin{center}
\includegraphics[width=100mm]{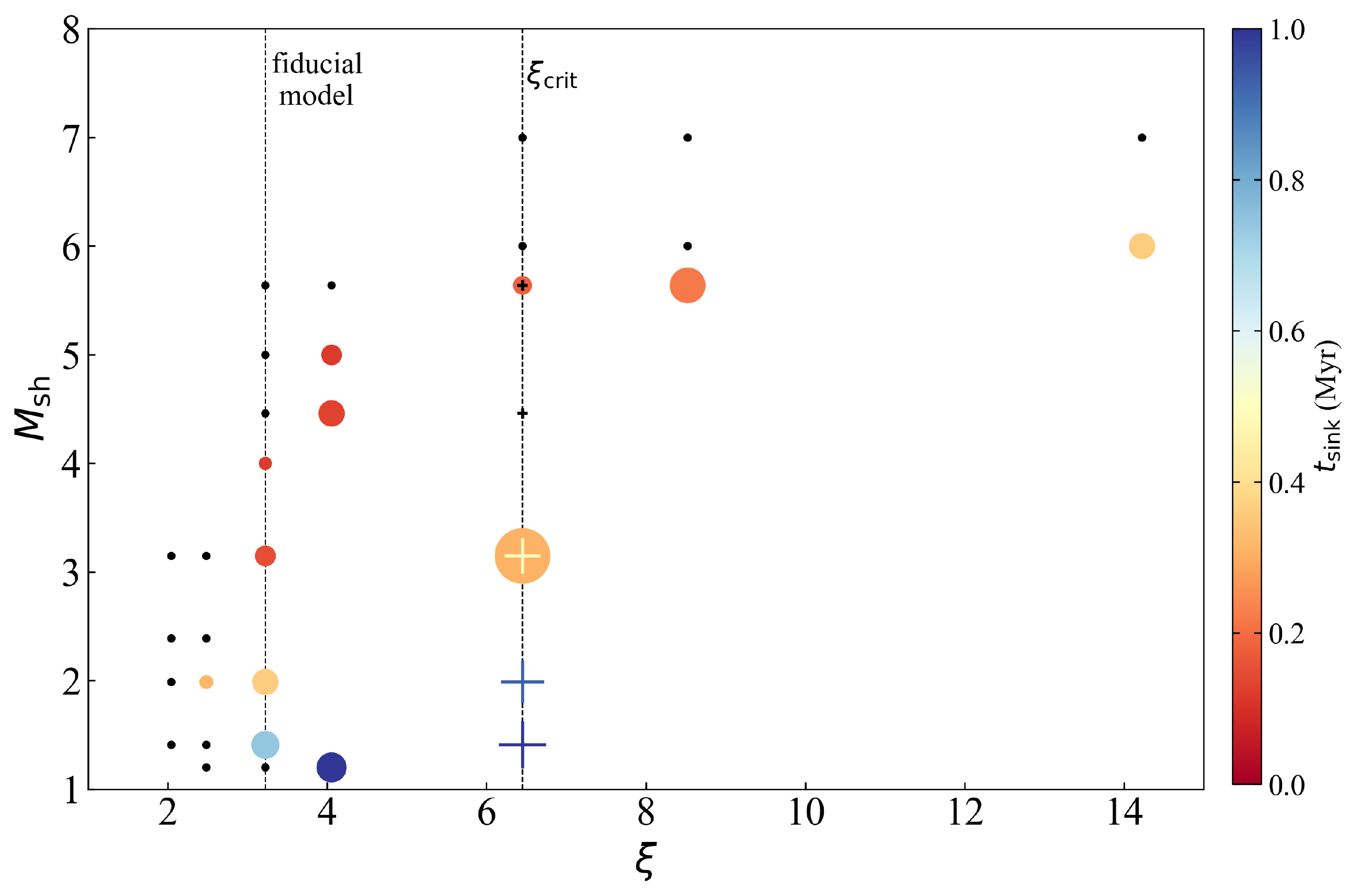}
\begin{flushleft}
\caption{\edit1{Formation} time and the final mass of sink particles in the initial Mach number vs. cloud radius plane. The position of each scattering point indicates the initial condition as \edit1{in} Figure \ref{fig:parameter}. The colors of points correspond to the time interval from when the shock wave reaches the cloud until the sink particles are introduced. For non-collapse cases, the point color is black. Sizes of points correspond to the asymptotic mass of sink particles. Appendix \ref{app:Simulation results} also specifies results of simulations numerically.}
\label{fig:formed_result}
\end{flushleft}
\end{center}
\end{figure}

\begin{figure*}
  
    \begin{tabular}{c}

      \begin{minipage}{0.5\hsize}
        \begin{center}
          \includegraphics[clip, width=6.7cm]{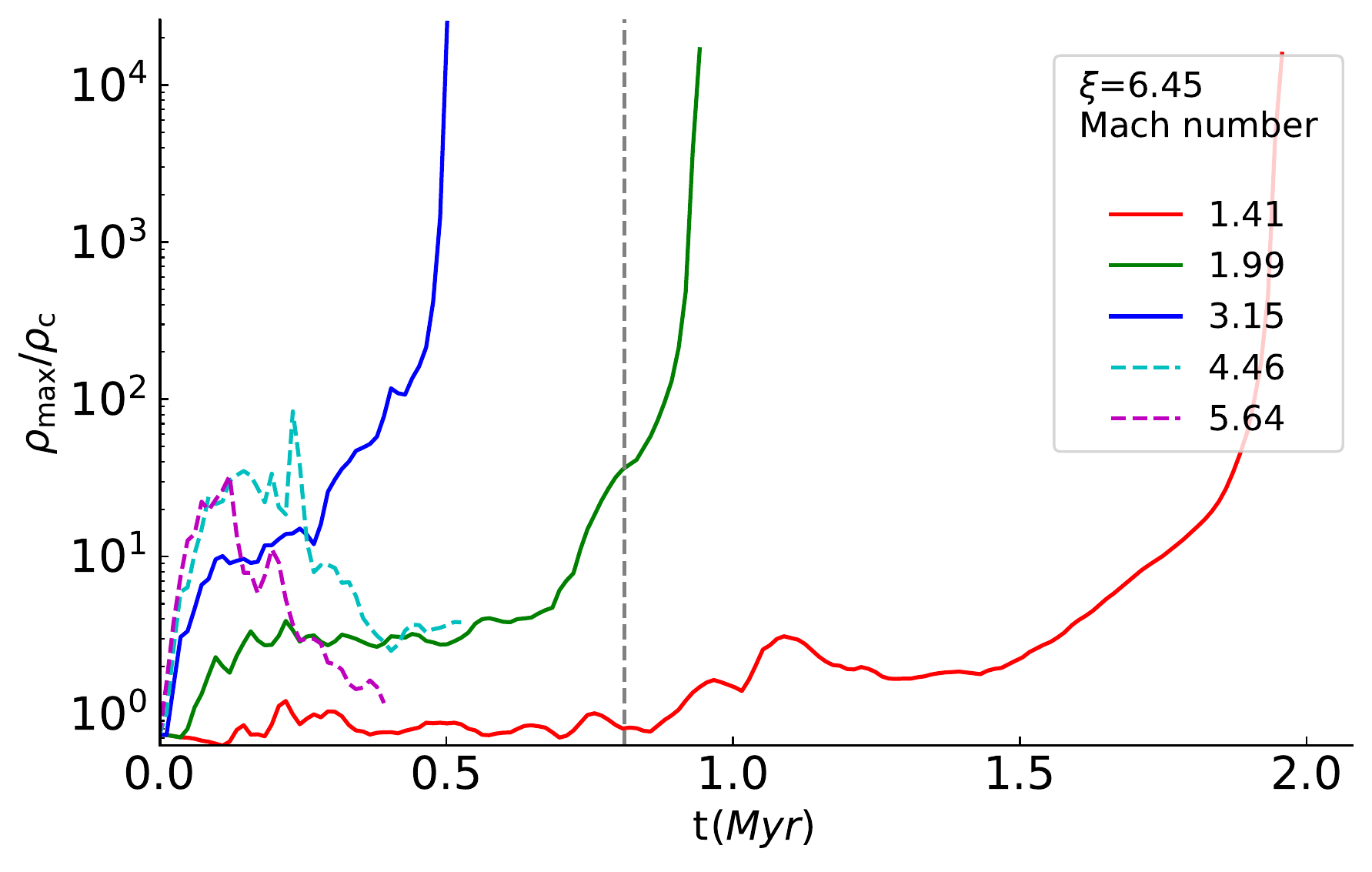}
          \hspace{2.0cm} (a) $\rho_{\rm max}/\rho_{\rm c}$
        \end{center}
      \end{minipage}

      \begin{minipage}{0.5\hsize}
        \begin{center}
          \includegraphics[clip, width=6.7cm]{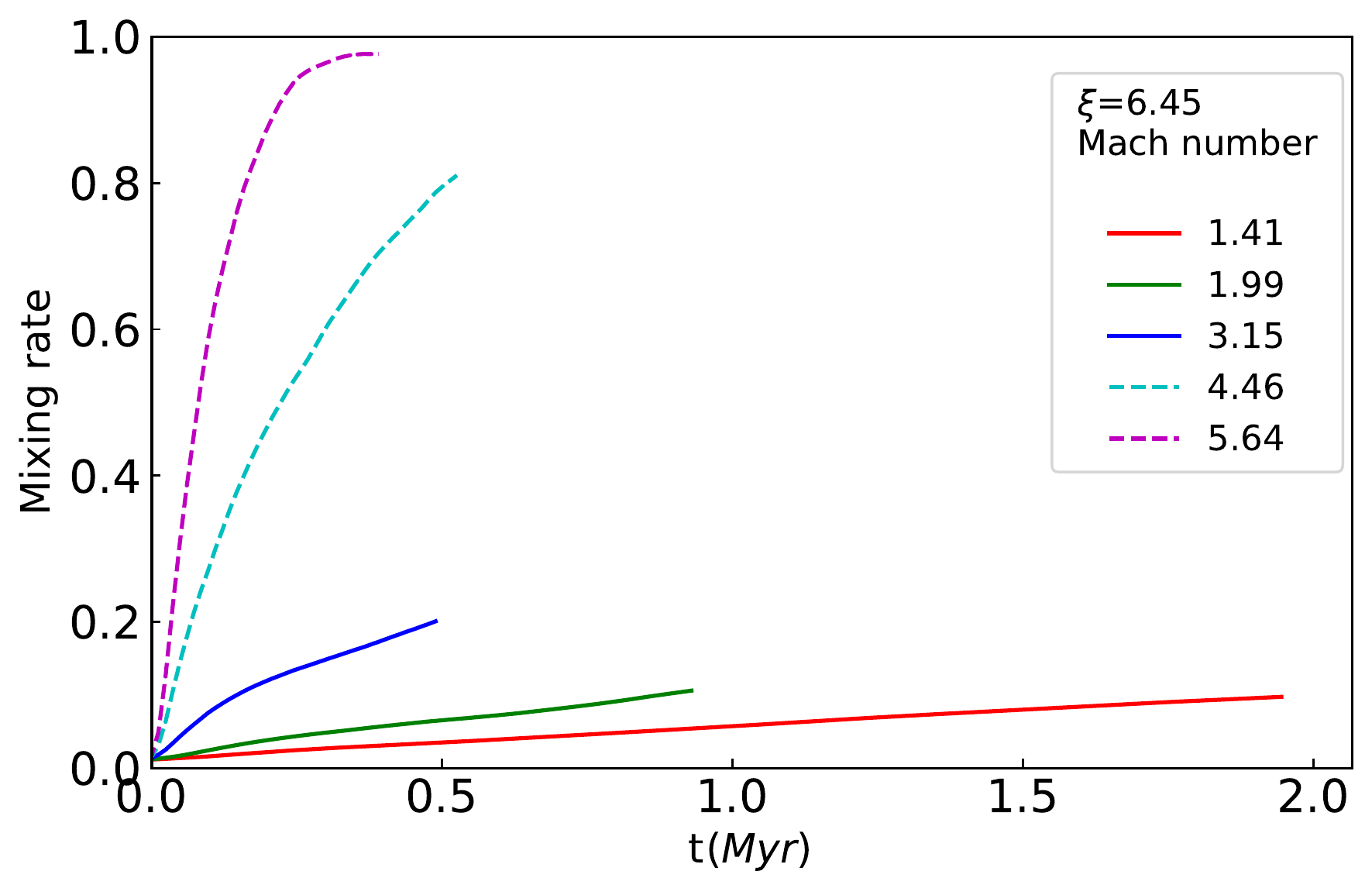}
          \hspace{2.0cm} (b) $M_{\rm mix}/M_{\rm cl}$
        \end{center}
      \end{minipage}

      \end{tabular}
    
       \begin{tabular}{c}

      \begin{minipage}{0.5\hsize}
        \begin{center}
          \includegraphics[clip, width=6.7cm]{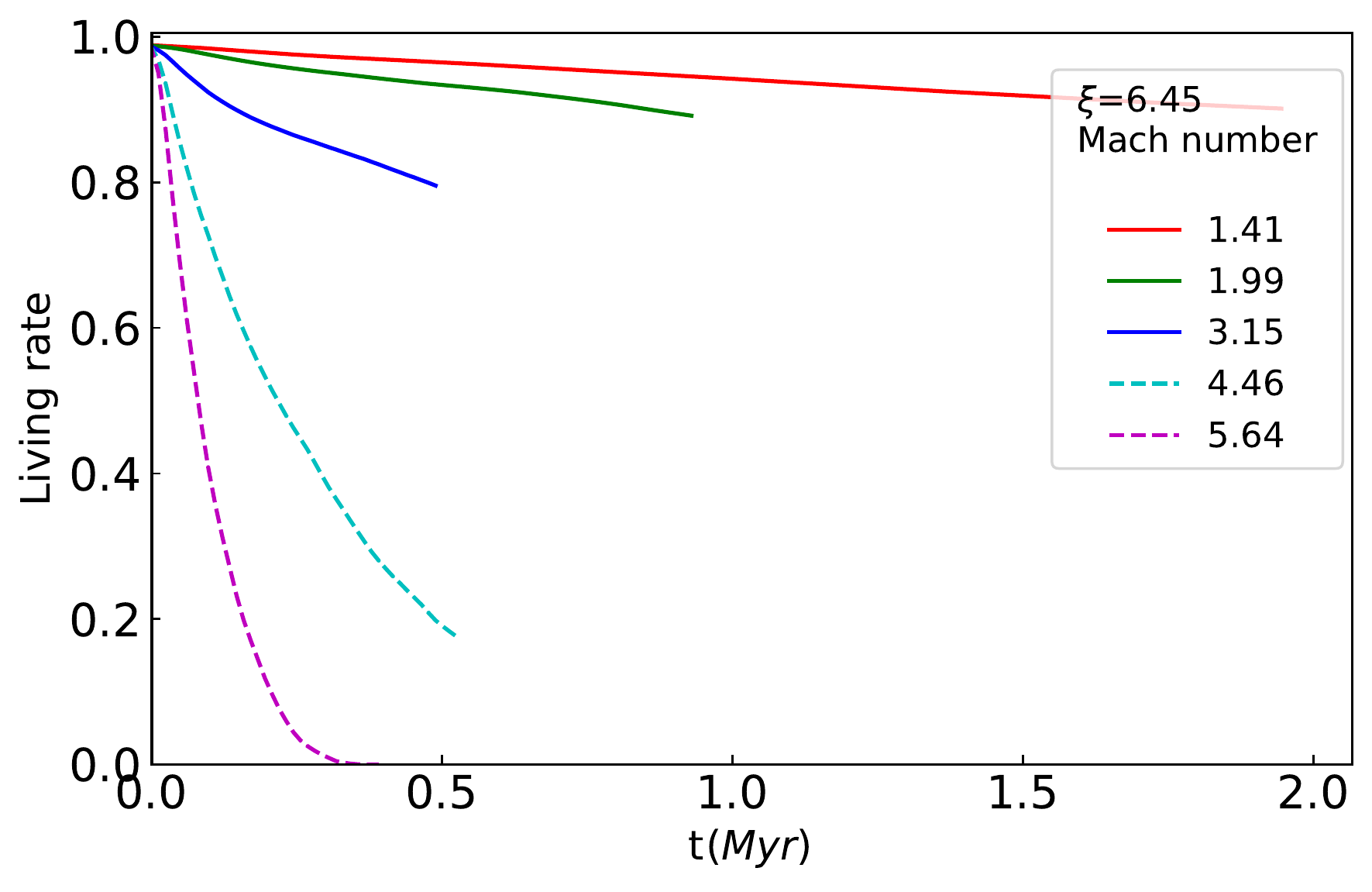}
          \hspace{2.0cm} (c) $M_{\rm live}/M_{\rm cl}$
        \end{center}
      \end{minipage}


      \begin{minipage}{0.5\hsize}
        \begin{center}
          \includegraphics[clip, width=6.7cm]{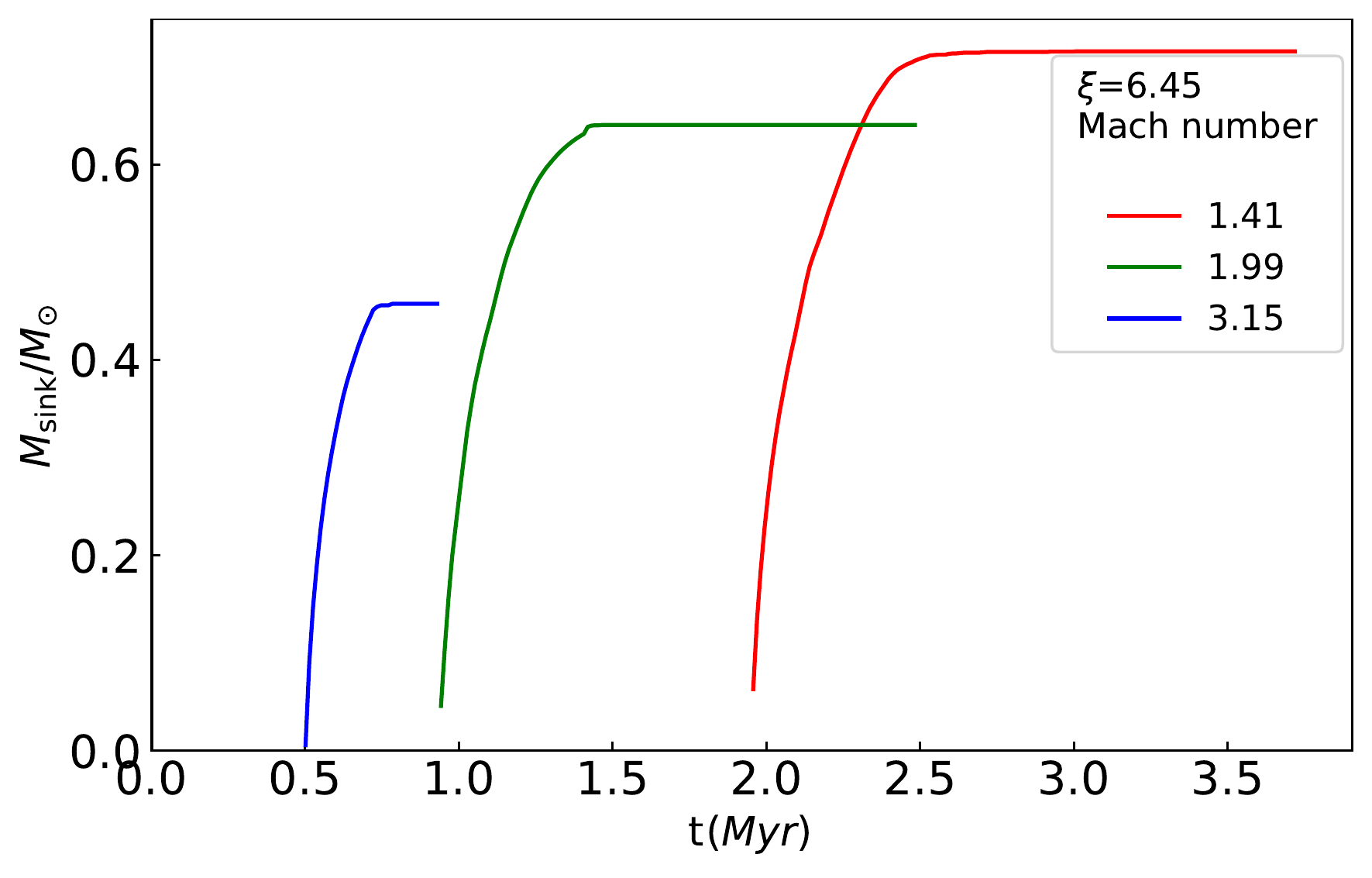}
          \hspace{2.0cm} (d) $M_{\rm sink}/M_{\rm cl}$
        \end{center}
      \end{minipage}
      \end{tabular}
      
      \begin{tabular}{c}
      
      \begin{minipage}{0.5\hsize}
        \begin{center}
          \includegraphics[clip, width=6.7cm]{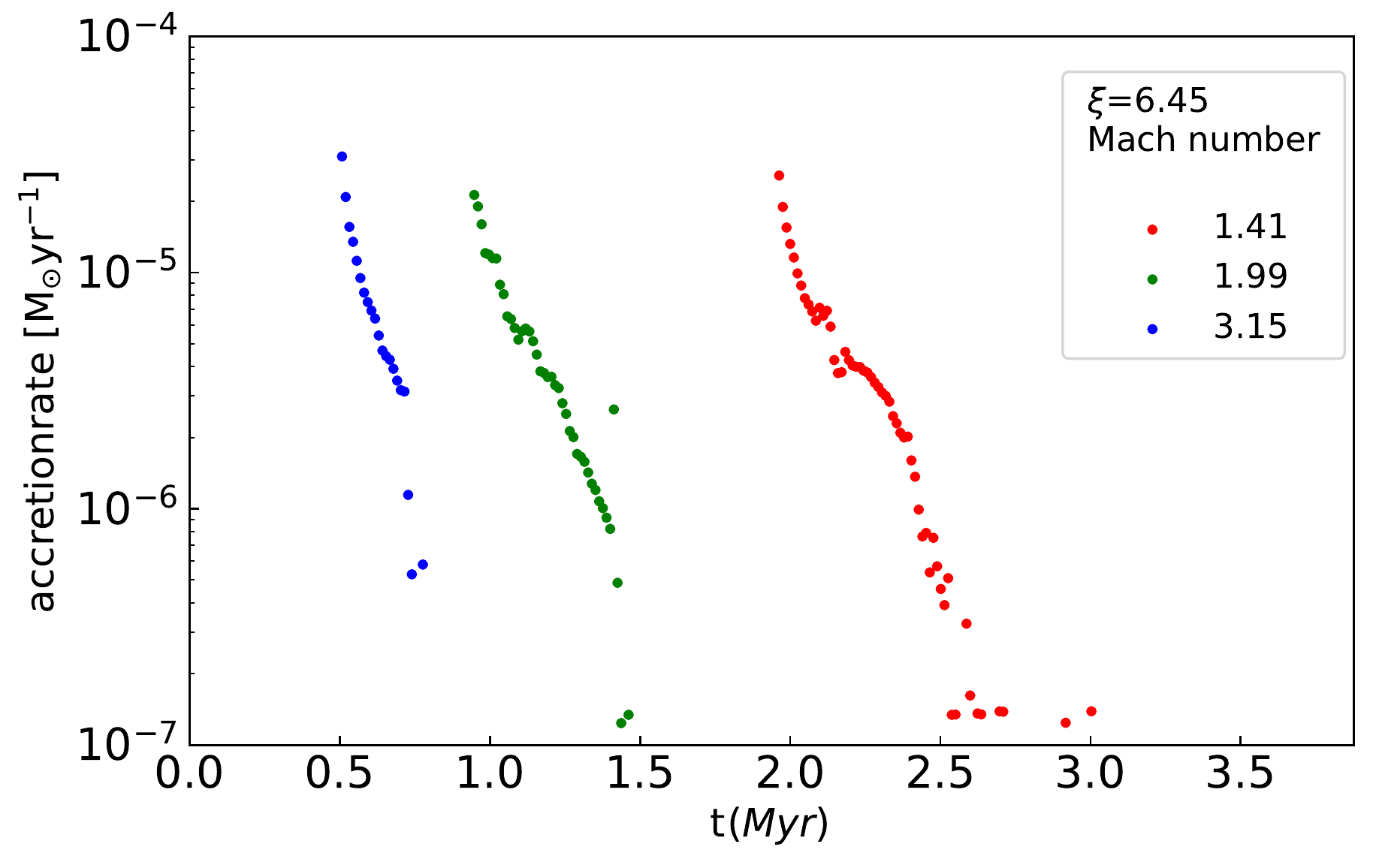}
          \hspace{2.0cm} (e) Accretion rate of sink particles
        \end{center}
      \end{minipage}

    \end{tabular}
    \vskip5pt  
    \caption{Time evolution of each physical quantities at different Mach numbers in $\xi$=6.45 $M_{\rm tur}=1.0$ (initial cloud is turbulent) models. 
     (a) As \edit1{in} Figure \ref{fig:rho_max_3.22_3}, ratio of the maximum density to the initial cloud central density $\rho_{\rm max}/\rho_{\rm c}$.
     (b) As \edit1{in} the top panel \edit1{of} Figure \ref{fig:mix_live_r_3.22_3}, mixing rate defined in Equation (\ref{eq:mixing rate}) .
     (c) As \edit1{in} the bottom panel \edit1{of} Figure \ref{fig:mix_live_r_3.22_3}, living rate defined in Equation (\ref{eq:living rate}).
     (d) As \edit1{in} the top panel \edit1{of} Figure \ref{fig:particle_3.22_3}, ratio of sink particle mass and initial cloud mass $M_{\rm sink}/M_{\rm cl}$.
     (e) As \edit1{in} the bottom panel \edit1{of} Figure \ref{fig:particle_3.22_3}, evolution of mass accretion rate of sink particles. 
     }
    \label{fig:turbulent}

\end{figure*}

\begin{figure}[hbtp]
\begin{center}
\includegraphics[width=150mm]{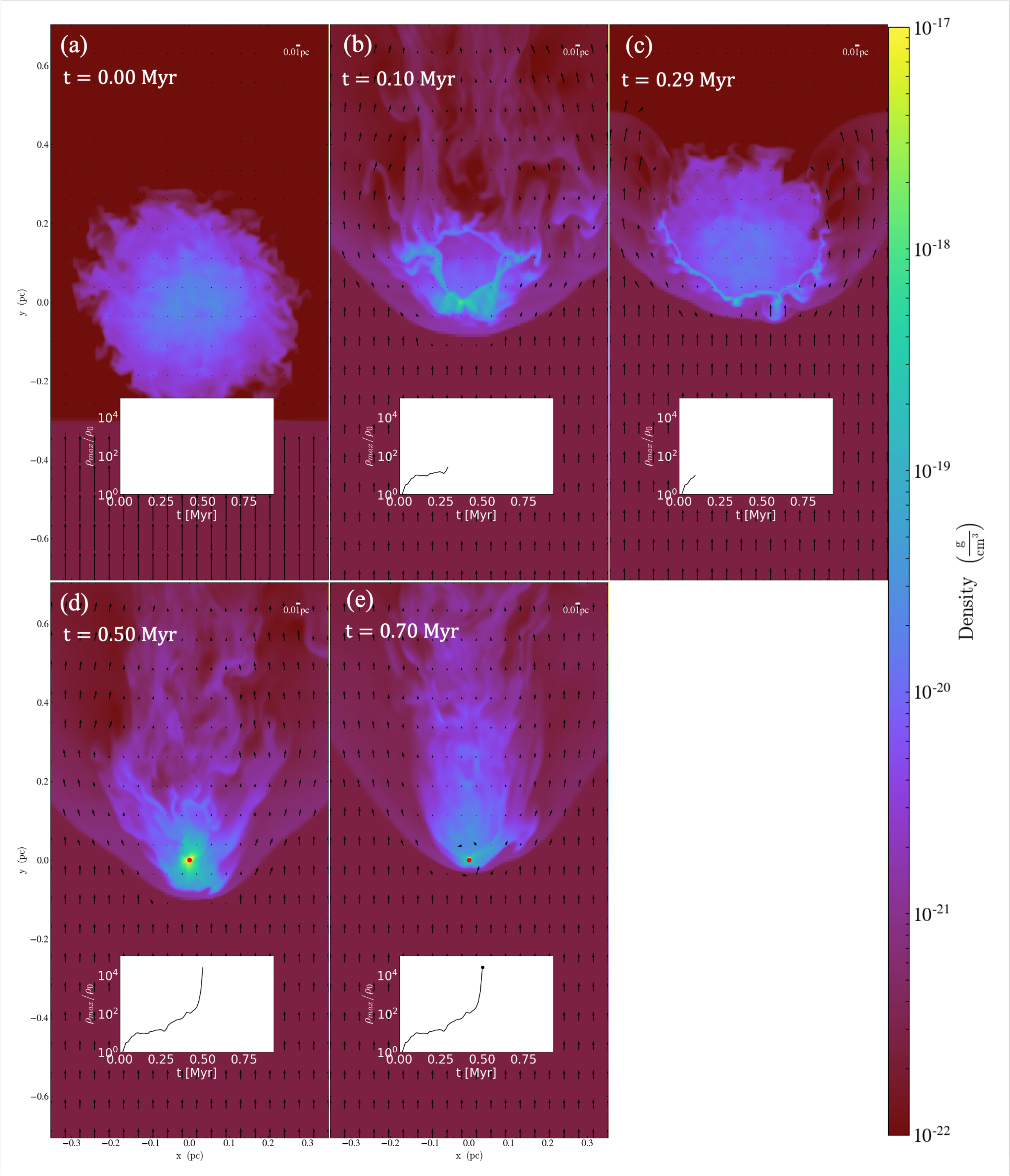}
\begin{flushleft}
\caption{As \edit1{in} Figure \ref{fig:slice_05_1_41} for $M_{\rm sh}=3.15$, $\xi=6.45$\edit1{,} and $M_{\rm tur}=1.00$ (initial cloud is turbulent). 
    }
    \label{fig:slice_1p_3_15}
\end{flushleft}
\end{center}
\end{figure}

\begin{figure}[hbtp]
\begin{center}
\includegraphics[width=150mm]{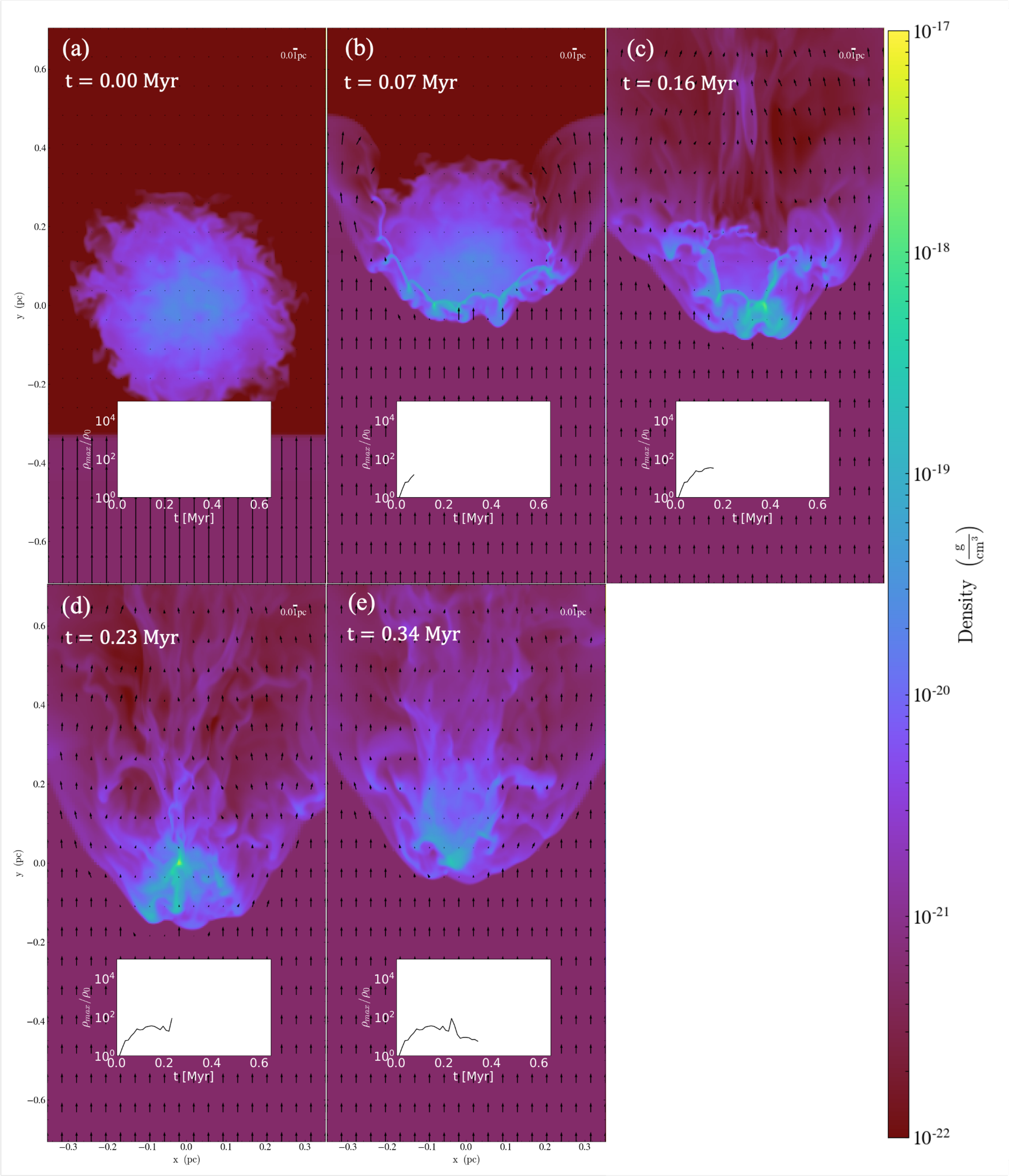}
\begin{flushleft}
\caption{As \edit1{in} Figure \ref{fig:slice_05_1_41} for $M_{\rm sh}=4.46$, $\xi=6.45$\edit1{,} and $M_{\rm tur}=1.00$ (initial cloud is turbulent). 
    }
    \label{fig:slice_1p_4_46}
\end{flushleft}
\end{center}
\end{figure}

Here, we \edit1{present} results of turbulent cloud models. We followed \edit1{five} models of turbulent ($M_{\rm tub}=1.00$) clouds with a critical Bonnor-Ebert radius by changing the shock Mach number of $1.41-5.64$. 
Figure \ref{fig:turbulent} shows the time evolution of $\rho_{\rm max}/\rho_{\rm c}$,  $M_{\rm mix}/M_{\rm cl}$,  $M_{\rm live}/M_{\rm cl}$,  $M_{\rm sink}/M_{\rm cl}$, and accretion rates of sink particles. 
Figure \ref{fig:turbulent} (a) indicates that $M_{\rm sh}=1.41-3.15$ shocks \edit1{induced} cloud collapse, \edit1{whereas} relatively stronger $M_{\rm sh}=4.46$ and $5.64$ shocks \edit1{could} not induce cloud collapse. 
Figure \ref{fig:slice_1p_3_15} and \ref{fig:slice_1p_4_46} show the \edit1{magnified} slices of the mass density distribution for $M_{\rm sh}$=3.15 and 4.46 as \edit1{in} Figure \ref{fig:slice_05_1_41}. \edit1{For} $M_{\rm sh}$=3.15, a strongly compressed layer \edit1{was} formed and then the cloud \edit1{collected} and \edit1{collapsed}. \edit1{In contrast}, \edit1{for} $M_{\rm sh}$=4.46, a strongly compressed layer \edit1{was} formed initially \edit1{and} the cloud \edit1{was} later \edit1{gradually} destroyed by the shock and mixed with the ambient gas.
Figure \ref{fig:turbulent} (b) and (c) indicate that the larger the Mach number \edit1{was}, the shorter the mixing timescale with the ambient gas \edit1{became}, \edit1{resulting in} higher living rates.  
Similar to the non-turbulent models, intermediate shocks \edit1{induced} cloud collapse, \edit1{whereas} \edit1{excessively} strong shocks destroyed \edit1{entire} clouds before \edit1{the} collapse and mixed clouds with ambient gas faster. 
Figure \ref{fig:turbulent} (d) and (e) show the evolution of mass and mass accretion rates of sink particles. Cross points in Figure \ref{fig:formed_result} shows results of turbulent cloud models. The larger the Mach number, the shorter the span of effective accretion rate, and the lower the asymptotic mass of the sink particle.

\begin{figure*}
  
    \begin{tabular}{c}

      \begin{minipage}{0.5\hsize}
        \begin{center}
          \includegraphics[clip, width=6.7cm]{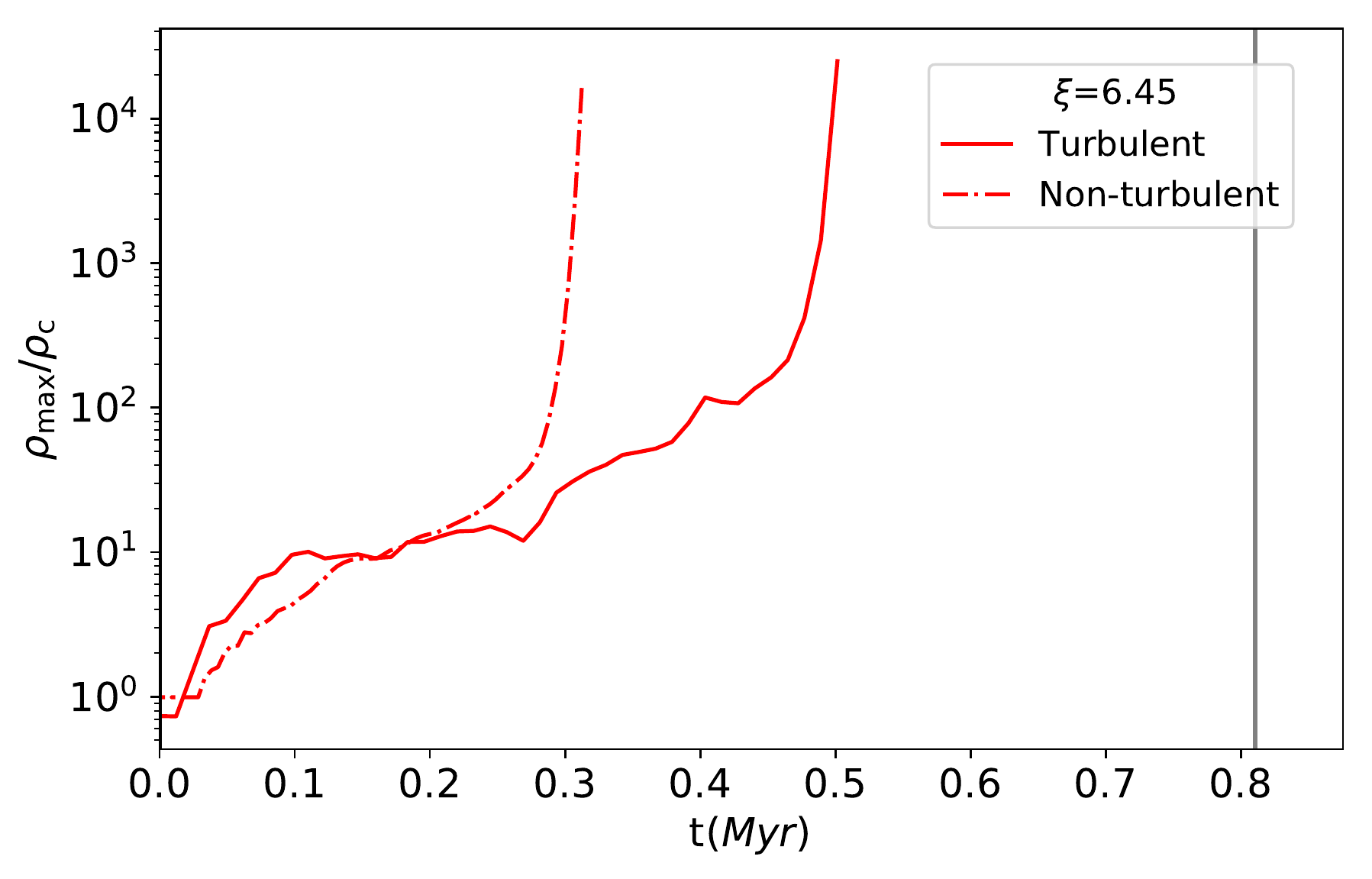}
          \hspace{2.0cm} (a) $\rho_{\rm max}/\rho_{\rm c}$
        \end{center}
      \end{minipage}

      \begin{minipage}{0.5\hsize}
        \begin{center}
          \includegraphics[clip, width=6.7cm]{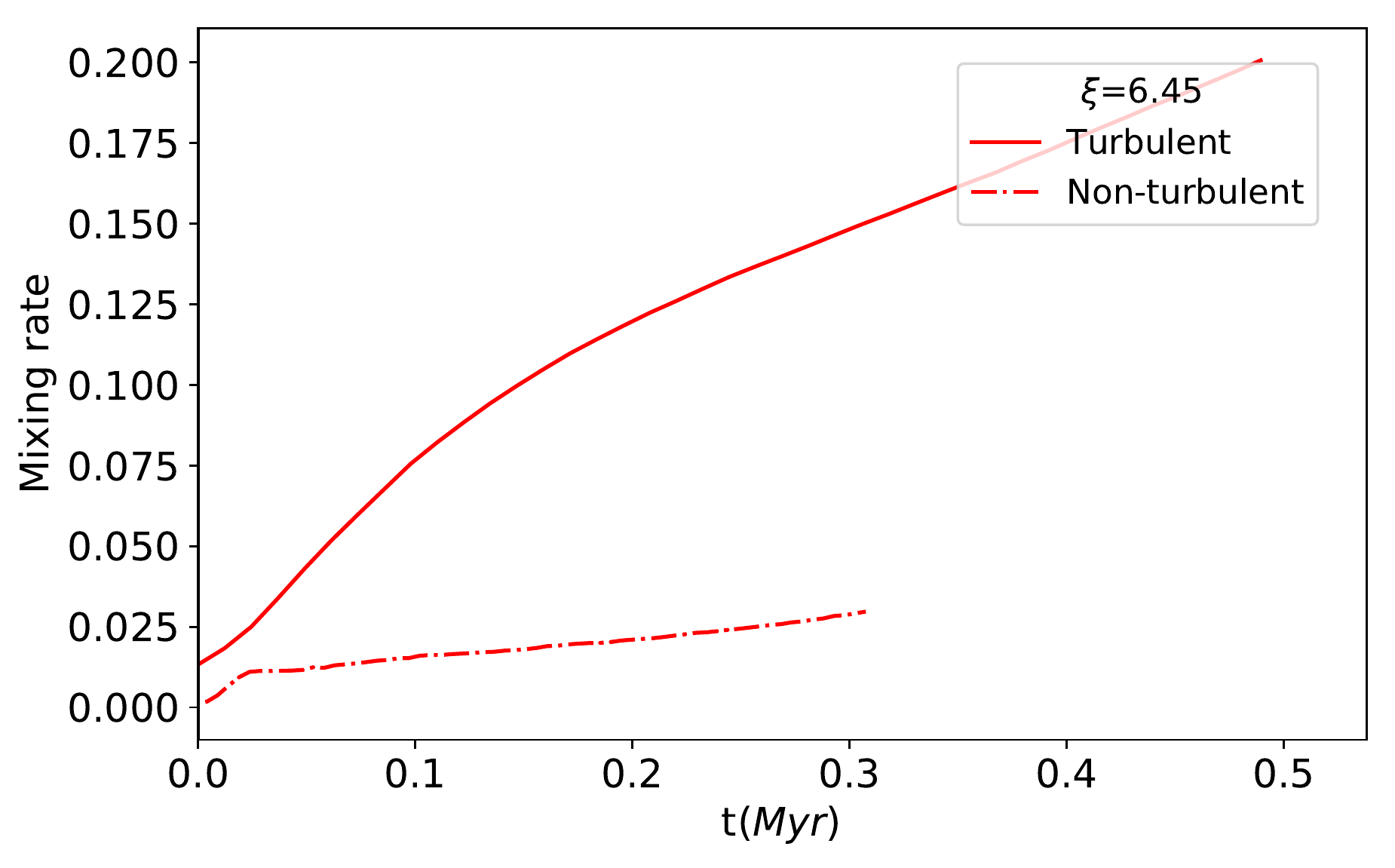}
          \hspace{2.0cm} (b) $M_{\rm mix}/M_{\rm cl}$
        \end{center}
      \end{minipage}

      \end{tabular}
    
       \begin{tabular}{c}

      \begin{minipage}{0.5\hsize}
        \begin{center}
          \includegraphics[clip, width=6.7cm]{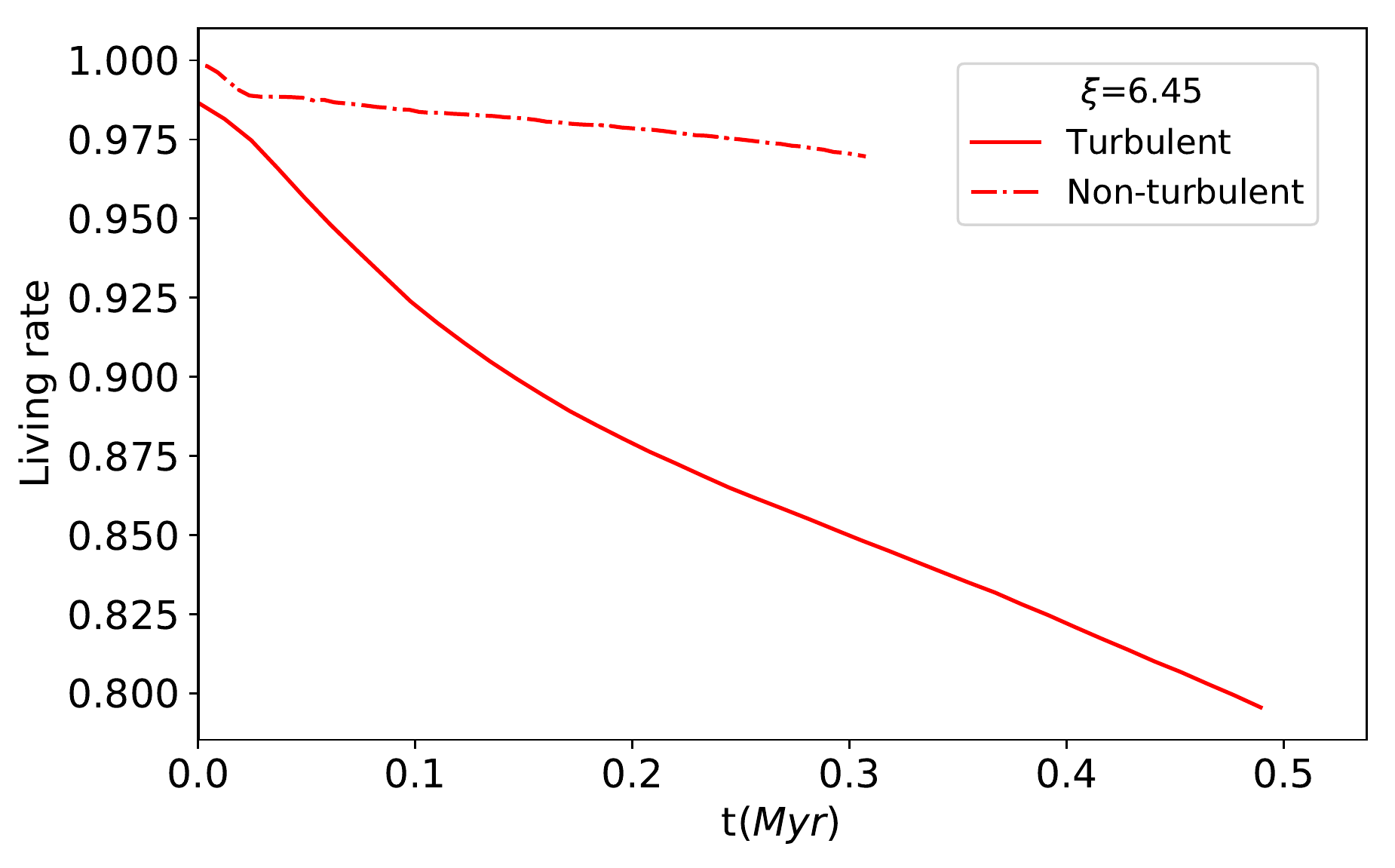}
          \hspace{2.0cm} (c) $M_{\rm live}/M_{\rm cl}$
        \end{center}
      \end{minipage}


      \begin{minipage}{0.5\hsize}
        \begin{center}
          \includegraphics[clip, width=6.7cm]{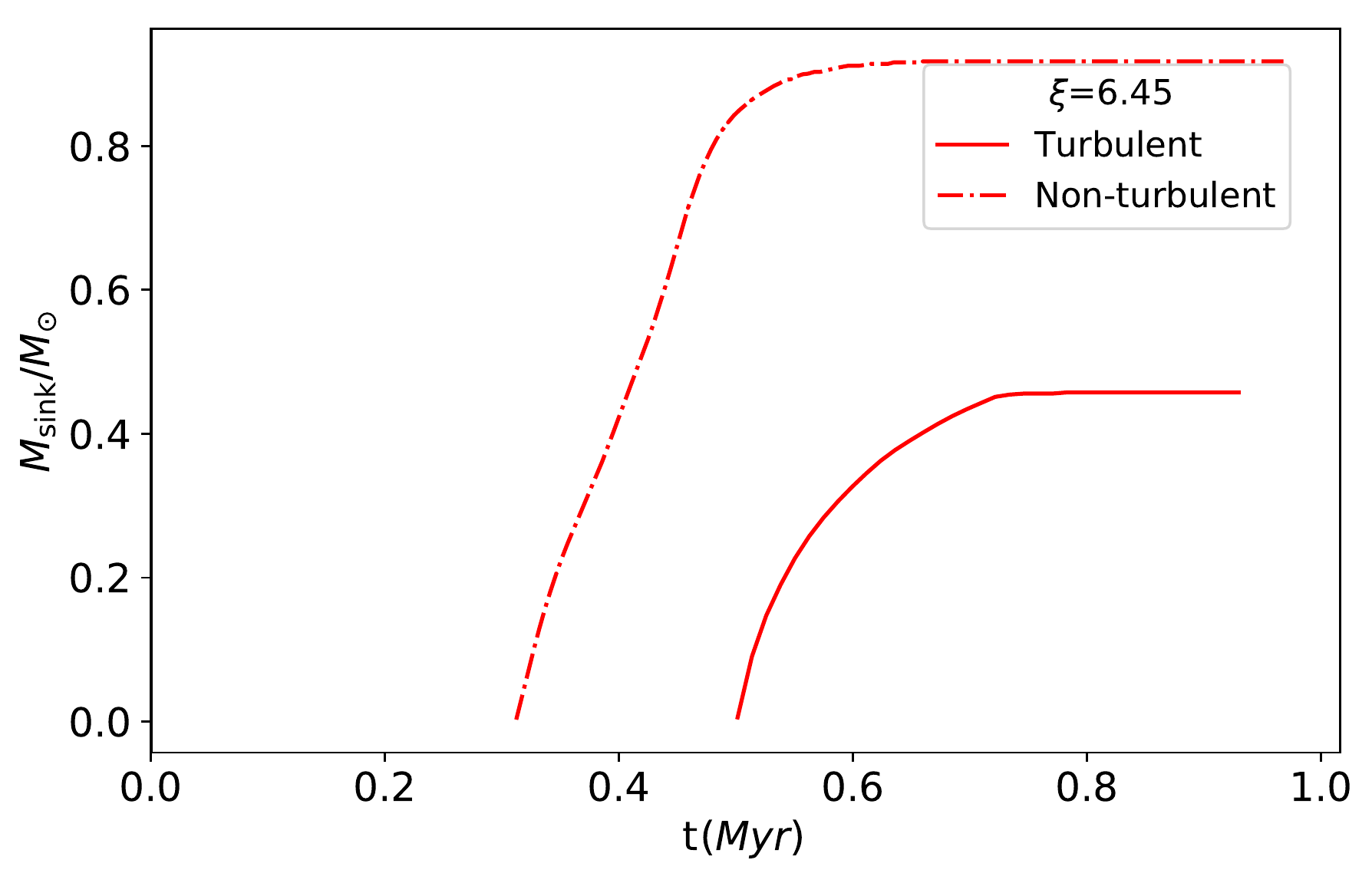}
          \hspace{2.0cm} (d) $M_{\rm sink}/M_{\rm cl}$
        \end{center}
      \end{minipage}
      \end{tabular}
      
      \begin{tabular}{c}
      
      \begin{minipage}{0.5\hsize}
        \begin{center}
          \includegraphics[clip, width=6.7cm]{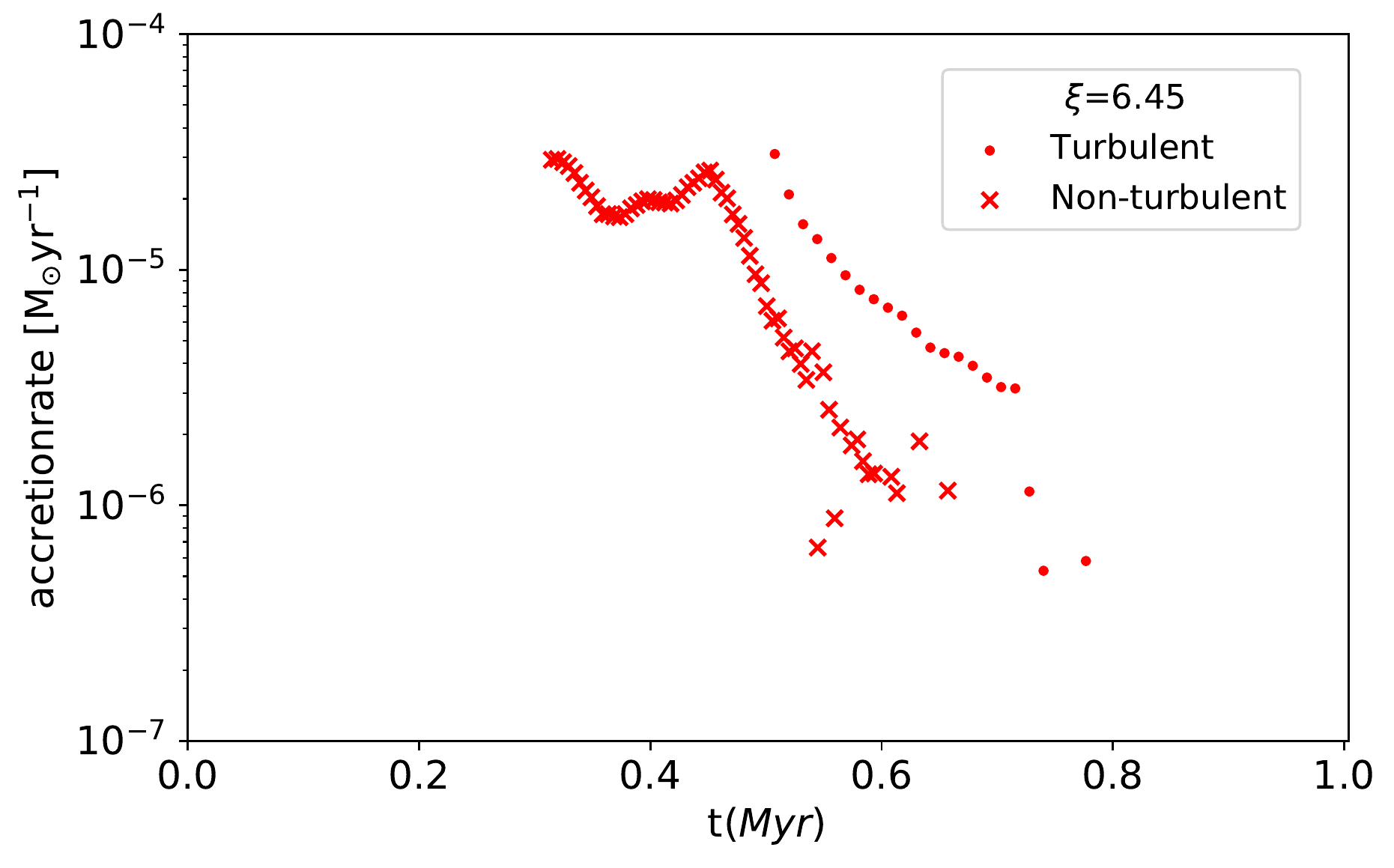}
          \hspace{2.0cm} (e) Accretion rate of sink particles
        \end{center}
      \end{minipage}

    \end{tabular}
    \vskip5pt  
    \caption{Time evolution of each physical quantities in both turbulent and non-turbulent models for $\xi=6.45$ and $M_{\rm sh}=3.15$.
    \edit1{The solid and dashed-dot} lines \edit1{indicate the results} for initially turbulent \edit1{and non-turbulent clouds, respectively}.
     (a) As in Figure \ref{fig:rho_max_3.22_3}, the ratio of the maximum density to the initial cloud central density $\rho_{\rm max}/\rho_{\rm c}$.
     (b) As \edit1{in} the top panel \edit1{of} Figure \ref{fig:mix_live_r_3.22_3}, mixing rate defined in Equation (\ref{eq:mixing rate}) .
     (c) As \edit1{in} the bottom panel \edit1{of} Figure \ref{fig:mix_live_r_3.22_3}, living rate defined in Equation (\ref{eq:living rate}).
     (d) As \edit1{in} the top panel \edit1{of} Figure \ref{fig:particle_3.22_3}, ratio of sink particle mass and initial cloud mass $M_{\rm sink}/M_{\rm cl}$.
     (e) As \edit1{in} the bottom panel \edit1{of} Figure \ref{fig:particle_3.22_3}, evolution of mass accretion rate of sink particles. 
     }
    \label{fig:turbulent_com}

\end{figure*}

To investigate the effect of cloud turbulence on shock-cloud evolution, here, we will compare turbulent \edit1{cloud} models with corresponding non-turbulent models for the same cloud radius and shock Mach number.
Figure \ref{fig:turbulent_com} shows \edit1{the} time \edit1{evolutions} of some physical quantities in both turbulent and non-turbulent models for $\xi=6.45$ and $M_{\rm sh}=3.15$. Figure \ref{fig:turbulent_com} (a) \edit1{shows the} maximum density evolution $\rho_{\rm max}/\rho_{\rm c}$. Up to $t \sim 0.2 $ Myr\edit1{, the} density evolution in two cases \edit1{were} similar. \edit1{Thereafter}, the turbulent cloud \edit1{had} a slower increase in density. As shown in Appendix \ref{app:Simulation results}, for the turbulent model, the sink particle was formed by 0.49 Myr after the shock touched the cloud, \edit1{whereas} for the non-turbulent clouds model, it was formed by 0.31 Myr. That is, the turbulent cloud \edit1{had} a slower increase in density and slower sink particle formation than \edit1{for} the non-turbulent cloud. 
Figure \ref{fig:turbulent_com} (b) and (c) show the evolution of mixing rates and living rates. For the turbulent cloud, the living rate \edit1{declined} faster and \edit1{the} mixing rate \edit1{increased} faster than that of non-turbulent \edit1{cloud}. That is, the turbulent cloud \edit1{mixed} faster with the surrounding ambient gas and \edit1{was} destroyed. 
Figure \ref{fig:turbulent_com} (d) and (e) show the evolution of sink particle mass and accretion rates. The asymptotic sink particle mass in the turbulent cloud model \edit1{was} lower than non-turbulent counterparts. The accretion rate \edit1{exceeded} $10^{-5} M_{\odot}$ by $t=0.06 \rm Myr$ \edit1{for} the turbulent cloud, \edit1{whereas} this occurs at $t=0.18$ Myr for the non-turbulent \edit1{cloud}. For the turbulent cloud, accretion time \edit1{was} shorter and the asymptotic mass of the sink particle \edit1{became} lower.

As above, the turbulence \edit1{prevented} cloud contraction, \edit1{promoted} the destruction of the cloud, and \edit1{reduced} the mass of the formed stars. We can \edit1{conclude} that turbulence in the dense cloud has the effect of suppressing star formation by shocks.

\clearpage
\section{Discussion}

\ifpdf
    \graphicspath{{Chapter3/Figs/Raster/}{Chapter3/Figs/PDF/}{Chapter3/Figs/}}
\else
    \graphicspath{{Chapter3/Figs/Vector/}{Chapter3/Figs/}}
\fi

\label{sec:Discussion}

\subsection{Conditions for cloud collapse by shocks}

\begin{figure*}
  
    \begin{tabular}{c}

      \begin{minipage}{0.45\hsize}
        \begin{center}
          \includegraphics[clip, width=7.0cm]{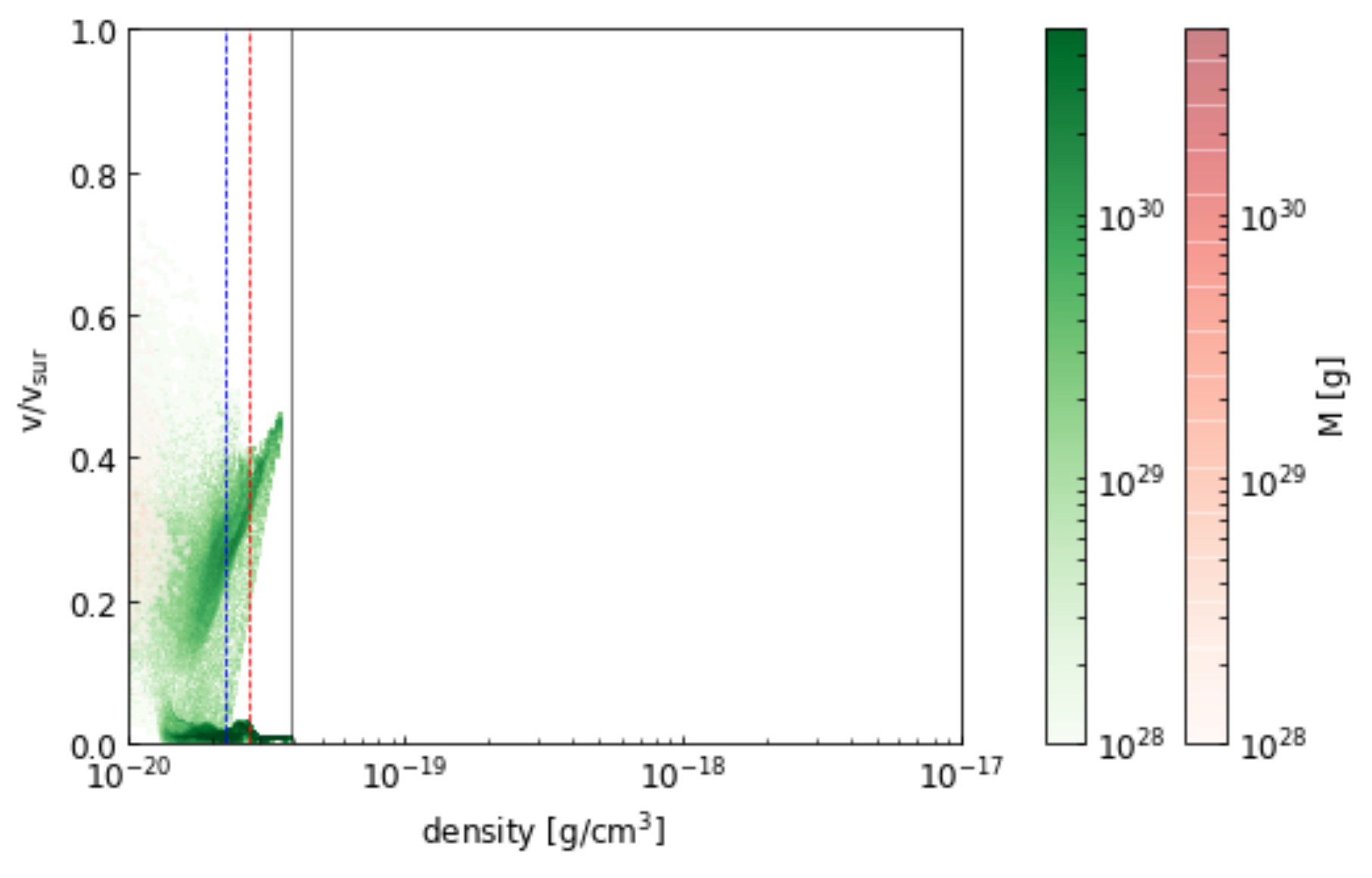}
          \hspace{2.0cm} (a) $M_{\rm sh}=1.20$, t = 0.16 Myr (=0.27 $t_{\rm cc}$)
        \end{center}
      \end{minipage}

      \begin{minipage}{0.45\hsize}
        \begin{center}
          \includegraphics[clip, width=7.0cm]{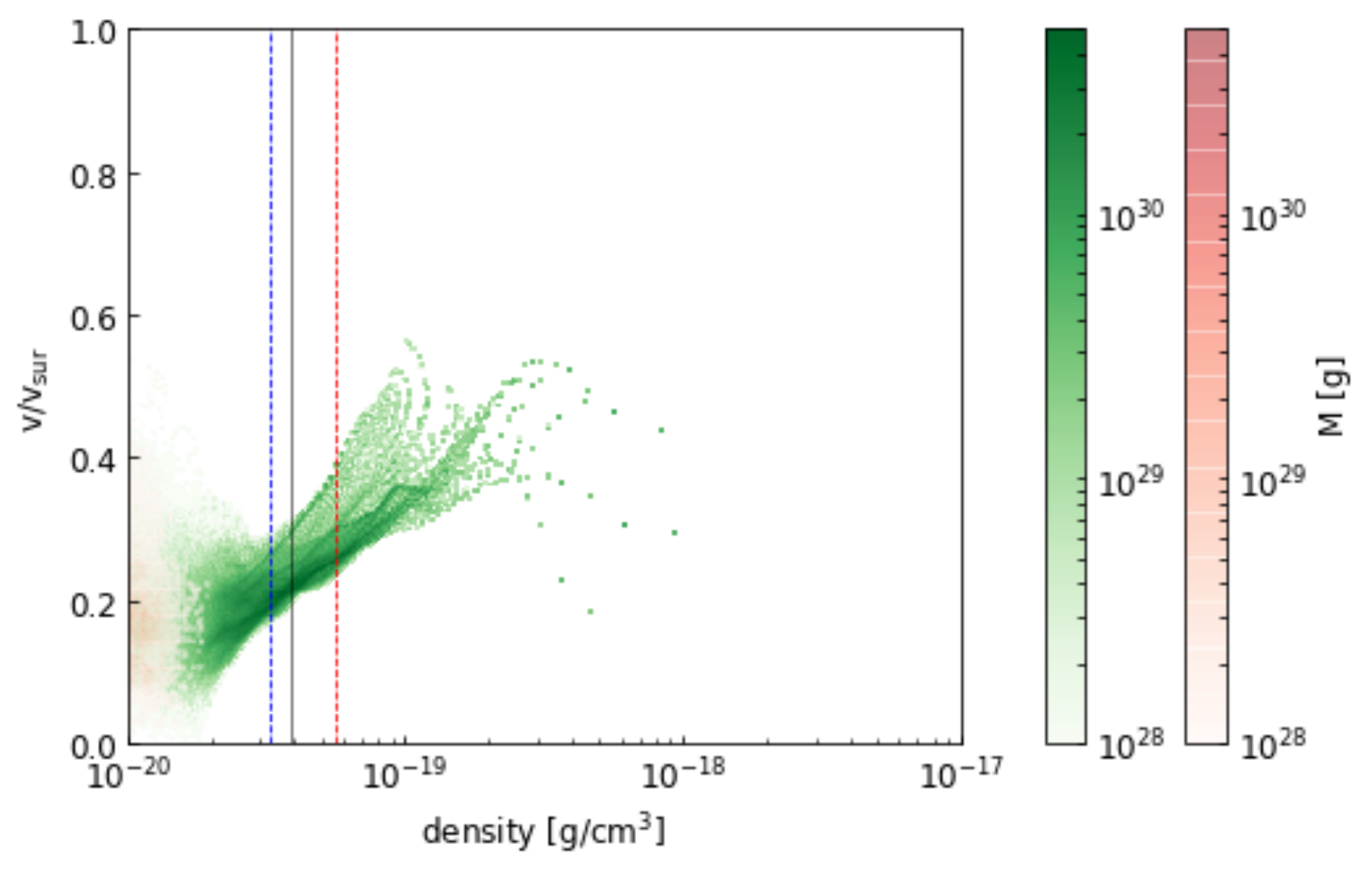}
          \hspace{2.0cm} (b) $M_{\rm sh}=1.20$, t = 0.49 Myr (=0.83 $t_{\rm cc}$)
        \end{center}
      \end{minipage}

      \end{tabular}

      \begin{tabular}{c}

      \begin{minipage}{0.45\hsize}
        \begin{center}
          \includegraphics[clip, width=7.0cm]{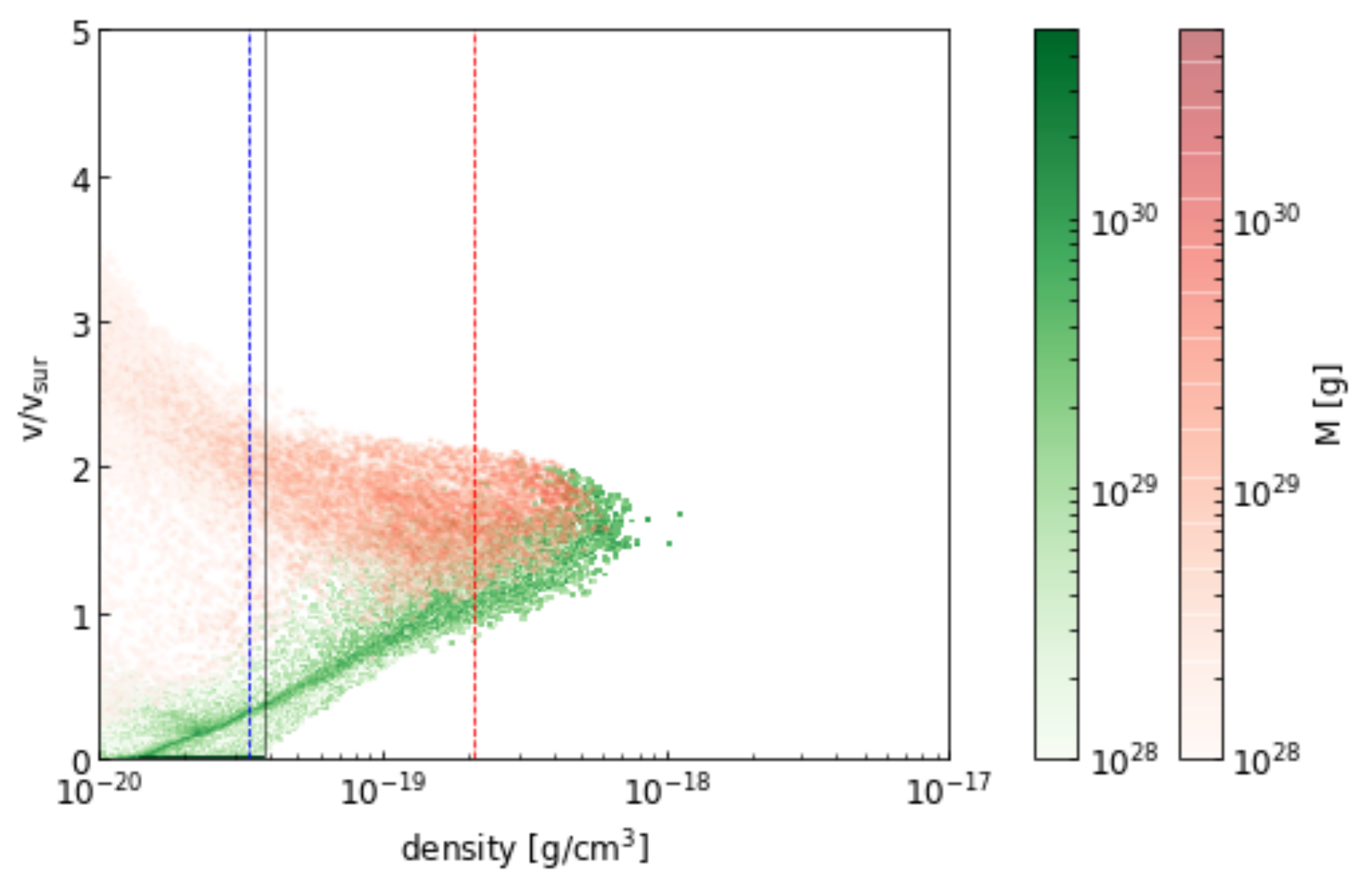}
          \hspace{2.0cm} (c) $M_{\rm sh}=3.15$, t = 0.06 Myr (=0.27 $t_{\rm cc}$)
        \end{center}
      \end{minipage}


      \begin{minipage}{0.45\hsize}
        \begin{center}
          \includegraphics[clip, width=7.0cm]{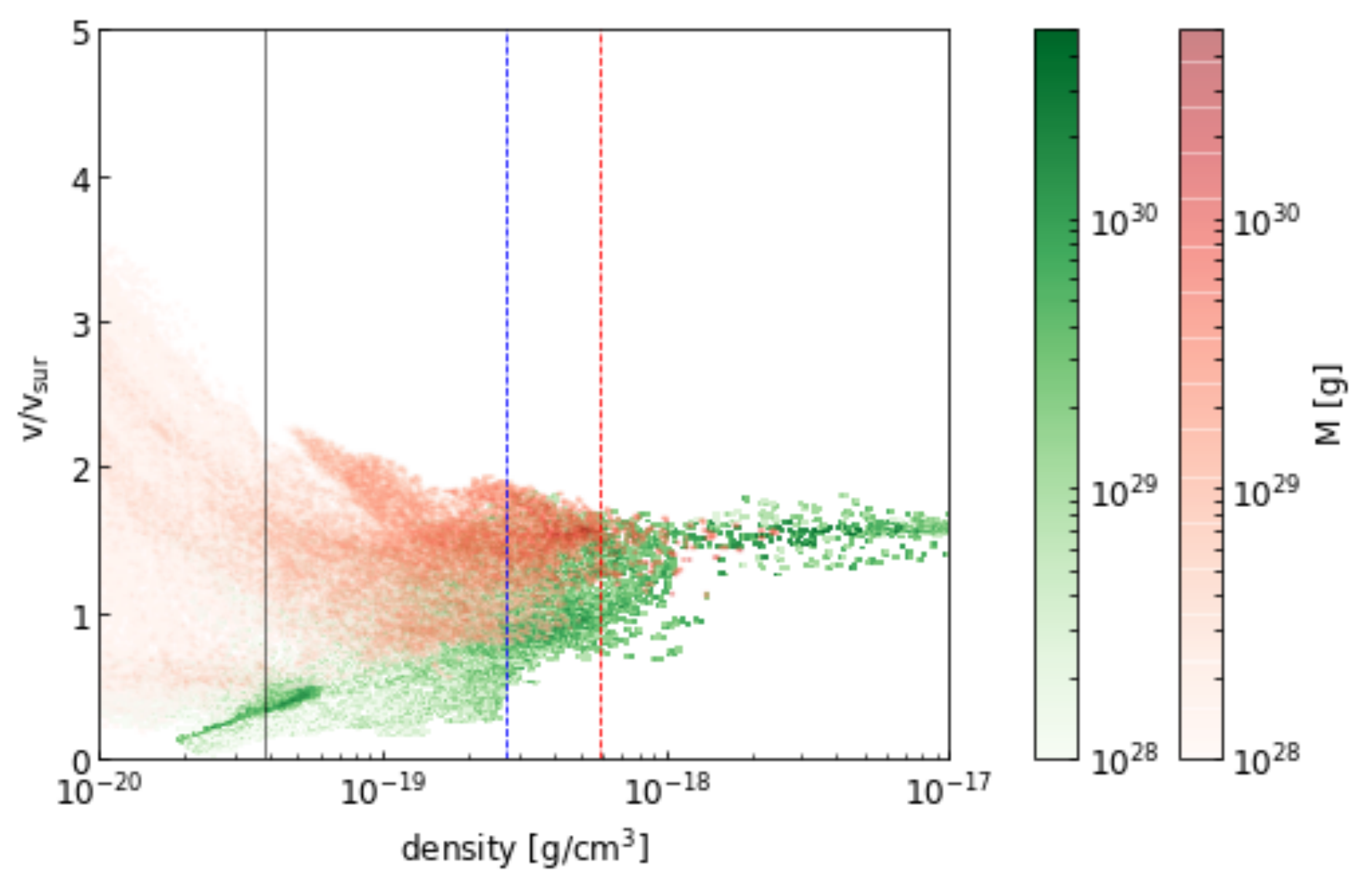}
          \hspace{2.0cm} (d) $M_{\rm sh}=3.15$, t = 0.15 Myr (=0.67 $t_{\rm cc}$)
        \end{center}
      \end{minipage}
      
      \end{tabular}

      \begin{tabular}{c}

      \begin{minipage}{0.45\hsize}
        \begin{center}
          \includegraphics[clip, width=7.0cm]{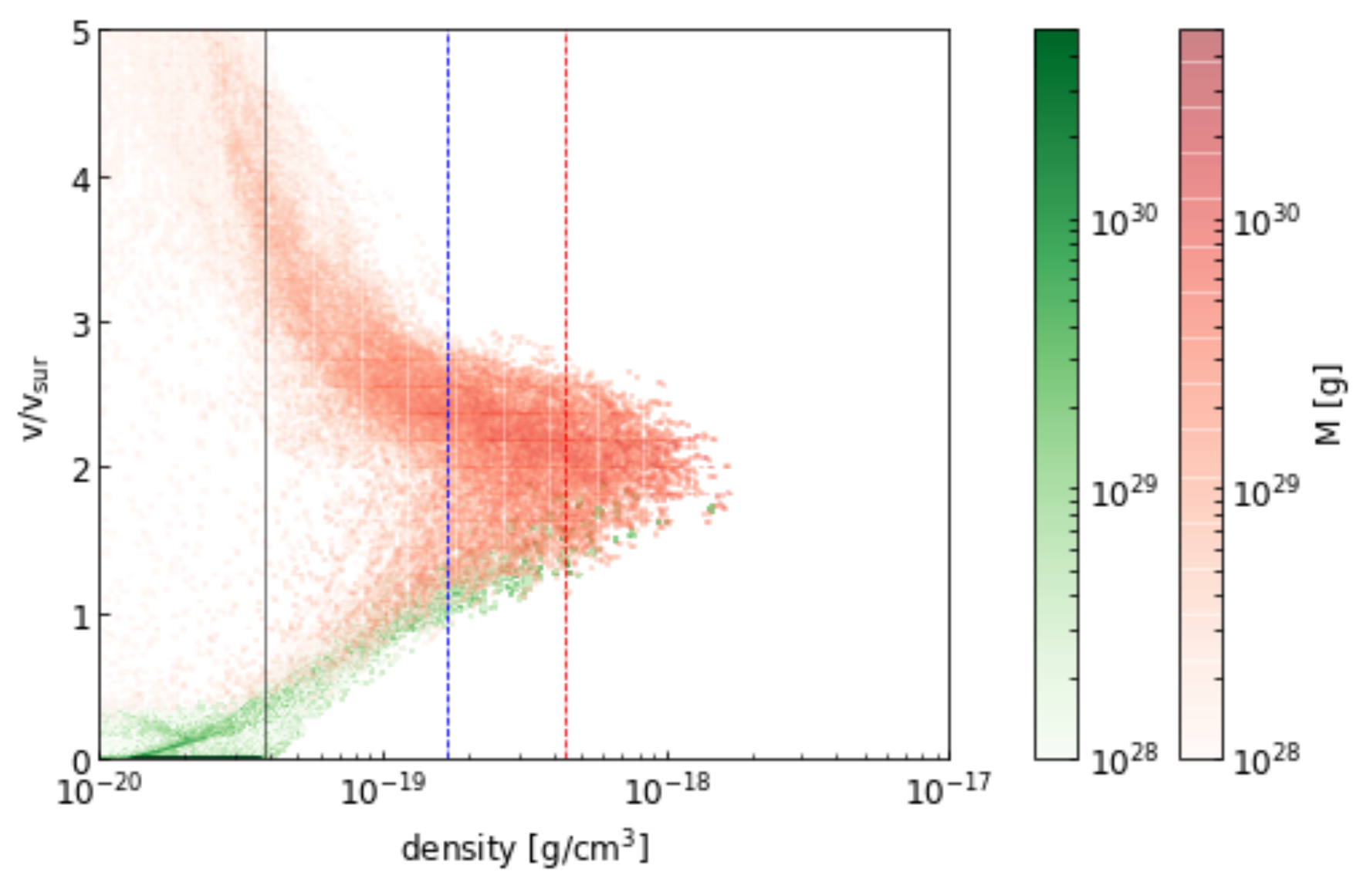}
          \hspace{2.0cm} (e) $M_{\rm sh}=5.0$, t = 0.03 Myr (=0.27 $t_{\rm cc}$)
        \end{center}
      \end{minipage}
      
      \begin{minipage}{0.45\hsize}
        \begin{center}
          \includegraphics[clip, width=7.0cm]{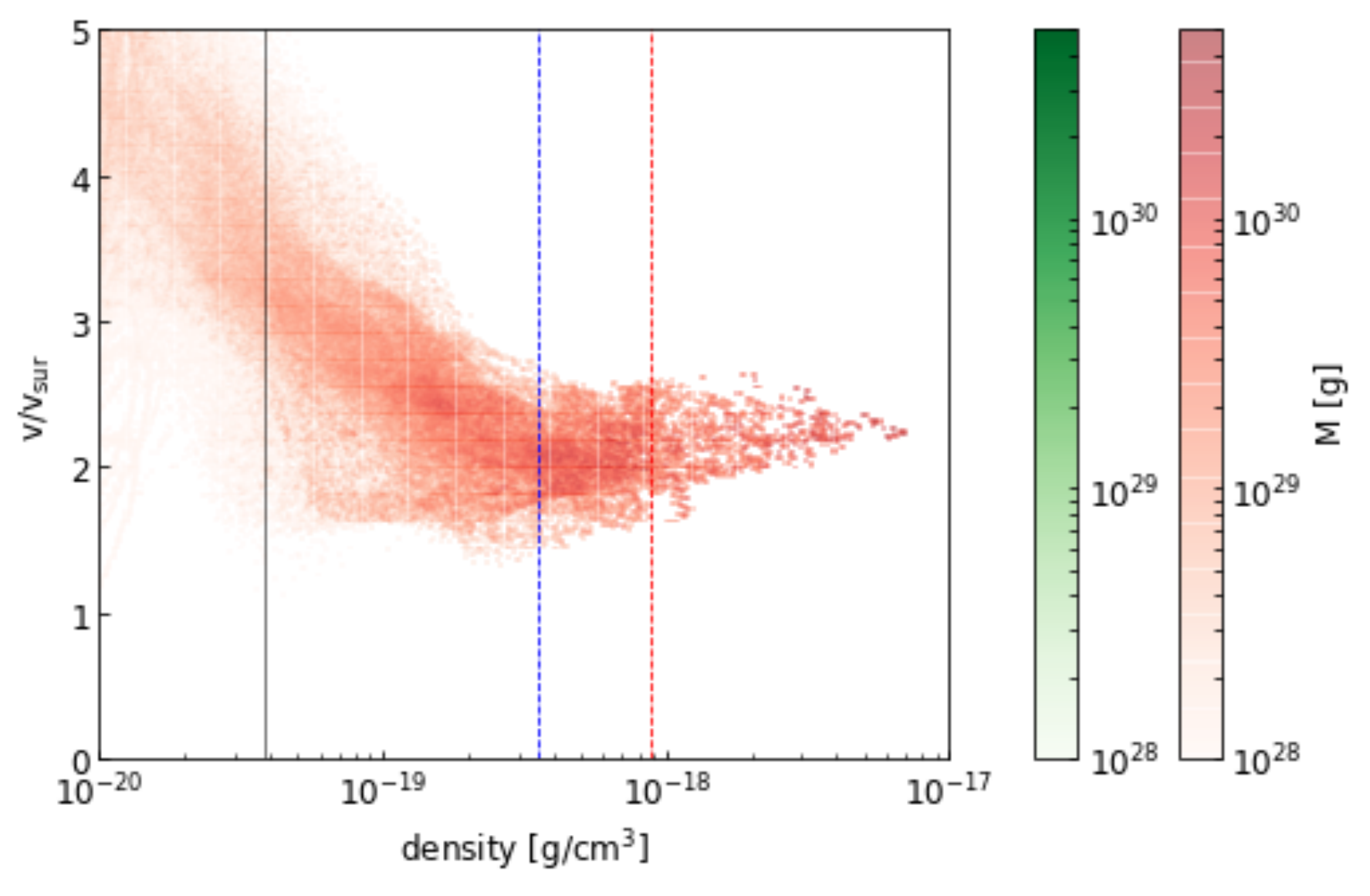}
          \hspace{2.0cm} (f)  $M_{\rm sh}=5.0$, t = 0.08 Myr (=0.56 $t_{\rm cc}$)
        \end{center}
      \end{minipage}

    \end{tabular}
    \vskip5pt  
    \caption{\edit1{Mass per unit velocity and density interval} for the $\xi=3.22$ models as \edit1{in} Figure \ref{fig:pdf-rho-v_1_41}. The vertical axis shows the value of the velocity normalized by $v _{\rm sur}$ ($=v_{\rm sh}\chi_{\rm s}^{-1/2}$). \edit1{Gases} \edit1{with} the color variable $C>0.99$ region \edit1{are indicated} in green, while those of $0.1<C<0.99$ \edit1{are indicated} in red. (a) and (b) For $M_{\rm sh}=1.20$, at t=0.16 Myr and t = 0.49 Myr, respectively. (c) and (d) For $M_{\rm sh}=3.15$, at t=0.06 Myr and t = 0.15 Myr, respectively. (e) and (f) For $M_{\rm sh}=5.00$, at t=0.03 Myr and t = 0.56 Myr , respectively. (a),(c)\edit1{,} and (e) are the same time when expressed in $t_{\rm cc}$. (b) and (f)  correspond to rebounding phase. (d) corresponds to \edit1{immediately} before the sink particle \edit1{was} formed.}
    \label{fig:pdf-rho-two}

\end{figure*}

As shown in Figure \ref{fig:formed_result}, shocks that are \edit1{excessively} strong or weak cannot induce cloud collapse. This is because \edit1{excessively} strong shocks tend to destroy the cloud \edit1{through} hydrodynamic instabilities and \edit1{excessively} weak shocks do not compress the cloud \edit1{sufficiently} to \edit1{cause them to} become unstable.
This implies the existence of parameter windows for cloud collapse.
In this subsection, we \edit1{discuss} the conditions for gravitational collapse triggered by shocks.

Figure \ref{fig:pdf-rho-two} shows the \edit1{mass per unit velocity and density interval} for the $\xi=3.22$ models as \edit1{in} Figure \ref{fig:pdf-rho-v_1_41}. The vertical axis shows the value of the velocity normalized by estimated \edit1{the} propagating shock velocity at the cloud surface $v _{\rm sur}=v_{\rm sh}\chi_{\rm s}^{-1/2}$ (see Equation (\ref{eq:v_cl_in_the_cloud})). \edit1{Gases} with the color variable $C>0.99$ \edit1{are indicated} in green, while those of $0.1<C<0.99$ were shown in red. The panels in the upper row are for the $M_{\rm sh}=1.20$ cases corresponding to $t=0.27t_{\rm cc}$ and the rebounding phase. 
For $M_{\rm sh}=1.20$, \edit1{the} gas accounting for more than 50$\%$ of the mass \edit1{was} compressed almost without being stripped. However, as mentioned in Section \ref{subsec:1.20 case}, the clouds \edit1{were} not sufficiently compressed to develop gravitational instability.
The panels in the middle row are for the $M_{\rm sh}=3.15$ cases corresponding to $t=0.27t_{\rm cc}$ and \edit1{immediately} before sink particle creation. In panel (c), the distribution is shaped like a horizontal V with the root at $v/v_{\rm sur}\sim1$. The gas distributed from the root of V toward the origin of the figure corresponds to the gas compressed by the shock progressing to the center of the cloud. Gas distributed from the base of V to the upper left of the figure corresponds to gas stripped and dissipated by the shock. For $M_{\rm sh}=3.15$, \edit1{the} compressed gas \edit1{became} dense, \edit1{resulting in} gravitational collapse.
The panels in the lower row are for the $M_{\rm sh}=5.00$ cases corresponding to $t=0.27t_{\rm cc}$ and the rebounding phase. In panel (e), the distribution is shaped like a V as in panel (c), but the V is wider and more gas is dissipated. As shown in Section \ref{subsec:1.20 case}, eventually, the entire cloud \edit1{was} destroyed.

\edit1{Based on} these results, two conditions must be \edit1{satisfied} for the cloud to contract.
The first condition \edit1{involves} the degree of compression by shocks. 
For low \edit1{Mach numbers}, most of the gas does not become \edit1{sufficiently} dense and gravitational collapse cannot be induced. It is assumed that for cloud collapse, \edit1{sufficient} compression to make Bonnor-Ebert spheres unstable is required. This condition would determine the lower limit of the Mach number.
The second condition \edit1{involves} the destruction of clouds. Even if clouds are strongly \edit1{compressed}, shocks with Mach numbers that \edit1{are excessively large} \edit1{cause} the destruction of \edit1{entire} clouds. \edit1{For clouds} to collapse, the timescale for the development of gravitational instability must be shorter than that of destruction. This condition would determine the upper limit of the Mach number. Hereafter, based on the above two conditions, we will estimate the Mach number parameter window for the cloud collapse.

First, we consider the lower limit of $M_{\rm sh}$. 
Considering an isothermal shock with a Mach number $M_{\rm sh}$, the ambient gas pressure behind the shock is $\edit1{P_{0}}M_{\rm sh}^{2}$.

Considering conditions for a Bonnor-Ebert sphere to become unstable, we expect
\begin{eqnarray}
\label{eq:lower_condition}
\edit1{P_{0}}M_{\rm sh}^{2}\geq P_{\rm crit},
\end{eqnarray}
where $P_{\rm crit}$ is the critical pressure of Bonnor-Ebert sphere (see Equation (\ref{eq:bonnor_critical_P})).
This relation implies
\begin{eqnarray}
\label{eq:lower_M}
M_{\rm sh}>\left(\frac{1.4c_{\rm cl}^{6}}{\rho_{\rm s}G^{3}m_{\rm cl}^{2}}\right).
\end{eqnarray}
Substituting Equation (\ref{eq:bonnor_set}),(\ref{eq:bonnor_set-2})\edit1{,} and (\ref{eq:bonnor-mass-4}), we \edit1{obtain} a dimensionless expression:
\begin{eqnarray}
\label{eq:lower_M_nond}
M_{\rm sh}> (5.6\pi)^{\frac{1}{2}}\left[ \xi^{2}\frac{d\psi}{d\xi}\rm exp(-\frac{\psi}{2})\right]^{-1}.
\end{eqnarray}

Since the \edit1{right-hand} side of Equation (\ref{eq:lower_M_nond}) is a function of $\xi$, it can also be expressed \edit1{as} a function of $\rho_{\rm c}/\rho$.
\edit1{Calculating} the third degree polynomial regression of the \edit1{right-hand} side of Equation (\ref{eq:lower_M_nond}) in the range of $2.0\leq\rho_{\rm c}/\rho\leq14.1$, we \edit1{obtain} the approximate equation of lower limit of $M_{\rm sh}$\edit1{:}
\begin{eqnarray}
\label{eq:lower_M_nond_fit}
M_{\rm low} \approx 0.96+0.34\left(\frac{\rho_{\rm c}}{\rho}\right)^{-1}+2.10\left(\frac{\rho_{\rm c}}{\rho}\right)^{-2}+5.16\left(\frac{\rho_{\rm c}}{\rho}\right)^{-3}~~~~\left(2.0\leq\frac{\rho_{\rm c}}{\rho}\leq14.1\right),
\end{eqnarray}
\edit1{where} $\rho_{\rm c}/\rho=14.1$ corresponds to $\xi=\xi_{\rm crit}$. When $\rho_{\rm c}/\rho>14.1$, a Bonnor-Ebert sphere is unstable. 

Next, we consider \edit1{the} upper limit of $M_{\rm sh}$.
\citet{Iwasaki_2008} \edit1{observed} that the timescale of the gravitational instability of isothermal layers bounded by \edit1{a} shock wave with the Mach number $M_{\rm sh}$ is an order of $t_{\rm ff}/\sqrt{M_{\rm sh}}$\edit1{,}
where $t_{\rm ff}$ is the free-fall time scale of the preshock region.
In our model, the sum of the timescale on which the shock propagates through the cloud and the timescale on which gravitational instability behind the shock \edit1{increases} is estimated \edit1{as}

\begin{eqnarray}
\label{eq:instability}
t_{\rm cc}+t_{\rm ff}/\sqrt{M_{\rm sh}}=t_{\rm cc}+\left(\frac{3\pi}{32G\langle\rho_{\rm cl}\rangle M_{\rm sh}}\right)^{1/2},
\end{eqnarray}
where $t_{\rm ff}$ is the free fall time scale of spherical cloud.

It is known that the timescale of the cloud destruction is \edit1{in} the order of the cloud crushing time $t_{\rm des}$=$\alpha t_{\rm cc}$ (see Appendix \ref{sec:timescales}). Previous numerical hydrodynamical simulations \edit1{indicated} that $\alpha\sim 2.5-4$ (e.g.,  \citealp{Klein_1994}; \citealp{Poludnenko_2002}). \edit1{For clouds to be collapsed} by shocks, gravitational instability must \edit1{increase rapidly} within the timescale of cloud destruction $t_{\rm des}$. Hence, we expect 
\begin{eqnarray}
\label{eq:t_upper_con}
  t_{\rm cc}+t_{\rm ff}/\sqrt{M_{\rm sh}}<t_{\rm des} = \alpha t_{\rm cc}.
\end{eqnarray}
This implies
\begin{eqnarray}
\label{eq:M_upper}
  M_{\rm sh}<\left[(\alpha-1)\left(\frac{32G}{3\pi}\right)^{1/2}\frac{r_{\rm cl}\langle\rho_{\rm cl}\rangle}{c_{\rm ism}\rho_{\rm ism}^{1/2}}\right]^{2}.
\end{eqnarray}
This equation can be rewritten dimensionlessly as 
\begin{eqnarray}
\label{eq:M_upper_}
  M_{\rm sh}<(\alpha-1)^{2}\left(\frac{72}{3\pi^{2}}\right)\rm exp(\psi)\left(\frac{d\psi}{d\xi}\right)^{2},
\end{eqnarray}

\edit1{Calculating} the the minus third degree polynomial regression of the \edit1{right-hand} side of Equation (\ref{eq:M_upper_}) in the range of $2\leq\rho_{\rm c}/\rho\leq100$, we \edit1{obtain} the approximate equation of upper limit of $M_{\rm sh}$\edit1{:}
\begin{eqnarray}
\label{eq:lower_M_upper_fit}
M_{\rm upp} \approx (\alpha-1)^{2}\left\{-0.78\left[\rm log\left(\frac{\rho_{\rm c}}{\rho}\right)\right]^{3}+1.61\left[\rm log\left(\frac{\rho_{\rm c}}{\rho}\right)\right]^{2}+3.36\left[\rm log\left(\frac{\rho_{\rm c}}{\rho}\right)\right]+0.06\right\}~~~~\left(2\leq\frac{\rho_{\rm c}}{\rho}\leq100\right).
\end{eqnarray}

\begin{figure}[hbtp]
\begin{center}
\includegraphics[width=110mm]{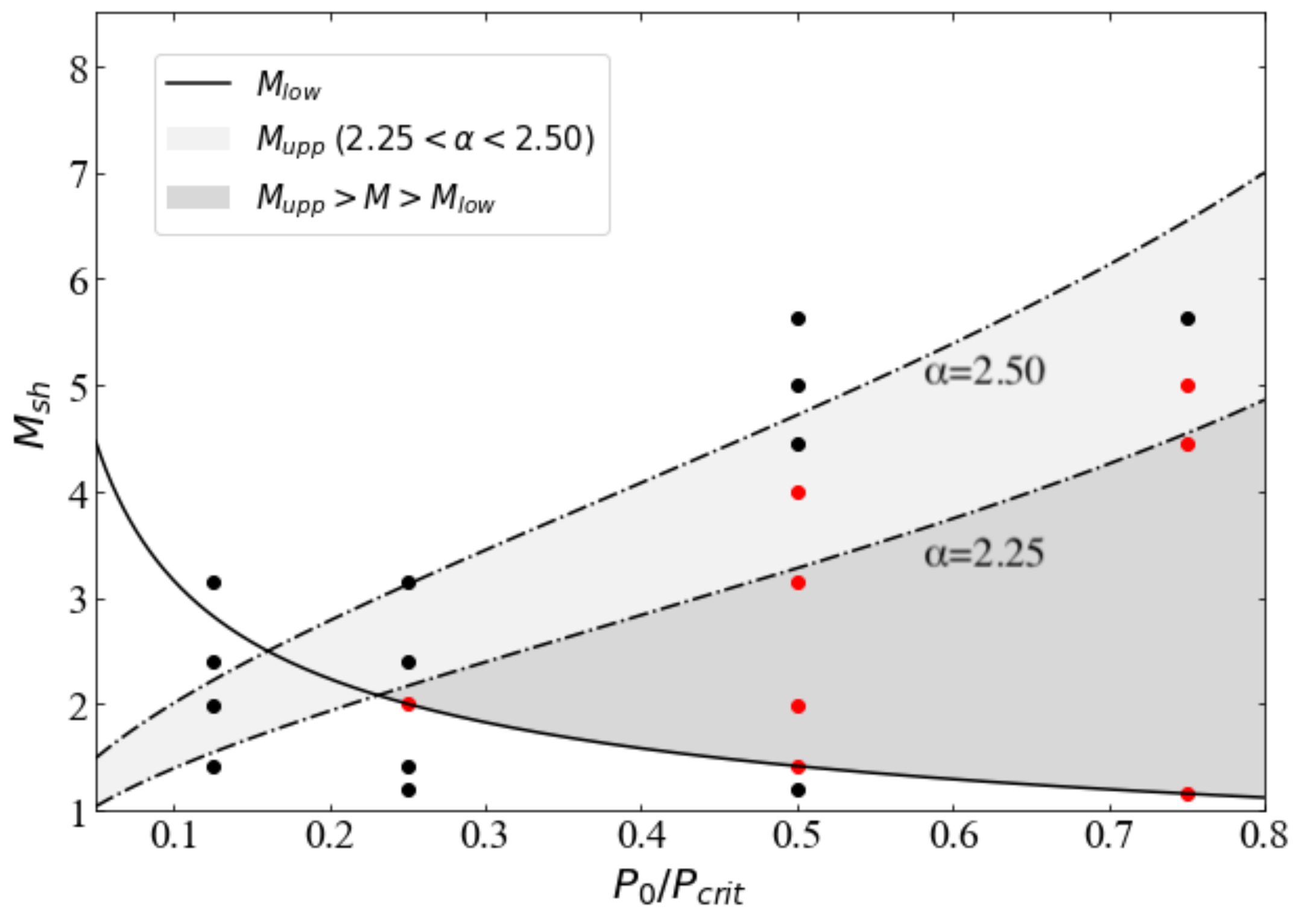}
\begin{flushleft}
\caption{Conditions on the pressure ratio $P_{\rm 0}/P_{\rm crit}$ versus Mach number for initially stable clouds. \edit1{The black} solid line \edit1{indicates} the lower limit of Mach number $M_{\rm low}$ represented by Equation (\ref{eq:lower_M_nond}). The light gray shaded area \edit1{indicates} the upper limit of Mach number $M_{\rm upp}$ represented by Equation (\ref{eq:M_upper_}) applying $2.00<\alpha<2.50$. The dashed-dot lines show the $M_{\rm upp}$ applying $\alpha=2.25$ and 2.5. The dark gray shaded area is the region \edit1{in which} cloud collapse can be induced by \edit1{a} shock. \edit1{The red} and black points indicate initial cloud states for which clouds will collapse or not\edit1{,} respectively (see also Figure \ref{fig:formed_result}.)}
\label{fig:M_range_p}
\end{flushleft}
\end{center}
\end{figure}

Figure \ref{fig:M_range_p} shows the pressure ratio $(P_{\rm 0}/P_{\rm crit})$ vs  $(M_{\rm sh})$ parameter space for an initially stable Bonnor-Ebert sphere. The solid line demarcates the Mach number estimate based on Equation (\ref{eq:lower_M_nond})\edit1{,} above which the cloud will become unstable by the shock.
The light gray shaded area shows the upper limit estimate for Mach number based on Equation (\ref{eq:M_upper_}) applying $2.25<\alpha<2.50$\edit1{,} above which the cloud will be destroyed by hydrodynamic instabilities before the collapse.
In the dark gray region, clouds are expected to be induced to collapse by shocks.
Based on simulation results, \edit1{the} red points indicate initial cloud parameter pairs corresponding to eventual collapse, while black points indicate cloud parameter pairs corresponding to predicted non-collapse. \edit1{We can observe} that red points are in dark or light gray regions and black points are outside of the light gray region. Hence, the simulation results are in \edit1{close} agreement with our estimates of conditions. However, the upper limit estimate for Mach numbers has \edit1{a} small range in terms of $\alpha$. This upper limit is derived based on \edit1{an} estimate, and hydro instability is a complex process with uncertainty. It may be difficult to \edit1{perfectly} set an upper limit with a single line. \edit1{However,} the range of $\alpha$ used is not \edit1{excessively} large. The criteria we have derived are useful for \edit1{crude} estimates of stable cloud collapse.

\begin{figure}[hbtp]
\begin{center}
\includegraphics[width=145mm]{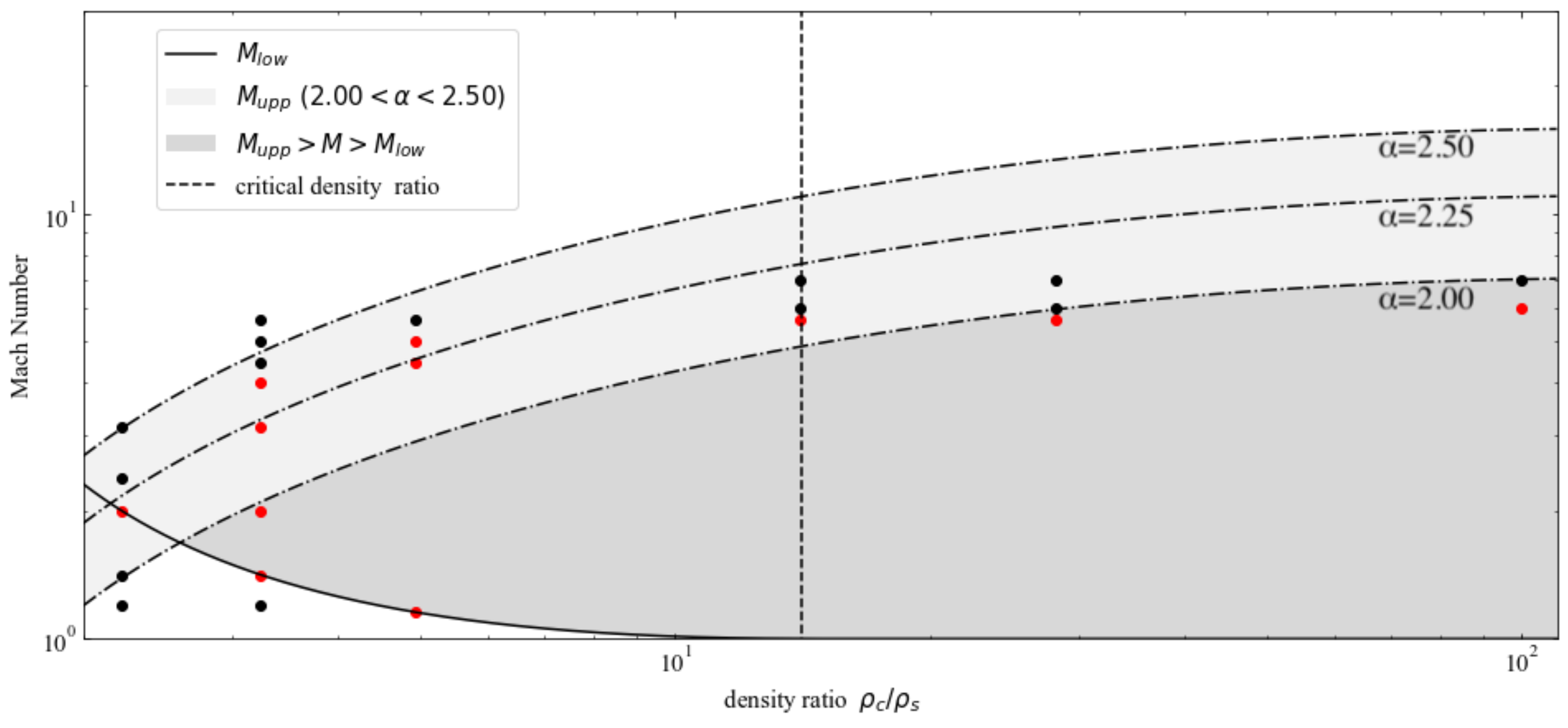}
\begin{flushleft}
\caption{As Figure \ref{fig:M_range_p}. Conditions on the density ration $\rho_{\rm c}/\rho_{\rm 0}$ versus Mach number for all models. The dashed \edit1{lines} indicate the critical density \edit1{ratio} above which cloud is unstable. The dashed-dot lines \edit1{indicate} the upper limit of the Mach number applying $\alpha=2.0, 2.25$, and $\alpha=2.5$.}
\label{fig:M_range_rho}
\end{flushleft}
\end{center}
\end{figure}

Figure \ref{fig:M_range_rho} shows the density ratio $(\rho_{\rm c}/\rho_{\rm 0})$ vs $(M_{\rm sh})$ parameter space similar to Figure \ref{fig:M_range_p} with all parameter pairs shown in Figure \ref{fig:formed_result}. 
For initially unstable clouds (i.e.\edit1{,} in the right-side region of the critical density \edit1{ratio} line), when applying smaller $\alpha$ ($\sim$ 2.0), results of initially unstable clouds consist with the upper limit estimate. The reason for this gap in $\alpha$ range between stable and unstable clouds is that for unstable clouds, the propagating shock speed estimated \edit1{using} Equation (\ref{eq:v_cl_in_the_cloud}) and actual one \edit1{are not} an exact match.
For unstable clouds, the density ratio $\rho_{\rm c}/\rho_{\rm 0}$ is larger and a density gradient along the radial direction affects the shock velocity propagating in the cloud.
Therefore, the timescale for shock propagation and cloud contraction expressed by the \edit1{left-hand} side of Equation (\ref{eq:t_upper_con}) is affected, and the $\alpha$ range changes. 
However, our upper limit estimate is useful for an order discussion. Although the upper limit has some uncertainty for larger unstable clouds, the general tendency of results is consistent with our estimate.  

In our calculation, for clouds initially at $\xi>6.45$, at the time when propagating shock \edit1{begins} to compress clouds, little gravitational collapse is in progress. That is, the density profile of the clouds is almost identical between the initial conditions and when the shocks arrive. We note that \edit1{for} $\xi>6.45$ unstable clouds\edit1{,} in which collapse has progressed further by the time the shock arrives, shock-cloud evolution may change\edit1{,} and ranges of $M_{\rm upp}$ \edit1{may differ} from our results.

\subsection{Asymptotic mass of sink particles}

\begin{figure*}
      \begin{tabular}{c}
      \begin{minipage}{0.45\hsize}
        \begin{center}
          \includegraphics[clip, width=8.0cm]{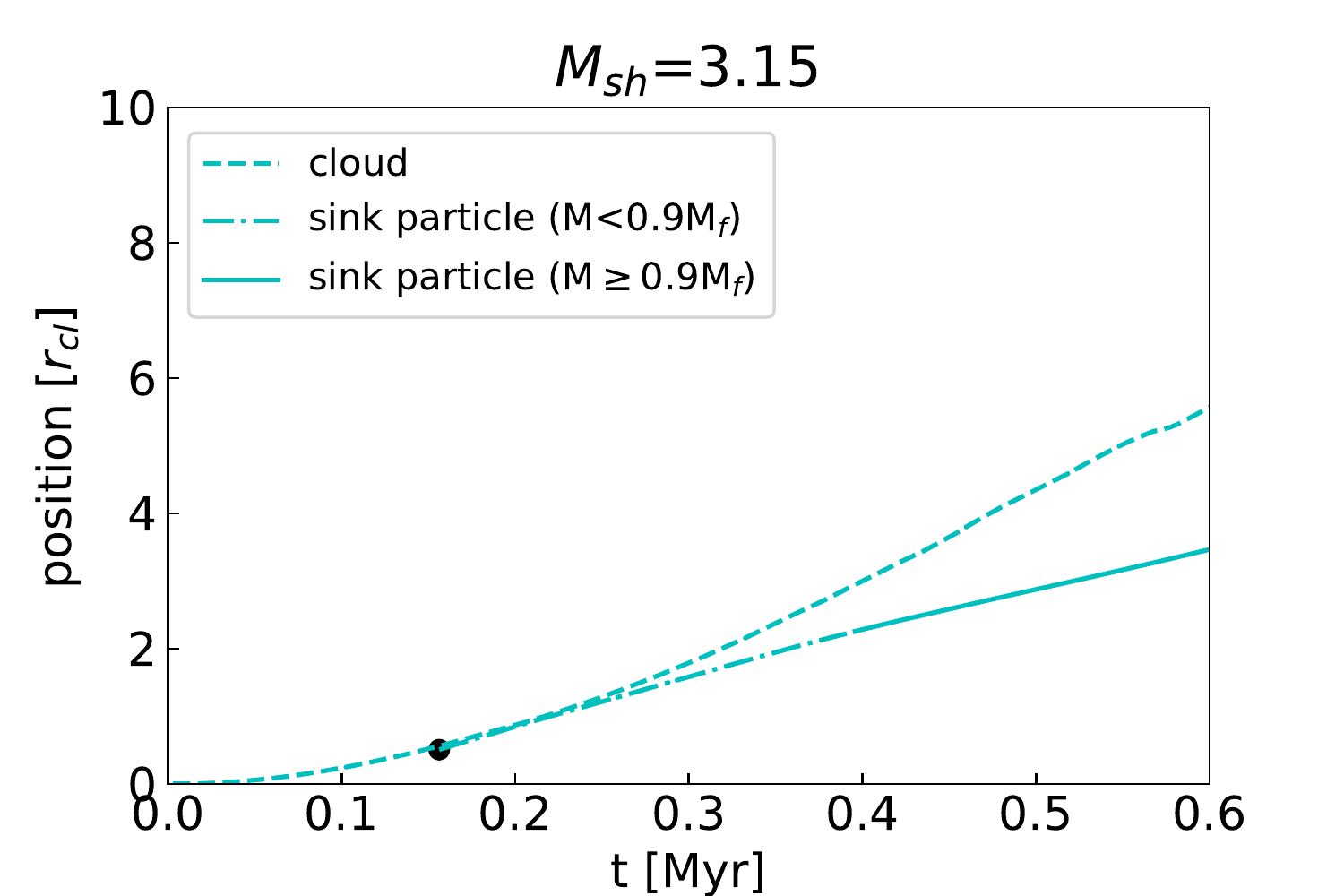}
          \hspace{1.0cm} (a) $M_{sh}=3.15$
        \end{center}
      \end{minipage}

      \begin{minipage}{0.45\hsize}
        \begin{center}
          \includegraphics[clip, width=8.0cm]{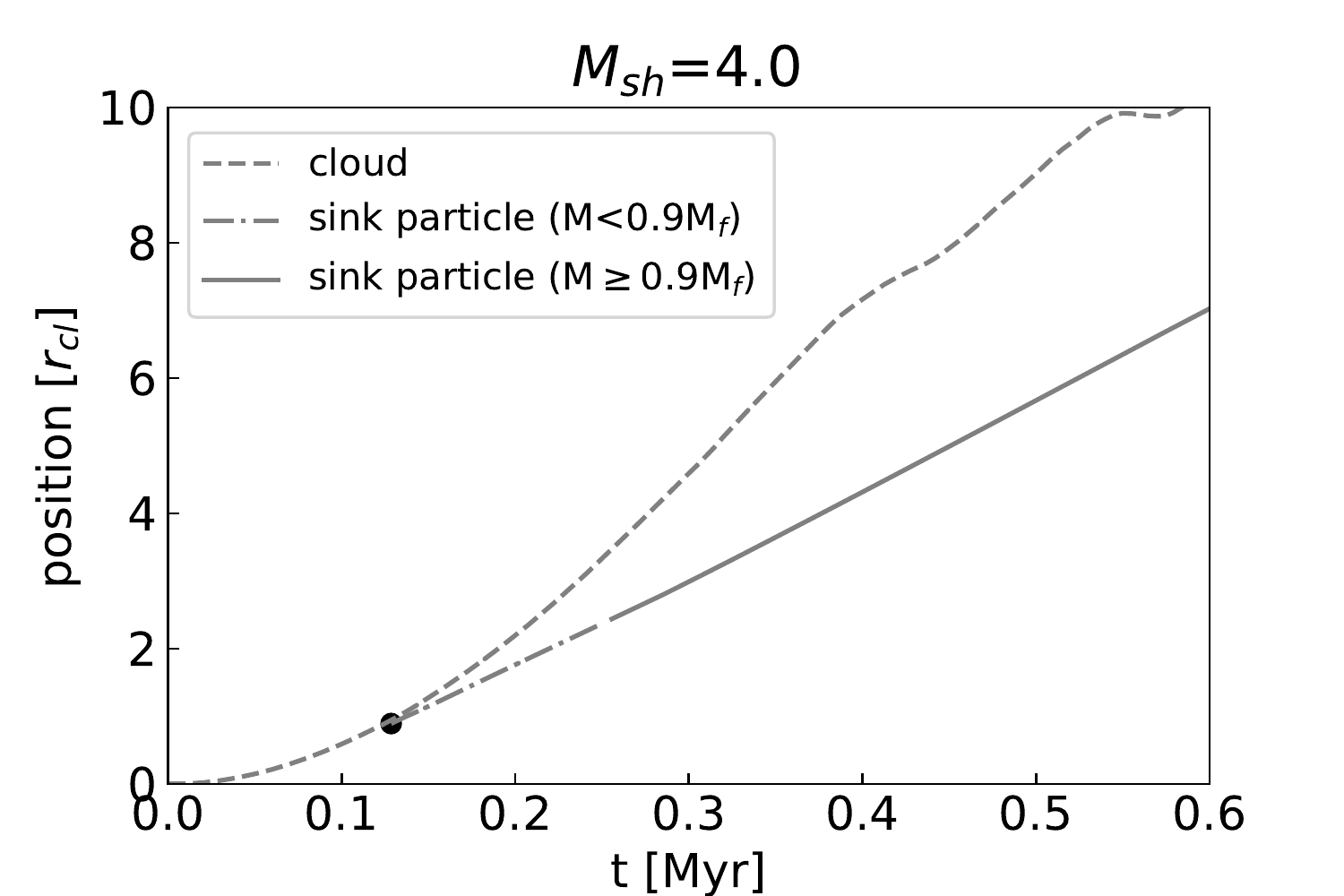}
          \hspace{1.0cm} (b) $M_{sh}=4.00$
        \end{center}
      \end{minipage}

      \end{tabular}

      \vskip5pt  
    \caption{(a): Position of sink particles and stripped clouds for $\xi=3.22$ and $M_{\rm sh}=3.15$. The dashed line indicates the $\langle z\rangle$ defined by the Equation (\ref{eq:z}). When the mass of the sink particle is less than 90$\%$ of the asymptotic mass, the position of the $z$ coordinate of the sink particle is \edit1{indicated} by the dashed-dot line. \edit1{Moreover}, when it is more than 90$\%$ of the asymptotic mass, that is \edit1{indicated} by the solid line. The unit of the shown position is $r_{\rm cl}$ with the origin the position of the center of the initial cloud. The black point indicates the position when the sink particle is created. (b): As (a) for $M_{\rm sh}=4.00$.}
    \label{fig:sink_pos}
\end{figure*}

As shown in Section \ref{subsub:Evolution of sink particles}, comparing results with the same initial radii, the higher \edit1{the} Mach number of the shock is, the lower \edit1{the} asymptotic sink particle mass \edit1{becomes} and the shorter \edit1{the} accretion time is. This would be because the higher the Mach number is, the faster the shock \edit1{strips} the cloud around sink particles.
As shown in Figure \ref{fig:slice_05_3_15}, the cloud around the sink particle is stripped and mixed with ambient gas. Cloud material is accelerated and displacement of cloud material and the sink particle \edit1{expands}. \edit1{Therefore}, the accretion timescale would be determined by how \edit1{rapidly} the incoming shock can accelerate and strip accreting clouds from sink particles.

Here, we will \edit1{quantitatively analyze} the moving of stripped clouds and sink particles. 
We use the $\langle z\rangle$ defined in Equation (\ref{eq:z}) to analyze the moving of stripped clouds.

Figure \ref{fig:sink_pos} shows the $\langle z\rangle$ and position of sink particles in the z-direction for $\xi=3.22$ and $M_{\rm sh}=3.15$ \edit1{and} $4.00$. Displacement of the cloud and sink particle is initially small because of the low mass of sink particles. \edit1{Subsequently}, the sink particle drifts behind and displacement becomes larger because accretion progresses and the sink particle mass \edit1{increases}.

\begin{figure*}
      \begin{tabular}{c}
      \begin{minipage}{0.5\hsize}
        \begin{center}
          \includegraphics[clip, width=8.0cm]{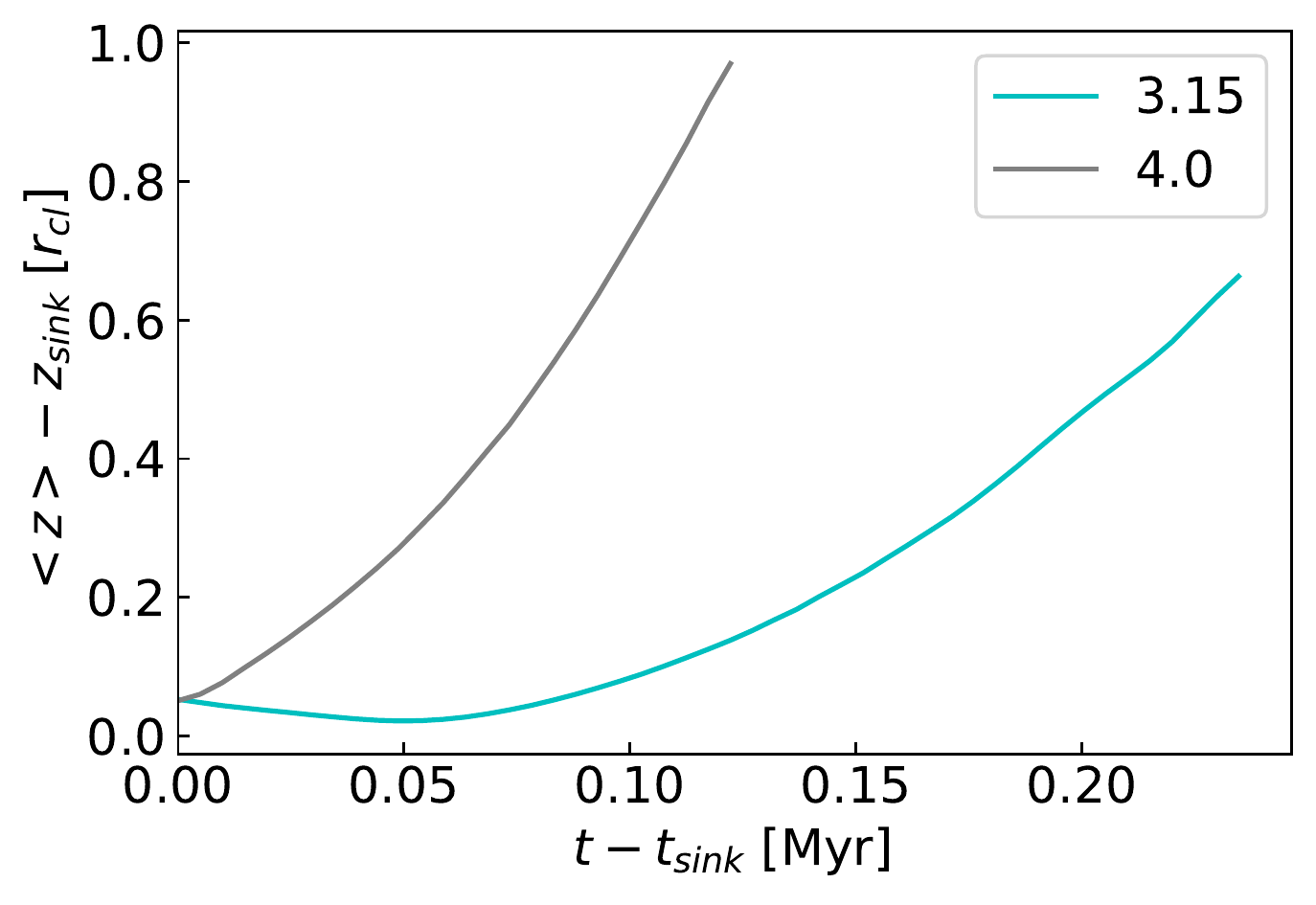}
          \hspace{1.6cm} (a) Relative \edit1{displacement}
        \end{center}
      \end{minipage}

      \begin{minipage}{0.5\hsize}
        \begin{center}
          \includegraphics[clip, width=8.5cm]{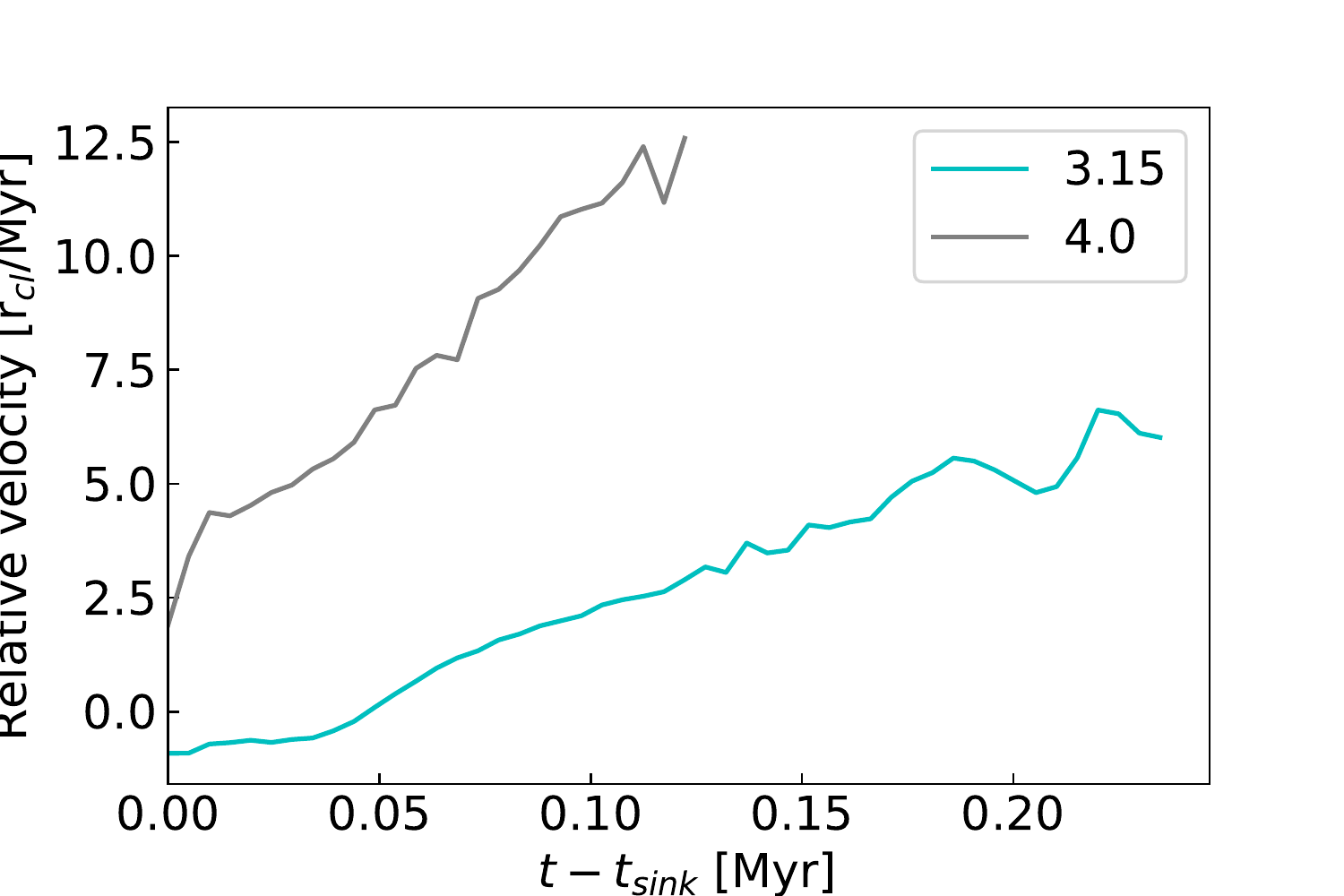}
          \hspace{1.6cm} (b) Relative velocity
        \end{center}
      \end{minipage}

      \end{tabular}
      \vskip5pt  
    \caption{(a): \edit1{Relative} \edit1{displacement between} sink particles and stripped clouds in the z-direction for $\xi=3.22$ and $M_{\rm sh}=3.15$ \edit1{and} $4.00$ \edit1{while} the mass of the sink particle is less than 90$\%$ of the asymptotic mass. The unit of the relative \edit1{displacement} is $r_{\rm cl}$. The horizontal line \edit1{indicates} the time after the formation of the sink particle. (b): As (a) for the relative velocity \edit1{between} sink particles and stripped clouds. The unit of the relative velocity is $r_{\rm cl}/\rm Myr$. }
    \label{fig:sink_relative}
\end{figure*}

Figure \ref{fig:sink_relative} shows the relative \edit1{displacement} and velocity \edit1{between} $\langle z\rangle$ and those of sink particles for $\xi=3.22$ and $M_{\rm sh}=3.15$ \edit1{and} $4.00$. The higher the Mach number of the shock is, the faster the relative \edit1{displacement} and velocity \edit1{increase}. This trend is also \edit1{observed} for other radii \edit1{values} (see Appendix \ref{app:Moving}).
Therefore, relative displacement and velocity of the particle and cloud \edit1{are} determined by the shock speed and \edit1{affect} the timescale of accretion and asymptotic mass. 

\subsection{Effect of turbulence}
\label{sec:Effect of turbulence existence}

\begin{figure*}
  
    \begin{tabular}{c}

      \begin{minipage}{0.33\hsize}
        \begin{center}
          \includegraphics[clip, width=6.5cm]{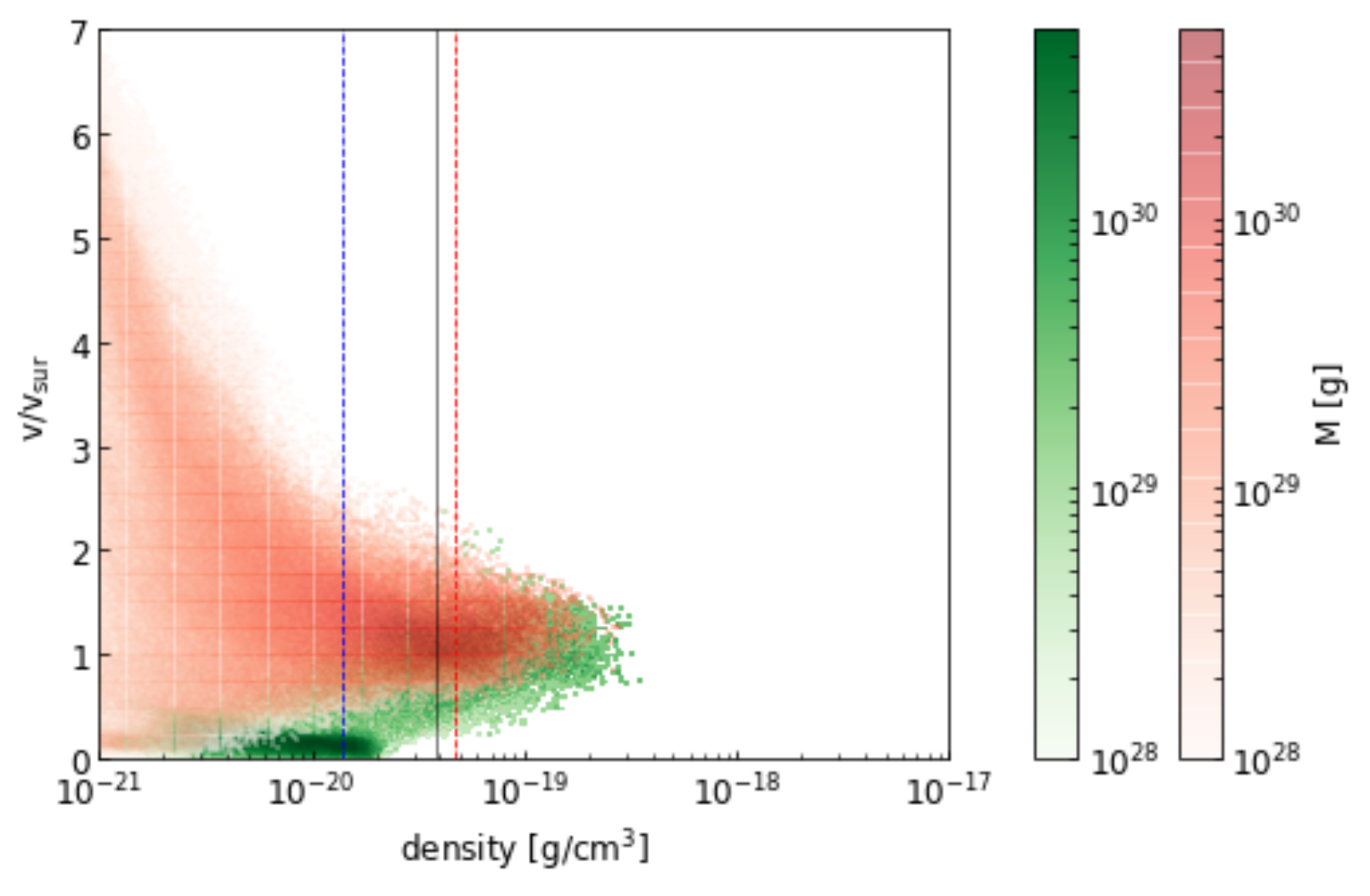}
          \hspace{2.0cm} (a) Turbulent cloud model, t = 0.16 Myr
        \end{center}
      \end{minipage}

      \begin{minipage}{0.33\hsize}
        \begin{center}
          \includegraphics[clip, width=6.5cm]{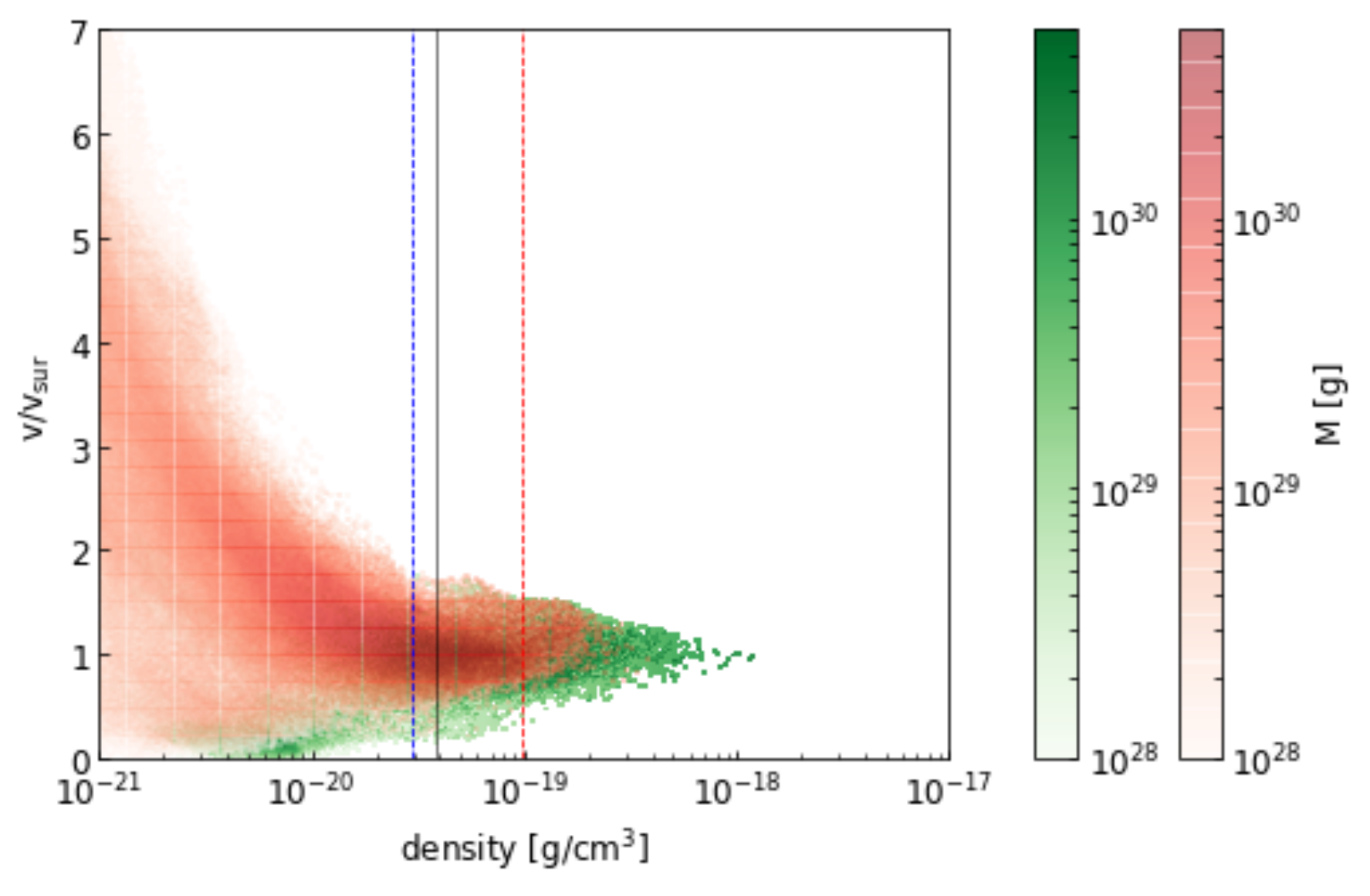}
          \hspace{2.0cm} (b) Turbulent cloud model, t = 0.31 Myr 
        \end{center}
      \end{minipage}

      \begin{minipage}{0.33\hsize}
        \begin{center}
          \includegraphics[clip, width=6.5cm]{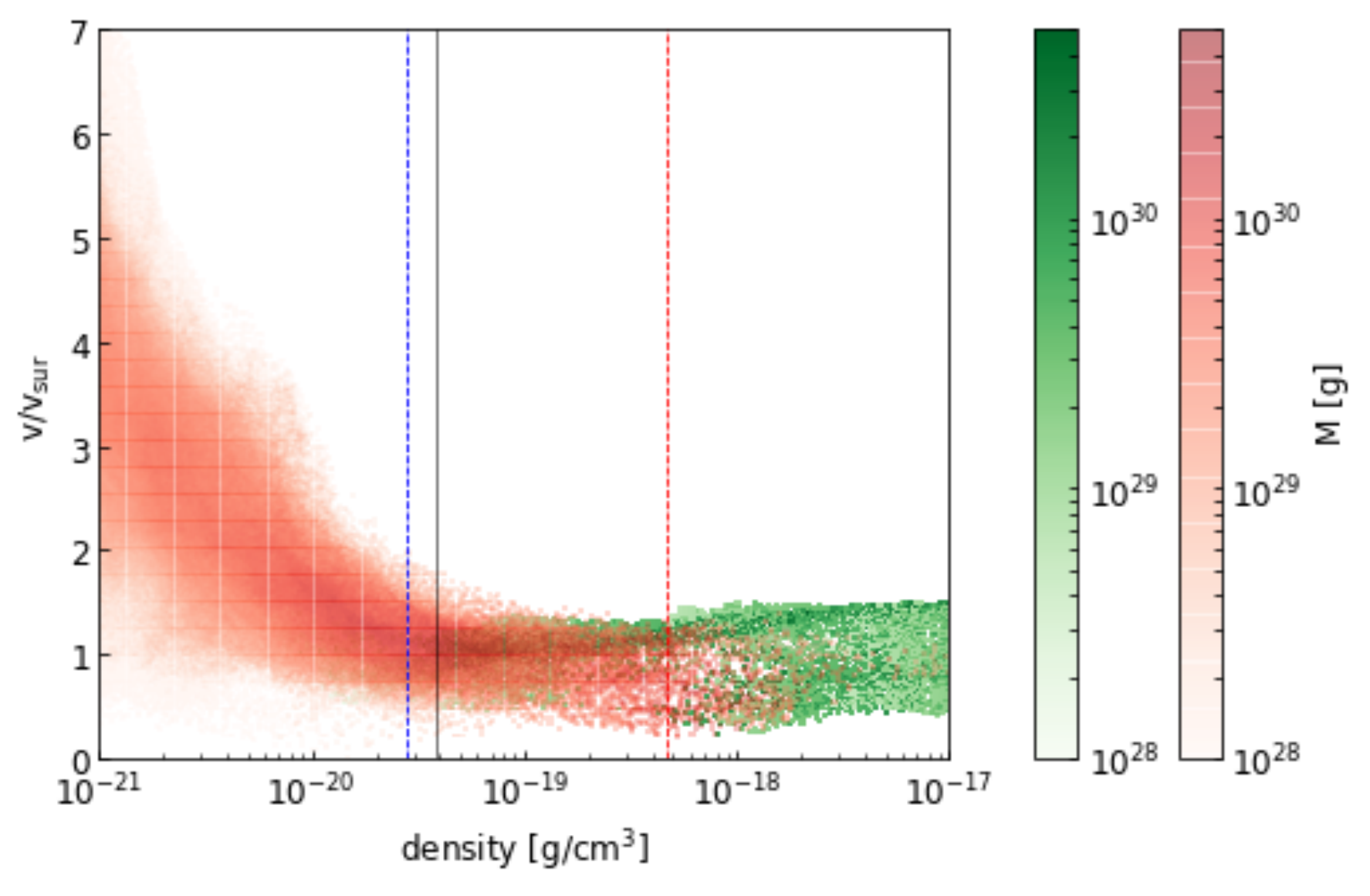}
          \hspace{2.0cm} (c) Turbulent cloud model, t = 0.49 Myr 
        \end{center}
      \end{minipage}
      
       \end{tabular}
      
      \begin{tabular}{c}
      \begin{minipage}{0.33\hsize}
        \begin{center}
          \includegraphics[clip, width=6.5cm]{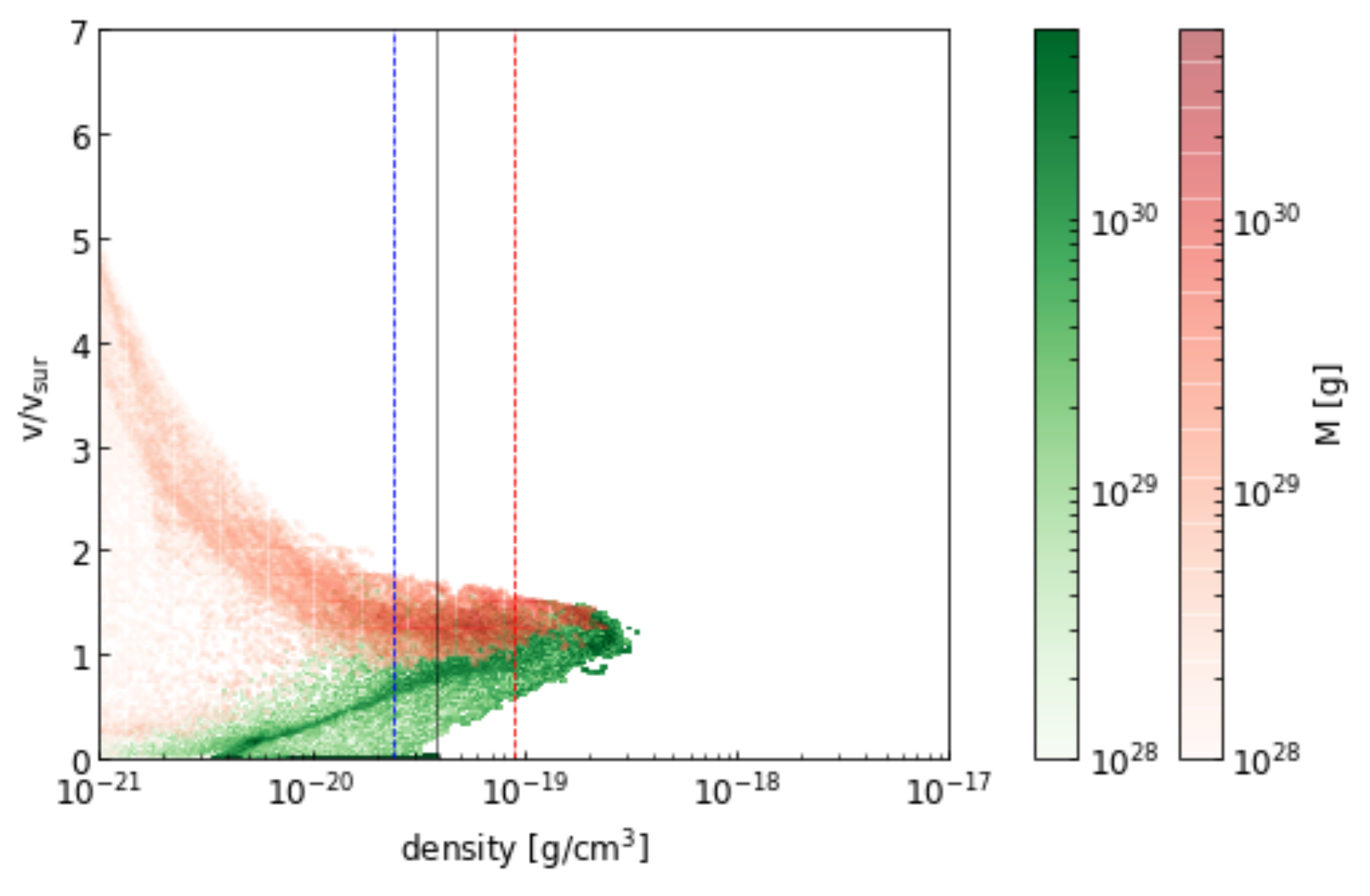}
          \hspace{2.0cm} (d) Fiducial cloud model, t = 0.16 Myr 
        \end{center}
      \end{minipage}

      \begin{minipage}{0.33\hsize}
        \begin{center}
          \includegraphics[clip, width=6.5cm]{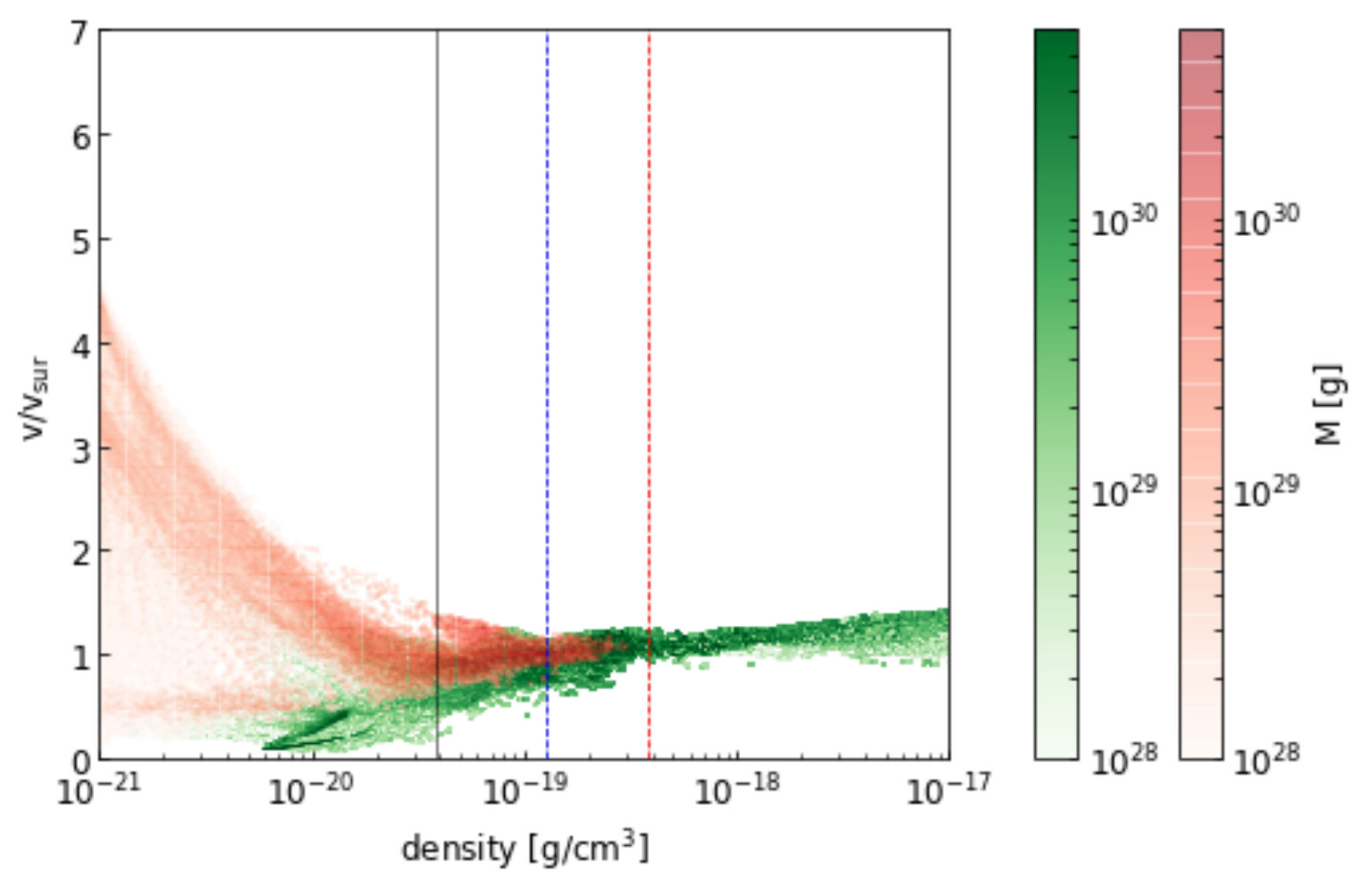}
          \hspace{2.0cm} (e) Fiducial cloud model, t = 0.31 Myr 
        \end{center}
      \end{minipage}

    \end{tabular}
    \vskip5pt  
    \caption{As Figure \ref{fig:pdf-rho-two}. (a),(b)\edit1{,} and (c): For $M_{\rm sh}=3.15$ and initially turbulent cloud model at t=0.16, 0.31, and 0.49 Myr, respectively. (c) corresponds to \edit1{immediately} before the sink particle is formed. (d) and (e) For $M_{\rm sh}=3.15$ and initially no-turbulent fiducial cloud model at t=0.16 \edit1{and} 0.31 Myr. (e) corresponds to \edit1{immediately} before the sink particle is formed.}
    \label{fig:pdf-rho-tur}

\end{figure*}

Figure \ref{fig:pdf-rho-tur} shows the \edit1{mass per unit velocity and density interval} as Figure \ref{fig:pdf-rho-two} for both turbulent and non-turbulent \edit1{clouds} ($M_{\rm sh}=3.15$ and $\xi=6.45$).  Figure \ref{fig:pdf-rho-tur} (a) and (d) are at the same time. Figure \ref{fig:pdf-rho-tur} (b) and (e) also at the same time, and Figure \ref{fig:pdf-rho-tur} (e) corresponds to the \edit1{time immediately} before \edit1{the creation of the} sink particle. Comparing Figure \ref{fig:pdf-rho-tur} (a) and (d), \edit1{the} cloud mass \edit1{is} less \edit1{distributed} on the high density side \edit1{on the former} compared \edit1{with the latter}. As \edit1{discussed} in Section \ref{sec:Initially turbulent cloud cases}, the turbulent cloud has a slower increase in density and a slower sink particle formation than the non-turbulent cloud. This slowness would be due to the effective pressure from turbulence in the cloud. Internal turbulence increases effective cloud pressure to the order of $\sim \rho \sigma^{2}$. This increased pressure enhances cloud diffusivity and suppresses cloud contraction by the external ram pressure due to \edit1{a} shock. Therefore, turbulence prevents rapid gravitational collapse and high density region is formed slowly.

\edit1{As shown in} Figure \ref{fig:pdf-rho-tur}, throughout the evolution \edit1{of} the turbulent cloud, the distribution of the gas shown in red ($C<0.99$) \edit1{differs} from that of the non-turbulent cloud. For example, \edit1{as shown} in Figure \ref{fig:pdf-rho-tur} (d), red and green gas shows near V-shaped distribution, while in Figure \ref{fig:pdf-rho-tur} (a), red gas shows a fan-like distribution. In other words, more gas is dispersed and mixed with the ambient gas. This trend is consistent with that of Figure \ref{fig:turbulent_com} (b) and (c).
Clouds with a turbulent velocity are more prone to Kelvin-Helmholtz instabilities than \edit1{their} non-turbulent counterparts. Therefore, \edit1{we can infer} that turbulent vortical motions in the clouds diffuse the cloud, and the timescale of the mixture becomes shorter. 

Figure \ref{fig:sink_relative} shows the relative \edit1{displacement} and velocity \edit1{between} $\langle z\rangle$ and those of sink particles for both turbulent and non-turbulent \edit1{clouds} ($M_{\rm sh}=3.15$ and $\xi=6.45$). For the turbulent case, throughout the entire evolution, the relative \edit1{displacement} is greater than that of the non-turbulent case. Most of the time, the relative velocity is also greater for the turbulent case. One reason for this trend is that the cloud with a turbulent velocity is \edit1{more easily} destroyed by the shocks and the gas around the sink particles is stripped faster downstream of the shock. Another reason is that, for the turbulent case, the sink particle \edit1{is} formed later, and by the time mass accretion \edit1{begins}, more gas \edit1{has} been accelerated. \edit1{Therefore}, the asymptotic sink particle mass in the turbulent clouds models is lower than the non-turbulent counterparts.

Comparing No.5 and No.8 models results in Appendix \ref{app:Simulation results}, the upper limit of the Mach number for cloud collapse \edit1{differed}. In turbulent cloud cases, the upper limit of the Mach number is $3.15<M_{\rm upp}<4.46$ \edit1{whereas} for the not turbulent cases, it is $5.64<M_{\rm upp}<6.00$. \edit1{That is}, \edit1{the} parameter window for the cloud collapse \edit1{become} narrower for turbulent cloud cases. 

\edit1{Thus}, turbulence in the cloud makes triggered star formation by shocks more difficult. Pressure due to turbulence \edit1{retards} cloud contraction and facilitates mixing with the ambient gases. The asymptotic stellar mass \edit1{decreases}.  
\edit1{Note} that our simulation models \edit1{did not} address magnetic fields. In a realistic ISM, magnetic fields exist and can alter the physical process of the shocked cloud. In this \edit1{study}, we \edit1{considered} purely hydrodynamic cases and investigate the effects of turbulence.

\begin{figure*}
      \begin{tabular}{c}
      \begin{minipage}{0.5\hsize}
        \begin{center}
          \includegraphics[clip, width=8.0cm]{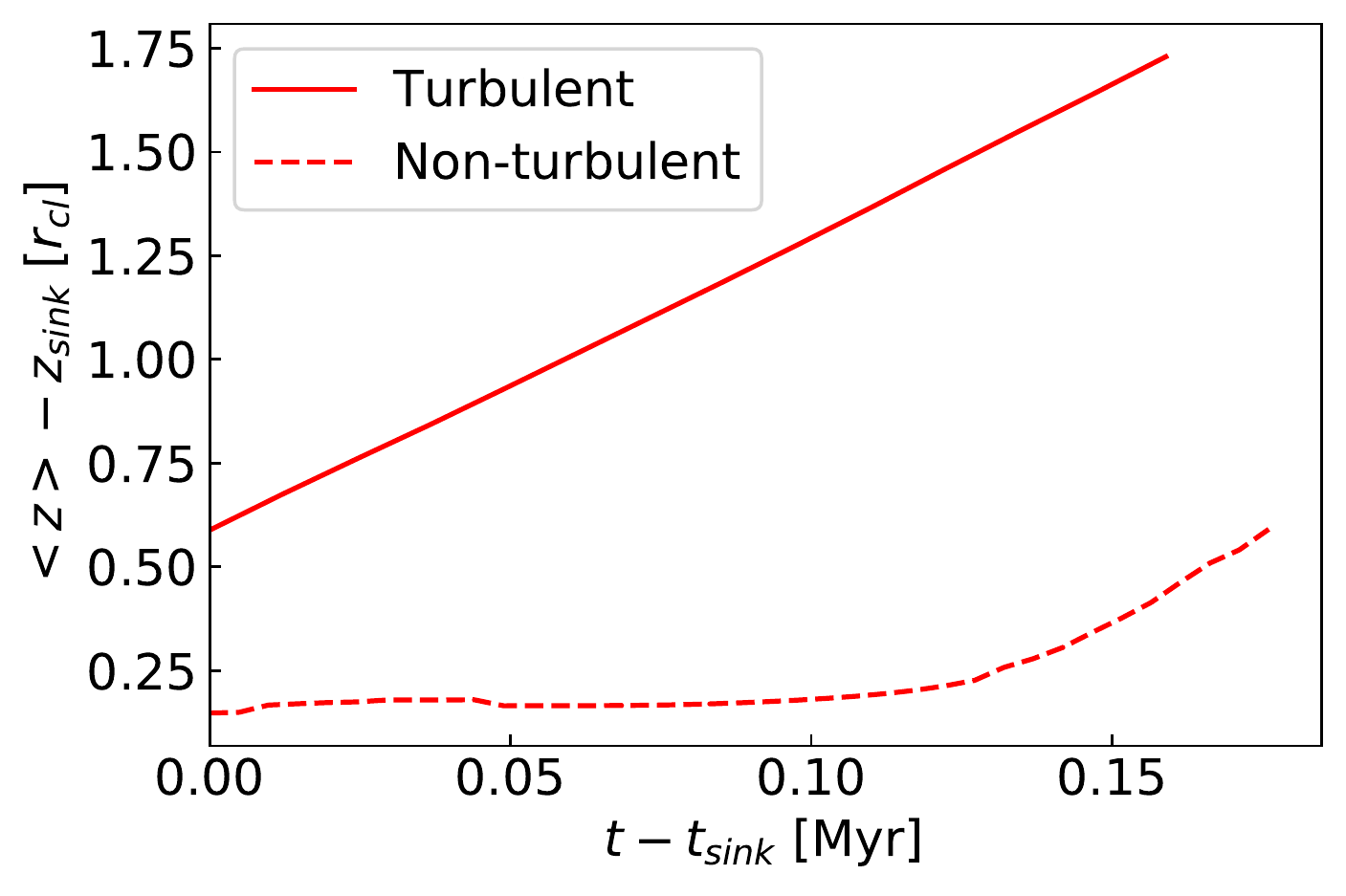}
          \hspace{1.6cm} (a) Relative \edit1{displacement}
        \end{center}
      \end{minipage}

      \begin{minipage}{0.5\hsize}
        \begin{center}
          \includegraphics[clip, width=8.5cm]{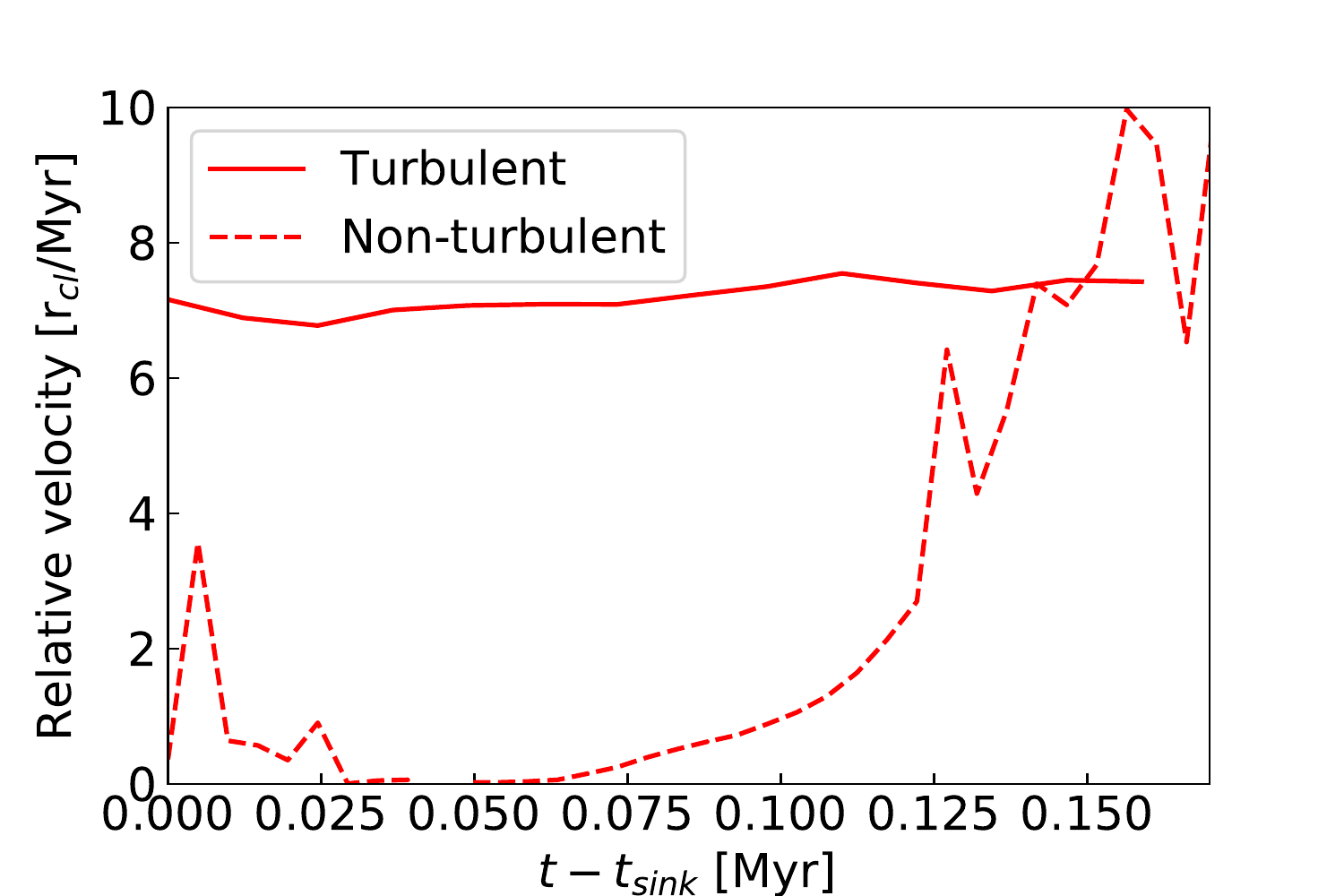}
          \hspace{1.6cm} (b) Relative velocity
        \end{center}
      \end{minipage}

      \end{tabular}
      \vskip5pt  
    \caption{As \edit1{in} Figure \ref{fig:sink_relative}, relative \edit1{displacement} and velocities \edit1{between} sink particles and stripped clouds. Both turbulent and non-turbulent models for $\xi=6.45$ and $M_{\rm sh}=3.15$ are shown. Solid lines are for initially turbulent cloud cases. Dashed-dot lines are for non-turbulent cloud cases. }
    \label{fig:sink_relative_tur}
\end{figure*}

\clearpage
\subsection{Comparison with observation}
\subsubsection{Globules toward Orion's veil bubble}
\edit1{The} Extended Orion Nebula (M42) is photoionized by \edit1{a} massive star in the Trapezium cluster, $\theta^{1}$ Ori C (e.g., \citealp{O'Dell_2001}; \citealp{2006A&A...448..351S}). 
Using the IRAM 30m telescope, \citet{Goicoechea_2020} \edit1{presented} $^{12}\rm CO$ and $^{13}\rm CO$ $(J=2-1)$ maps of \edit1{the} “Veil bubble" \edit1{driven} by the strong wind emanating from $\theta^{1}~$ Ori C. They \edit1{indicated} the presence of ten CO “globules" blueshifted from the OMC and embedded in the expanding shell that encloses the bubble. These CO globules are small ($R_{\rm g}\simeq7100$ AU), not massive ($M_{\rm g}\simeq0.3 M_{\odot}$) and are moderately dense: $n_{\rm H}\simeq4.0\times10^{4} \rm cm^{-3}$ (median values of the sample). \citet{Goicoechea_2020} \edit1{assumed} that they are either transient objects formed by hydrodynamic instabilities or pre-existing over-dense structures of the original molecular cloud. They are sculpted by the passing shock associated with the expanding shell and by UV radiation from the Trapezium. From the estimated masses of globules, \edit1{they} deduced that these globules will not easily form stars. For some globules, their masses are greater than the Bonnor-Ebert mass (see Equation (\ref{eq:bonnor_critical_M})) \edit1{but} less than Jeans mass. 

\begin{figure}[hbtp]
\begin{center}
\includegraphics[width=80mm]{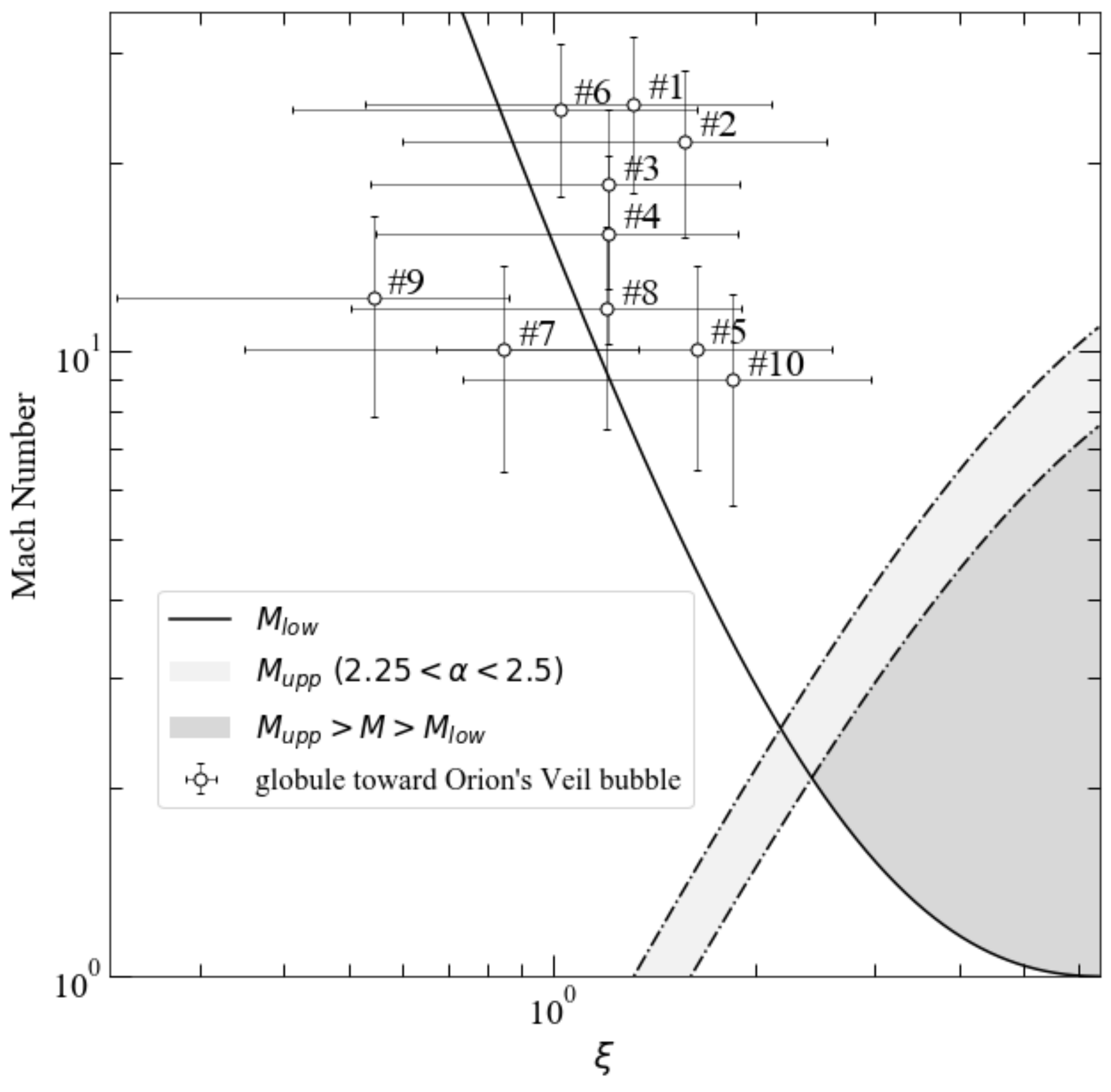}
\begin{flushleft}
\caption{As \edit1{in} Figure \ref{fig:M_range_p}. Conditions \edit1{of} the dimensionless radius $\xi$ versus Mach number. Parameter pairs of ten globules toward Orion's \edit1{Veil} bubble are plotted \citep{Goicoechea_2020}. The \# numbers near globules parameter pairs correspond to the numbers shown in Table 1 in \citet{Goicoechea_2020}. These parameter pairs \edit1{were} calculated using Table 1 values ($T_{\rm ex}(\rm CO)$, $R_{\rm g}$ and $n_{\rm H}$) and expanding shell speed $v_{\rm sh}=$13 $\rm km ~s^{-1}$ assuming \edit1{that the} initial gas density between globules and surrounding ISM is 10.}
\label{fig:globules}
\end{flushleft}
\end{center}
\end{figure}

We calculated \edit1{the} dimensionless radii and $M_{\rm sh}$ of these globules assuming that they are ideal Bonner-Ebert spheres and pressure equilibrium with the ambient gas. We note that our calculations and estimation \edit1{were} simplistic. In calculation, we \edit1{ignored} magnetic fields and turbulence. We also \edit1{assumed} the pre-shocked cloud sphere, but in \edit1{practice,} these globules have been compressed to some extent. \edit1{The estimation} here is \edit1{only} a \edit1{crude one} to investigate the general trend. In our calculation, we adopted estimated globule parameters ($T_{\rm ex}(\rm CO)$, $R_{\rm g}$\edit1{,} and $n_{\rm H}$) listed in Table 1 in \citet{Goicoechea_2020} and expanding shell speed $v_{\rm sh}$=13 km s$^{-1}$, assuming \edit1{that the} initial gas density between \edit1{the} globules surface and surrounding ISM is 10. Figure \ref{fig:globules} shows dimensionless radius $\xi$ versus Mach number parameter spaces. Derived parameter pairs of ten globules are plotted. As in Figure \ref{fig:M_range_p}, \edit1{the} estimated upper and lower limit of $M_{\rm sh}$ for collapse are also shown. 
Some globules are distributed in the parameter space above the $M_{\rm low}$, while all cores are distributed in the parameter space above the $M_{\rm upp}$. That is, most globules are strongly compressed and become dense, but they are destroyed by shocks\edit1{,} and star formation activities are limited. This prediction for the future star formation of these globules is consistent with the prediction by \citet{Goicoechea_2020}.

\begin{table}
\normalsize
\centering
\caption{Estimated parameters of globules toward Orion's Veil bubble}
\begin{tabular}{ccccc}

\toprule
globule $^{a}$& $\xi$ $^{b}$         & $M_{\rm sh}$$ ^{c}$    \\ \hline
\#1      & 1.32$\pm$0.79 & 24.9$\pm$6.99 \\
\#2      & 1.57$\pm$0.98 & 21.7$\pm$6.51 \\
\#3      & 1.21$\pm$0.68 & 18.5$\pm$5.88 \\
\#4      & 1.21$\pm$0.66 & 15.4$\pm$5.14 \\
\#5      & 1.63$\pm$0.96 & 10.1$\pm$3.64 \\
\#6      & 1.03$\pm$0.61 & 24.4$\pm$6.73 \\
\#7      & 0.85$\pm$0.50 & 10.1$\pm$3.65 \\
\#8      & 1.20$\pm$0.70 & 11.7$\pm$4.18 \\
\#9      & 0.54$\pm$0.32 & 12.2$\pm$4.32 \\
\#10     & 1.85$\pm$1.11 & 9.0$\pm$ 3.33  \\ \hline

\end{tabular}
\vskip5pt
\begin{tablenotes}

\item $^{a}$ Corresponding to \edit1{numbers in} Table 1 in \citet{Goicoechea_2020}.

$^{b}$ Estimated dimensionless radius.

$^{c}$ Estimated Mach numbers of the propagating shock.

\end{tablenotes}
\label{tab:orion core}
\end{table}

\subsubsection{Mass surface density PDFs}
\begin{figure*}
  
    \begin{tabular}{c}

      \begin{minipage}{0.33\hsize}
        \begin{center}
          \includegraphics[clip, width=6.0cm]{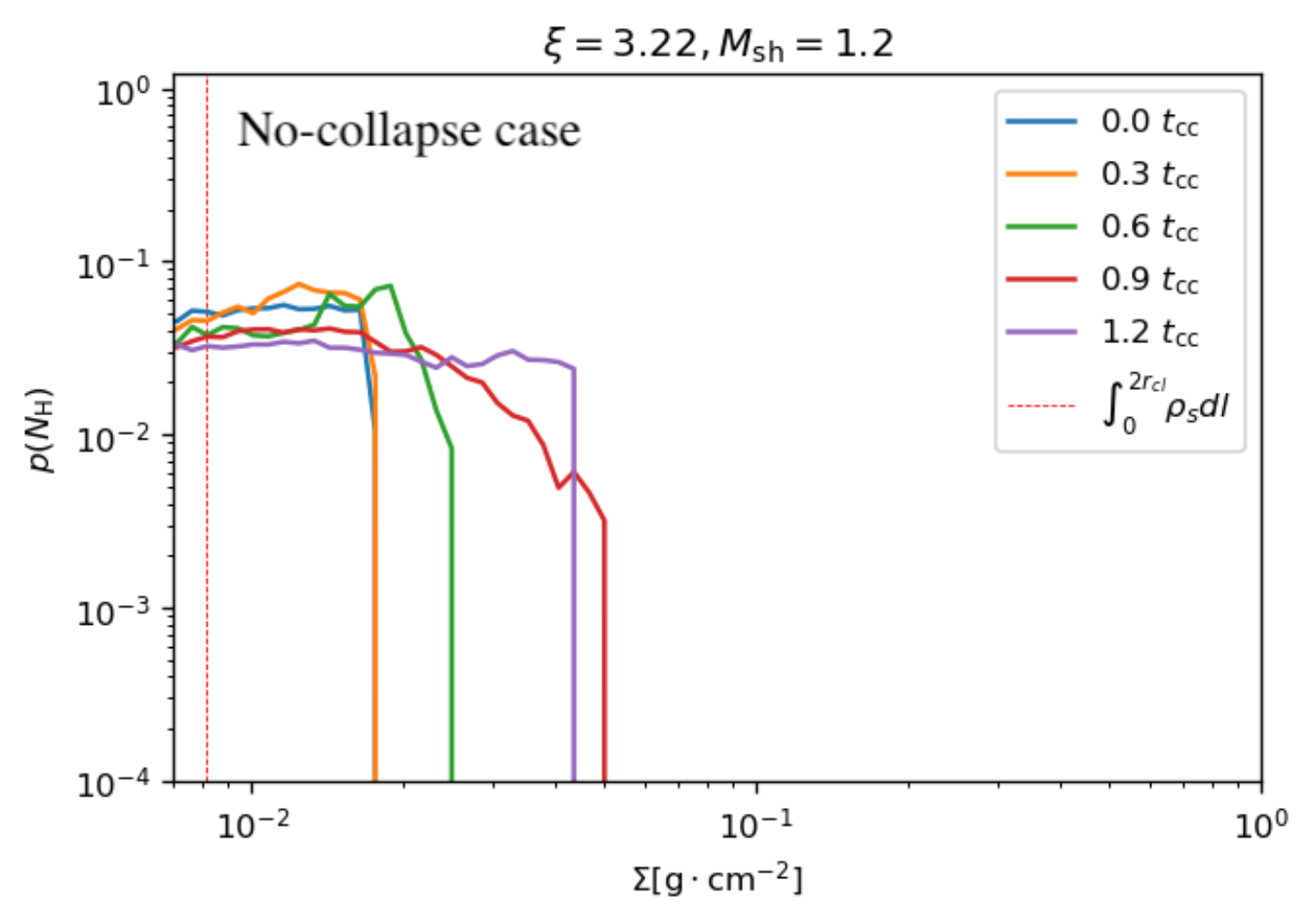}
          \hspace{2.0cm} (a) $M_{\rm sh}=1.20$, $\xi=3.22$
        \end{center}
      \end{minipage}

      \begin{minipage}{0.33\hsize}
        \begin{center}
          \includegraphics[clip, width=6.0cm]{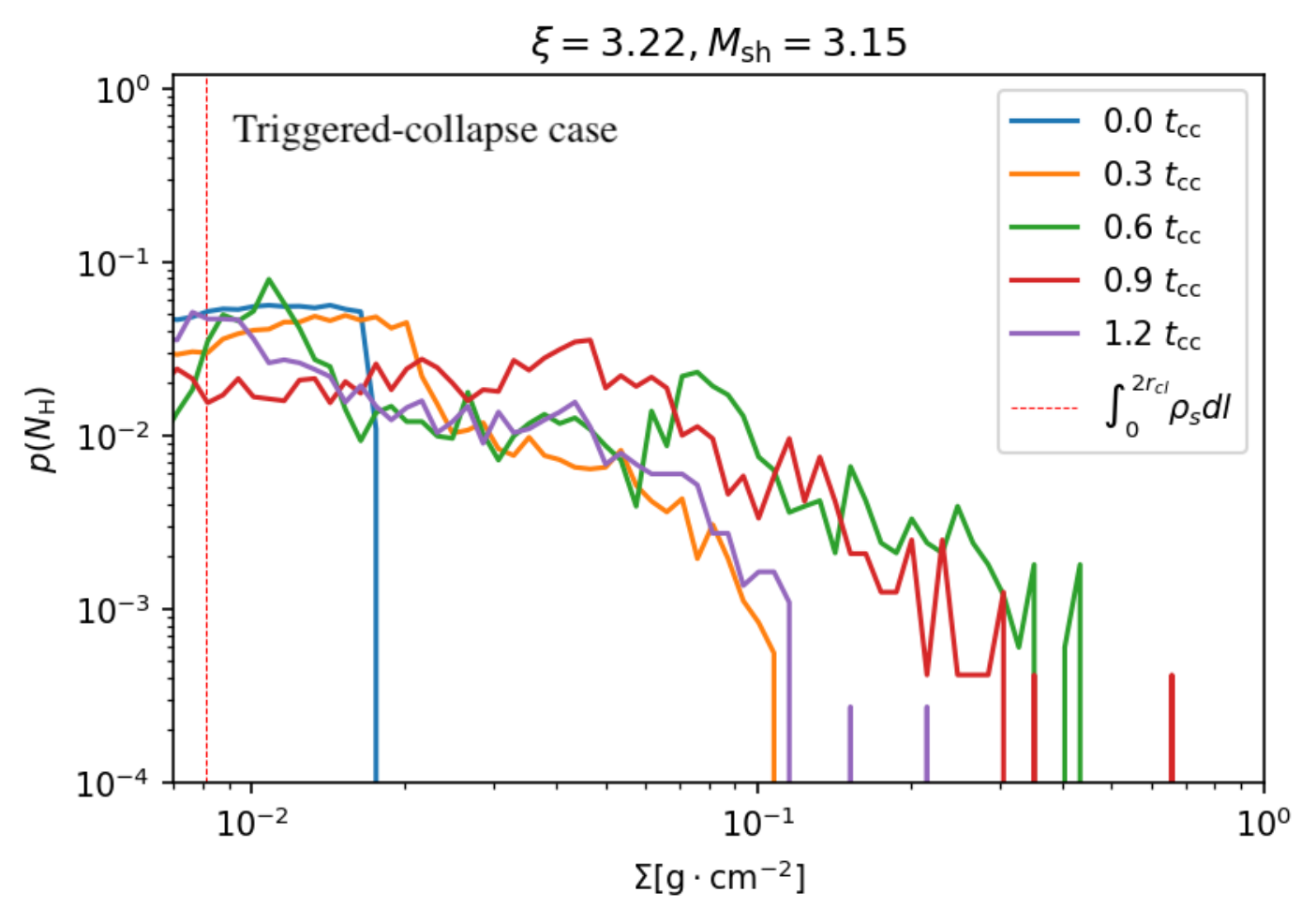}
          \hspace{2.0cm} (b) $M_{\rm sh}=3.15$, $\xi=3.22$
        \end{center}
      \end{minipage}

      \begin{minipage}{0.33\hsize}
        \begin{center}
          \includegraphics[clip, width=6.0cm]{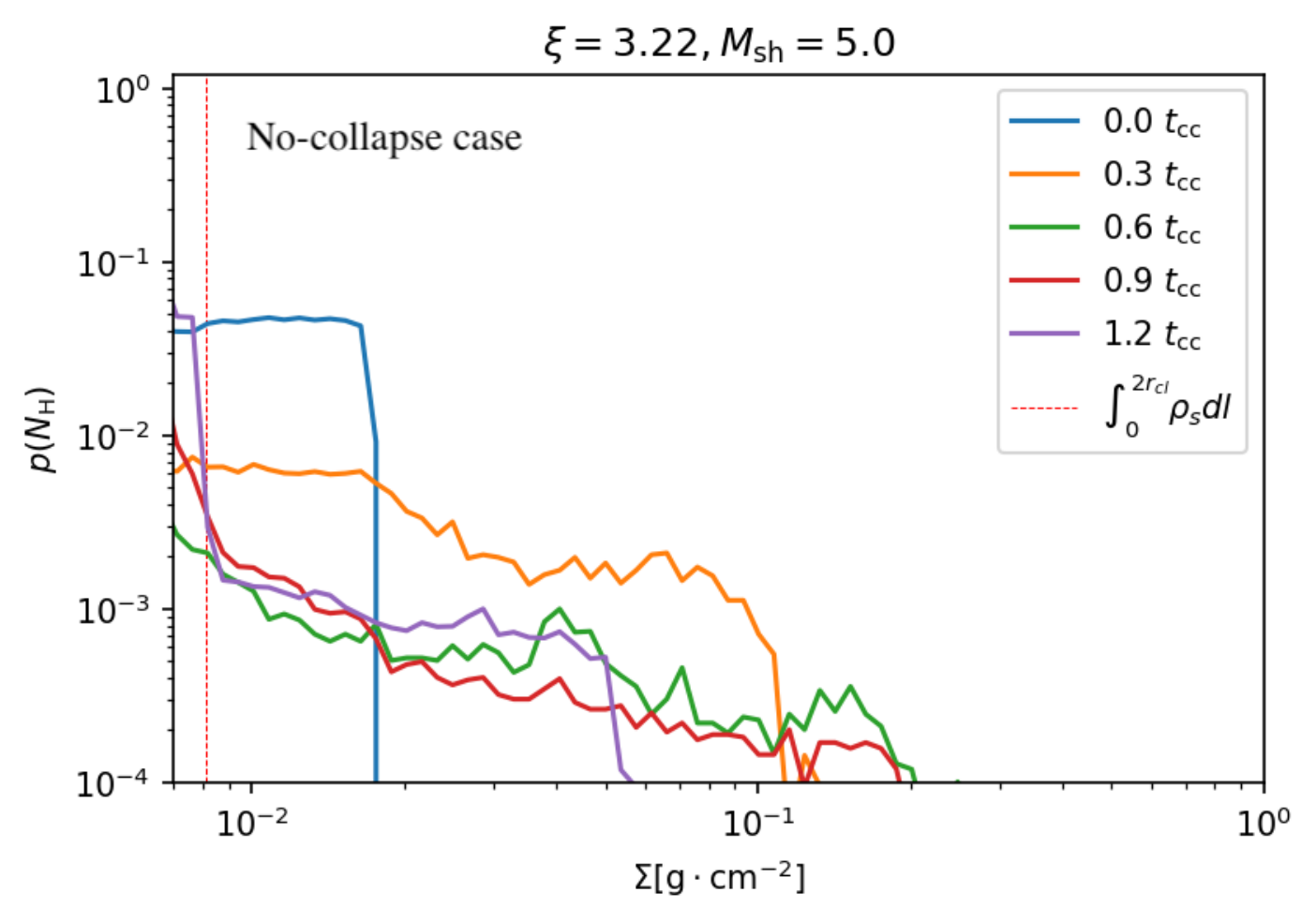}
          \hspace{2.0cm} (c) $M_{\rm sh}=5.00$, $\xi=3.22$
        \end{center}
      \end{minipage}
      
       \end{tabular}
    \vskip5pt  
    \caption{$\Sigma$-PDFs of regions from $\xi=3.22$ and $M_{\rm sh}=$1.20, 3.15, and 5.00 cases, respectively, as they evolve in time. For each case, $t=$0.0, 0.3, 0.6, 0.9, and 1.2 $t_{\rm cc}$ are shown. The $4r_{\rm cl}\times8r_{\rm cl}\times16r_{\rm cl}$ box data centered on the initial cloud is shown. The column density is derived assuming that the line of sight is parallel to the $x$-axis (see Figure \ref{fig:setup}). The vertical red dashed line shows the amount of integration of $ \rho_{\rm s}$ in the line of sight direction.}
    \label{fig:pdf-column-tur}

\end{figure*}

Figure \ref{fig:pdf-column-tur} shows time evolution of column \edit1{density} $\Sigma$-PDFs for $M_{\rm sh}=1.20,3.15$\edit1{,} and $5.00$ for $\xi=3.22$. These are the distribution of the column density when looking at the $4r_{\rm cl}\times8r_{\rm cl}\times16r_{\rm cl}$ box data centered on the initial cloud from the x-axis direction. $M_{\rm sh}=3.15$ corresponds to the triggered collapse case, while $M_{\rm sh}=1.20$ and $3.15$ correspond \edit1{to the} no-collapse cases. Although a sink particle is formed for $M_{\rm sh}=3.15$, only gas is included in this figure. 

For $M_{\rm sh}=1.20$, until \edit1{approximately} $t=0.9 t_{\rm cc}$ \edit1{the} PDFs \edit1{exhibit} a broadening distribution, with rebounding \edit1{occurring at approximately} $t=1.2 t_{\rm cc}$. The distribution \edit1{does} not \edit1{exceed} $10^{-1}$ $g\cdot \rm cm^{-2}$ and is not as broad as the other two PDFs.
For $M_{\rm sh}=3.15$, until \edit1{approximately} $t=0.6 t_{\rm cc}$ PDFs \edit1{exhibit} a broadening of distribution. The foot of high-density side increases to $\sim 10^{-1} g\cdot \rm cm^{-2}$. After the sink particle creation at $t\sim0.7 t_{\rm cc}$, \edit1{the} PDFs rebound. \edit1{For} $M_{\rm sh}=5.00$, the distribution also \edit1{becomes} wider and \edit1{the} foot of \edit1{the} high-density side \edit1{extends} above $\sim 10^{-1} g\cdot \rm cm^{-2}$. Compared \edit1{with} the other two PDFs, the value of the vertical axis $p(N_{\rm H})$ is generally smaller and the total amount of gas above $\int_{0}^{2 r_{\rm cl}}\rho_{\rm s}dl$ is smaller. After \edit1{approximately} $t=0.9 t_{\rm cc}$, the distribution rebounds. 

Depending on Mach numbers, each PDF \edit1{exhibits} different characteristics. However, since PDFs evolve over time, it is difficult to distinguish the presence or absence of collapse from only the $\Sigma$-PDFs. 
The distribution tail \edit1{extending} to the high-density side does not necessarily \edit1{indicate} \edit1{a} triggered collapse, and PDFs of observational data \edit1{should be addressed carefully}.

\section{Conclusion}
\label{sec:summary_1}
We studied the shock-cloud interaction using 3D hydrodynamical simulations with self-gravity and sink particles. We \edit1{demonstrated} that the evolution of the shocked clouds strongly depends on shock speeds and cloud radii. If the Mach number of the shock is \edit1{excessively} low, the shock cannot compress the cloud \edit1{sufficiently} to induce cloud collapse and the cloud \edit1{is} destroyed gradually. While, if the Mach number of the shock is \edit1{excessively} high, \edit1{the} shock destroys clouds \edit1{through the} hydrodynamical instability of the cloud surface before cloud collapse. Only \edit1{an} intermediate Mach number shock can trigger cloud collapse. In addition, even \edit1{when} clouds collapse, there are differences in cloud evolution such as the presence or absence of rebounding. We discuss that constraints of the Mach number for \edit1{the} collapse can be expressed as functions of dimensionless radii. The lower limit of the Mach number can be \edit1{got} by \edit1{comparing the} critical pressure of Bonnor-Ebert sphere and postshock pressure of ambient gas. The upper limit of the Mach number can be \edit1{got} by \edit1{comparing} the timescale of cloud collapse and cloud destruction. 

For the case in which cloud can collapse, the higher the Mach number of \edit1{the} shock is, the lower the asymptotic mass of the formed sink particle becomes. This is because \edit1{higher-Mach-number shocks strip cloud} gas around the sink particle faster and make effective accretion time shorter. We showed that the higher the Mach number of shock is, the faster the relative velocity and position increase. 

We also address cases in which initial clouds have turbulent velocity fields. 
\edit1{We observed} that turbulent clouds have the same trends as non-turbulent counterparts \edit1{on} evolution \edit1{differences} depending on Mach number. Some shocks can trigger cloud collapse, \edit1{whereas excessively} strong shocks destroy clouds faster and cannot induce cloud collapse. The turbulence itself suppresses cloud contraction and \edit1{decreases} formed asymptotic sink particles mass. 

These simulation results provide a general guide to the evolutionary process of dense cores or Bok globules impacted by shocks due to supernovae, stellar winds, and ionization fronts.

\acknowledgments

Numerical computations and analyses in this work were partly carried out on Cray XC50 and analysis servers at Center for Computational Astrophysics, National Astronomical Observatory of Japan.

Computations described in this work were performed using the
publicly-available \texttt{Enzo} code (http://enzo-project.org), which is
the product of a collaborative effort of many independent scientists from
numerous institutions around the world.  Their commitment to open science
has helped make this work possible.

%



\software{Enzo \citep{Bryan_2014},  
          Yt \citep{2011ApJS..192....9T}
          }



\appendix

\section{Bonnor-Ebert clouds}
\label{app:bonnor}
The Bonnor-Ebert sphere is an isothermal gas sphere 
remaining in hydrostatic equilibrium (\citealp{Ebert_1955}; \citealp{Bonnor_1956}).
The equation of hydrostatic equilibrium can be nondimensionalized to \edit1{obtain} the isothermal Lane-Emden equation \citep{chandrasekhar_1957}:
\begin{equation}
\label{eq:bonnor_lane}
\frac{1}{\xi^{2}}\frac{d}{d\xi}( \xi^{2}\frac{d\psi}{d\xi}) =\rm exp(-\psi),
\end{equation},
where $\xi$ is the dimensionless radius, by putting
\begin{eqnarray}
\label{eq:bonnor_set}
  \xi &=& \sqrt{\frac{4\pi G \rho_{\rm c}}{c_{\rm cl}^{2}}} r, \\
\end{eqnarray}
and
\begin{eqnarray}
\label{eq:bonnor_set-2}
  \rho &=&  \rho_{c} \rm exp(-\psi),
\end{eqnarray}
where $\rho_{c}$ is the central density of sphere, $c_{\rm cl}$ is the thermal speed of sound, $r$ is the characteristic radius of the cloud.

We can \edit1{obtain} a numerical solution with the boundary conditions
\begin{equation}
\label{eq:bonnor_boundary}
\psi (0)=\frac{d\psi (0)}{d\xi}=0.
\end{equation}

\begin{figure}[hbtp]
\begin{center}
\includegraphics[width=100mm]{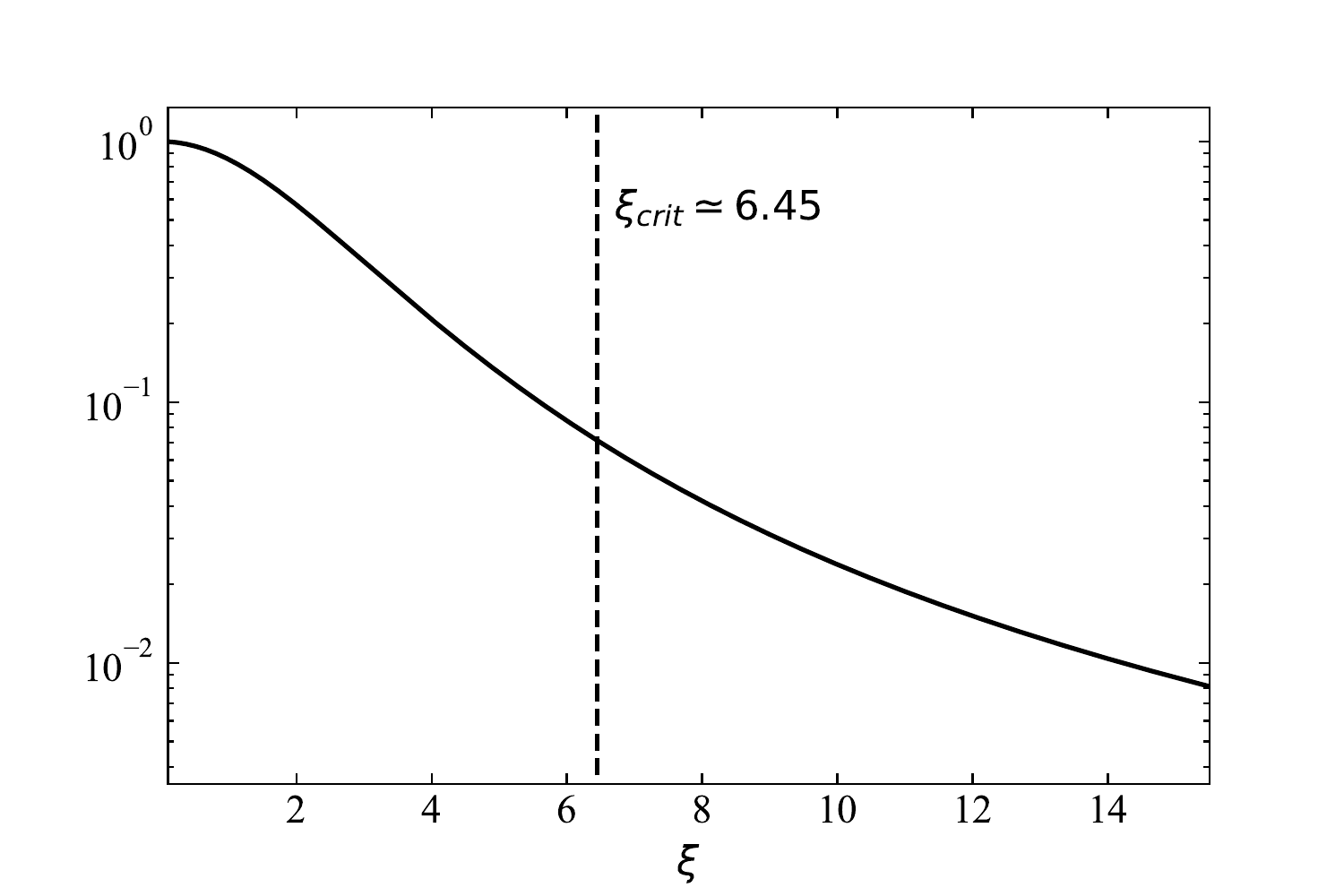}
\begin{flushleft}
\caption{\edit1{Density} profile of the Bonnor-Ebert sphere. The abscissa is the dimensionless radius $\xi$, and the ordinate is the density ratio $\rho/\rho_{c}$. \edit1{The vertical} dashed line indicates the critical dimensionless radius ($\xi_{\rm crit}=6.45$), above which \edit1{the} Bonner-Ebert sphere is unstable.
}
\label{fig:bonnor_radius}
\end{flushleft}
\end{center}
\end{figure}


Figure \ref{fig:bonnor_radius} shows the radial density profile of the Bonnor-Ebert sphere (i.e.\edit1{,} the numerical solution of Equation \ref{eq:bonnor_lane} and \ref{eq:bonnor_boundary} ).

The total mass within the Bonnor-Ebert sphere is expressed as

\begin{equation}
\label{eq:bonnor-mass-4}
  m_{\rm cloud}=4\pi\rho_{c}\left(\frac{c^{2}}{4\pi G\rho_{c}}\right)^{3/2}\xi^{2}\frac{d\psi}{d\xi}.
\end{equation}


The Bonnor-Ebert sphere is unstable when its dimensionless radius exceeds the critical dimensionless radius of  $\xi_{\rm crit}=6.45$. 
The Bonnor-Ebert critical mass and pressure corresponding this critical values are, respectively,  
\begin{eqnarray}
\label{eq:bonnor_critical_M}
  m_{\rm crit} & = & 1.18 \frac{c_{\rm s}^{4}}{G^{3/2}P_{0}^{1/2}},\\
\end{eqnarray}
and
\begin{eqnarray}
\label{eq:bonnor_critical_P}
  P_{\rm crit} & = & 1.40 \frac{c_{\rm s}^{8}}{G^{3}m_{\rm cloud}^{2}}.
\end{eqnarray}

\section{Timescales}
\label{sec:timescales}
Here, we define several important characteristic timescales to discuss the evolution of shocked clouds.
\subsection{Cloud crushing time}
\label{Cloud crushing time}
Consider a uniform spherical cloud of radius $r_{\rm cl}$ and density $\rho_{0}$ in pressure equilibrium with an ambient gas of density $\rho_{\rm ism}$. We focus on the case \edit1{in which} a planar shock of velocity $v_{\rm sh}$ interacts with an isothermal cloud without magnetic fields. When a shock encounters a cloud, an overpressure region and shock will be driven into the cloud.
If the shock is strong, the post-shock pressure is approximately $\rho_{\rm ism}v_{\rm sh}^{2}$. The post-shock pressure in the cloud is of \edit1{the} order $\rho_{0}v_{\rm cl}^{2}$, where $v_{\rm cl}$ is the velocity of the shock in the cloud.
Assuming that these two pressures must be comparable, we \edit1{obtain}
\begin{equation}
\label{eq:v_cl_in_the_cloud}
  v_{\rm cl}\simeq \left(\frac{\rho_{\rm 0}}{\rho_{\rm ism}}\right)^{1/2} v_{\rm sh}=v_{\rm sh}\chi^{-1/2},
\end{equation}
where $\chi$ is the ratio of cloud density to ambient gas density.
\edit1{The cloud} crushing time $t_{\rm cc}$ is the time for the shock to pass across the cloud:
\begin{equation}
\label{eq:tcc}
  t_{\rm cc} = \frac{r_{\rm cl}}{v_{\rm cl}}\simeq\frac{\chi^{1/2}r_{\rm cl}}{v_{\rm sh}}.
\end{equation}
This is the important time scale for the evolution of the shocked cloud. In this study, we \edit1{used} $t_{\rm cc}$ as $\chi =\langle\rho_{\rm cl}\rangle/\rho_{\rm ism}$.
\subsection{Drag timescale}
The shock wave accelerates \edit1{the} cloud until it is comoving with the postshock ambient gas. Let $v_{\rm c}$ is the mean velocity of the cloud, $v_{\rm p}$ the velocity of the postshock ambient gas, \edit1{and} $v_{\rm r}=|v_{\rm p}-v_{\rm c}|$ the magnitude of the velocity of the cloud relative to velocity of the postshock ambient gas. If we consider the momentum transfer from a ambient gas with cross-section $ \pi r_{\rm cl}^{2}$, the equation of motion of the cloud is
\begin{equation}
\label{eq:cloud_motion}
  M_{\rm cl}\frac{dv_{\rm r}}{dt}=-(\pi r_{\rm cl}^{2})\rho_{\rm ism}v_{\rm r}^{2},
\end{equation}
where $M_{\rm cl}$ is the cloud mass. This equation \edit1{yields} the characteristic drag timescale $t_{\rm acc}$ as  
\begin{equation}
\label{eq:drag_time}
  t_{\rm drag}\sim \chi^{1/2} t_{\rm cc}.
\end{equation}

\subsection{Destruction timescale}
After the shock wave has swept over the cloud, the shocked cloud is subject to Kelvin-Helmholtz and Rayleigh-Taylor instabilities. For $\chi>>1$, the time-scale for growth of the Kelvin-Helmholtz instability is $t_{\rm KH} =\chi^{1/2}/kv_{\rm rel}$ \citep{Chandrasekhar_1961} , where $k$ is the wave-number of perturbations\edit1{,} and $v_{\rm rel}$ is the relative velocity between \edit1{the} post-shock ambient gas and the cloud.
Since the clouds accelerate rather slowly, $v_{\rm rel}$ is approximately equal to the velocity behind the shock $v_{\rm p}=v_{\rm sh}(1-1/M^{2})$. Thus, the time-scale for growth of the Kelvin -Helmholtz instability is comparable to the cloud crushing time $t_{\rm cc}$\edit1{:}
\begin{equation}
\label{eq:t_KH}
  t_{\rm KH} \sim \frac{t_{\rm cc}}{kr_{\rm cl}}.
\end{equation}
The shortest wavelengths have the fastest growth, but wavelengths corresponding to  $kr_{\rm cl} \sim 1$ are most disruptive.

The \edit1{drag} timescale \edit1{results in} an acceleration $g\sim v_{\rm sh}/t_{\rm drag}\sim r_{\rm cl}/t_{\rm cc}^{2}$, corresponding to a growth timescale of Rayleigh-Taylor instabilities given by $t_{\rm RT}\simeq (gk)^{-1/2}$ \citep{Chandrasekhar_1961}. Thus, the Rayleigh-Taylor \edit1{growth} timescale is also comparable to the cloud crushing time\edit1{:}
\begin{equation}
\label{eq:t_RT}
  t_{\rm RT} \sim \frac{t_{\rm cc}}{(kr_{\rm cl})^{1/2}}.
\end{equation}

These timescales for instability suggest that a cloud will be destroyed in a time of the order of the cloud crushing time. Previous \edit1{studies have shown} that \edit1{the} time-scale for cloud destruction is indeed \edit1{in} the order of $t_{\rm cc}$ (e.g., \citealp{Klein_1994} ; \citealp{Nakamura_2006}).

\section{Values employed for each model}
\label{app:Employed Values for each model}
Table \ref{tab:simulation model} specifies the \edit1{values employed} for each model. 

\begin{table}

\caption{Values employed for each model}
\normalsize
\centering
\begin{tabular}{ccccccccc}
\toprule
No. $^{a}$ &  $\xi$ $^{b}$&   $\xi/\xi_{crit}$ $^{c}$ & $r_{\rm cl}$ $^{d}$ &$P_{0}/P_{\rm crit}$ $^{e}$&  $\rho_{c}/\rho_{\rm s}$ $^{f}$& $M_{\rm cl}^{g}$&  $M_{\rm sh}$ $^{h}$ &  $v_{\rm sh}^{i}$ \\
\hline
1  &   2.04 &  0.31  &   0.07    &  0.125 &                         1.78 &    0.56 & 1.41 &      2.67 \\
   &        &        &       &        &                              &         & 1.99 &      3.77 \\
   &        &        &       &        &                              &         & 2.39 &      4.53 \\
   &        &        &       &        &                              &         & 3.15 &      5.97 \\
2  &   2.48 &  0.38  &   0.08    &   0.25 &                         2.22 &    0.89 & 1.20 &      2.27 \\
   &        &        &       &        &                              &         & 1.41 &      2.67 \\
   &        &        &       &        &                              &         & 1.99 &      3.77 \\
   &        &        &       &        &                              &         & 2.39 &      4.53 \\
   &        &        &       &        &                              &         & 3.15 &      5.97 \\
3  &   3.22 &  0.5   &   0.11    &    0.5 &                         3.24 &    1.52 & 1.20 &      2.27 \\
   &        &        &       &        &                              &         & 1.41 &      2.67 \\
   &        &        &       &        &                              &         & 1.99 &      3.77 \\
   &        &        &       &        &                              &         & 3.15 &      5.97 \\
   &        &        &       &        &                              &         & 4.00 &     7.58  \\
   &        &        &       &        &                              &         & 4.46 &      8.45 \\
   &        &        &       &        &                              &         & 5.00 &      9.47 \\
   &        &        &       &        &                              &         & 5.64 &     10.68 \\
4  &   4.05 &  0.63  &     0.14  &   0.75 &                         4.94 &    2.31 & 1.20 &      2.27 \\
   &        &        &       &        &                              &         & 4.46 &      8.45 \\
   &        &        &       &        &                              &         & 5.00 &      9.47 \\
   &        &        &       &        &                              &         & 5.64 &     10.68 \\
5  &   6.45 &   1.00 &     0.22  &   1.00 &                        14.10 &    4.48 & 3.15 &     5.97  \\
   &        &        &       &        &                              &         & 5.64 &     10.68 \\
   &        &        &       &        &                              &         & 6.00 &     11.37 \\
   &        &        &       &        &                              &         & 7.00 &     13.26 \\
6  &   8.52 &  1.32  &    0.29   &  ---      &                        28.08 &    6.13 & 5.64 &     10.68 \\
   &        &        &       &        &                              &         & 6.00 &     11.37 \\
   &        &        &       &        &                              &         & 7.00 &     13.26 \\
7  &  14.22 &  2.20  &    0.49   &  ---      &                       100.00 &    9.78 & 6.00 &     11.37 \\
   &        &        &       &        &                              &         & 7.00 &     13.26 \\\hline
8  &   6.45 &  1.00  &   0.22    &     1.00 &                        14.10 &    4.48 & 1.41 &     2.67   \\
   &        &        &       &          &                              &         & 1.99 &     3.77   \\
   &        &        &       &          &                              &         & 3.15 &     5.97  \\
   &        &        &       &          &                              &         & 4.46 &     8.45  \\
   &        &        &       &          &                              &         & 5.64 &     10.68 \\\hline
\end{tabular}
\vskip5pt
\begin{tablenotes}

\item 
No.1$-$7: \edit1{Simulation} values \edit1{employed} for \edit1{non-turbulent} models. No.8: \edit1{Values employed} for turbulent models.

$^{a}$ ID of the initial cloud condition.

$^{b}$ \edit1{Dimensionless} radius of the Bonnor-Ebert sphere.

$^{c}$ Ratio of the dimensionless radius $\xi$ \edit1{to} the critical dimensionless radius $\xi_{\rm crit}$. 

$^{d}$ Radius of the Bonnor-Ebert sphere (pc)

$^{e}$ Ratio of the external pressure $P_{0}$ \edit1{to} the critical pressure $P_{\rm crit}$.
If the initial cloud is unstable ($\xi>6.45$), the value is not shown.

$^{f}$ Ratio of cloud central density $\rho_{\rm c}$ and cloud surface density $\rho_{\rm s}$.

$^{g}$ \edit1{Mass} of the initial cloud ($M_{\odot}$).

$^{h}$ Mach number of the propagating shock.

$^{i}$ \edit1{Propagating} shock speed (km s$^{-1}$).

\end{tablenotes}

\label{tab:simulation model}

\end{table}

\section{Simulation results} 
\label{app:Simulation results}
Table \ref{tab:simulation result} specifies results of simulations numerically. 
\begin{table}
\normalsize
\centering
\caption{Simulation results}
\begin{tabular}{ccccc}

\toprule
No${}^{a}$ &  $\xi{}^{b}$ &  $M_{\rm sh}{}^{c}$ & $t_{\rm sink}{}^{d}$ & $M_{\rm sink}{}^{e}$ \\
\hline
1  &   2.04 &      1.41 &         ---   &       ---  \\
   &        &      1.99 &         ---   &       ---  \\
   &        &      2.39 &         ---   &       ---  \\
   &        &      3.15 &         ---   &       ---  \\
2  &   2.48 &      1.20 &         ---   &       ---  \\
   &        &      1.41 &         ---   &       ---  \\
   &        &      1.99 &          0.32 &       0.71 \\
   &        &      2.39 &          ---   &      ---   \\
   &        &      3.15 &          ---   &      ---   \\
3  &   3.22 &      1.20 &          ---   &      ---   \\
   &        &      1.41 &          0.74 &       1.35 \\
   &        &      1.99 &          0.36 &       1.24 \\
   &        &      3.15 &          0.15 &       0.95 \\
   &        &      4.00 &          0.12 &       0.66 \\
   &        &      4.46 &          ---   &      ---   \\
   &        &      5.00 &          ---   &      ---   \\
   &        &      5.64 &          ---   &      ---   \\
4  &   4.05 &      1.20 &          1.14 &       1.49 \\
   &        &      4.46 &          0.13 &       1.25 \\
   &        &      5.00 &          0.12 &       0.94 \\
   &        &      5.64 &          ---     &    ---        \\
5  &   6.45 &      3.15 &          0.31 &       4.11 \\
   &        &      5.64 &          0.18 &       0.86 \\
   &        &      6.00 &          ---     &    ---     \\
   &        &      7.00 &          ---     &    ---     \\
6  &   8.52 &      5.64 &          0.22 &       1.93 \\
   &        &      6.00 &          ---   &      ---      \\
   &        &      7.00 &          ---   &      ---      \\
7  &  14.22 &      6.00 &          0.36 &       1.23 \\
   &        &      7.00 &          ---     &    ---        \\\hline
 8  &  6.45 &      1.41 &          1.94 &       3.21 \\
     &      &      1.99 &          0.93 &       2.87 \\
     &      &      3.15 &          0.49 &       2.05 \\
     &      &      4.46 &          ---     &    ---        \\
     &      &      5.64 &          ---     &    ---        \\ \hline
\end{tabular}
\vskip5pt
\begin{tablenotes}

\item $^{a}$ ID of the initial cloud condition.

$^{b}$ \edit1{Dimensionless} radius of the Bonnor-Ebert sphere.

$^{c}$ Mach number of the propagating shock.

$^{d}$ \edit1{Time} interval from when the shock wave reaches the cloud until the sink particles are introduced (Myr). If the sink particles is not formed, the value is not shown.

$^{e}$ Asymptotic mass of sink particles ($M_{\odot}$). If the sink particles is not formed, the value is not shown.

\end{tablenotes}
\label{tab:simulation result}
\end{table}

\section{Results of different dimensionless radii models} 

\label{app:model_results}
\begin{figure*}
  
    \begin{tabular}{c}

      \begin{minipage}{0.5\hsize}
        \begin{center}
          \includegraphics[clip, width=6.7cm]{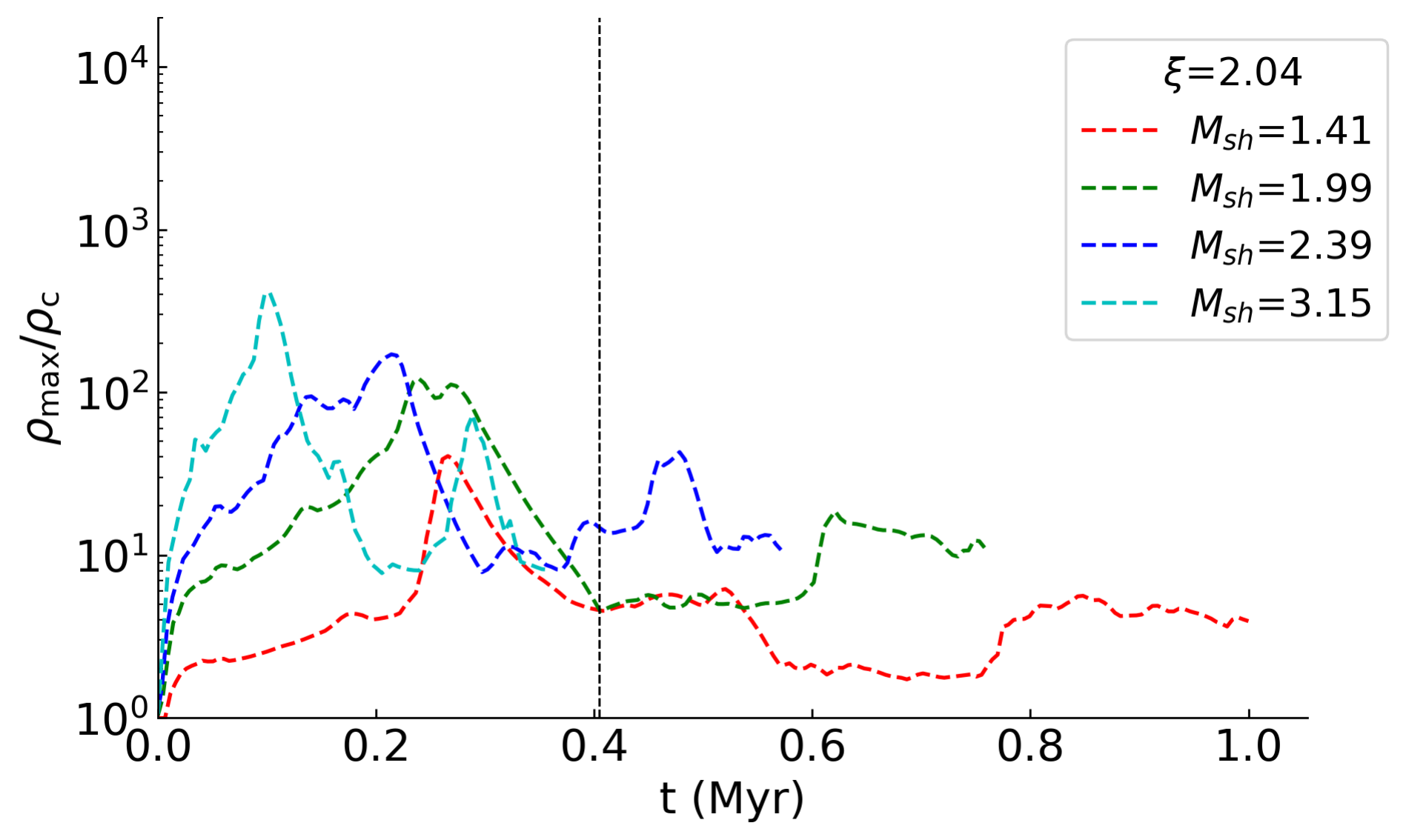}
          \hspace{2.0cm} (a) $\xi$=2.04
        \end{center}
      \end{minipage}

      \begin{minipage}{0.5\hsize}
        \begin{center}
          \includegraphics[clip, width=6.7cm]{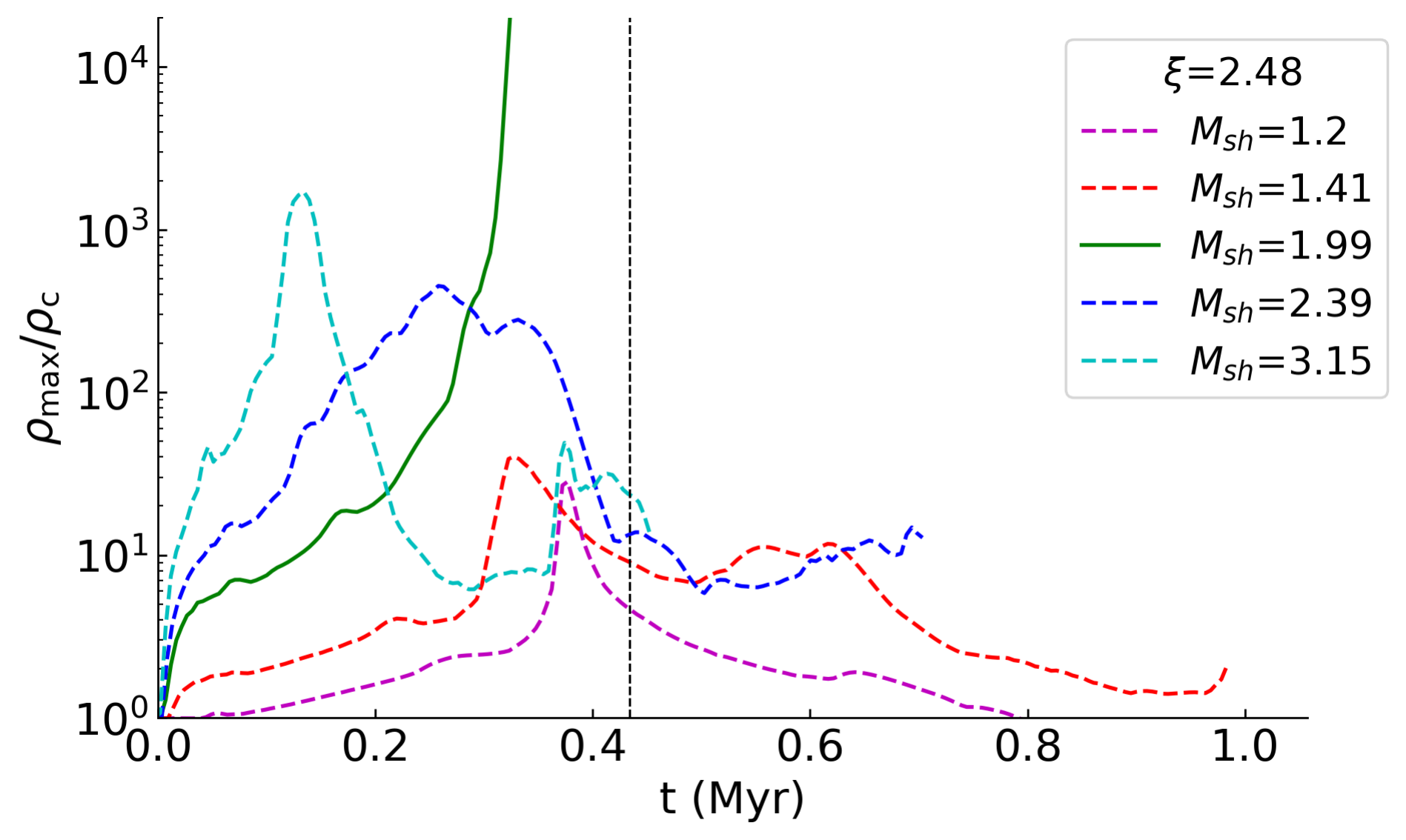}
          \hspace{2.0cm} (b) $\xi$=2.48
        \end{center}
      \end{minipage}
      
    \end{tabular}

    \begin{tabular}{c}

      \begin{minipage}{0.5\hsize}
        \begin{center}
          \includegraphics[clip, width=6.7cm]{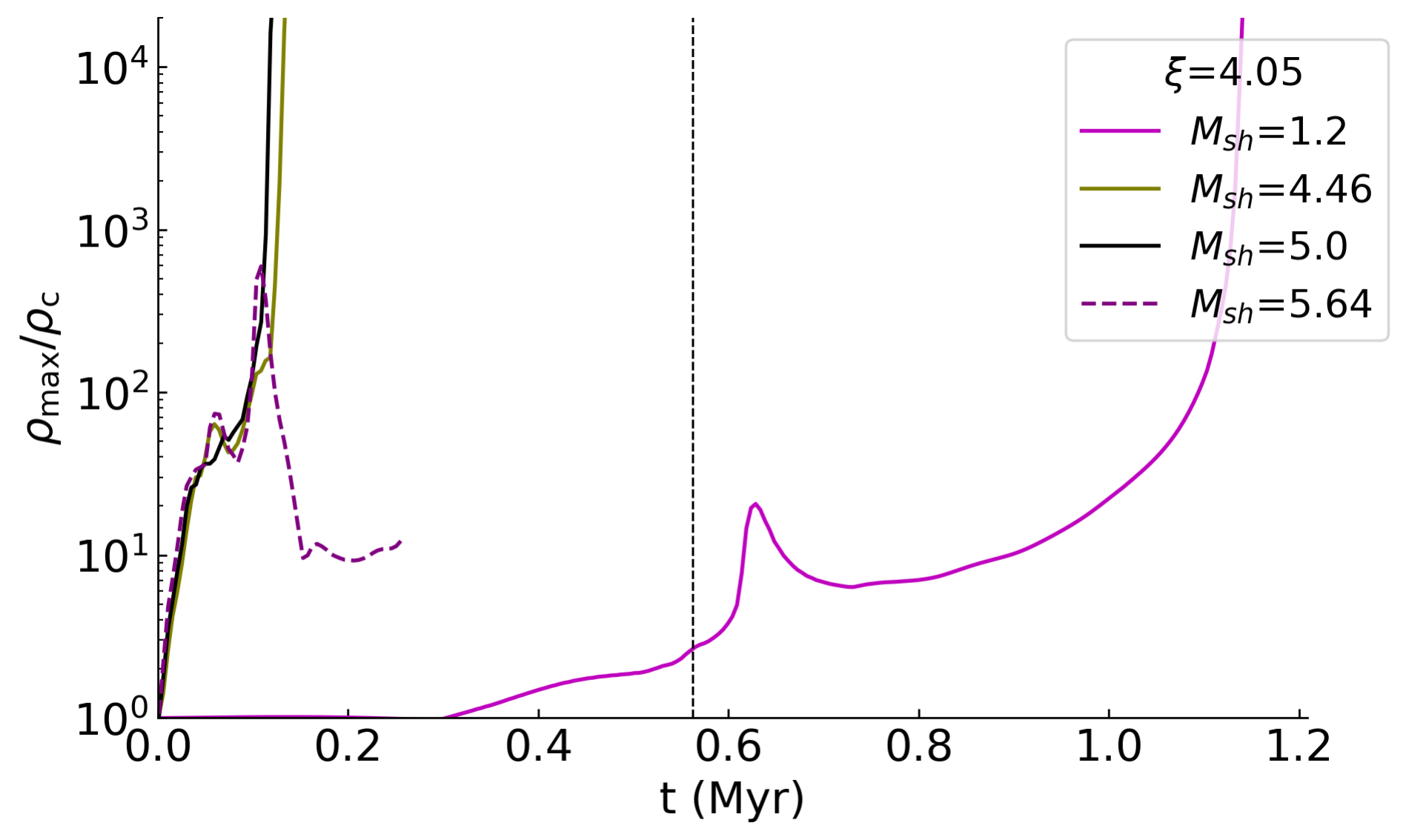}
          \hspace{2.0cm} (d) $\xi$=4.05
        \end{center}
      \end{minipage}

      \begin{minipage}{0.5\hsize}
        \begin{center}
          \includegraphics[clip, width=6.7cm]{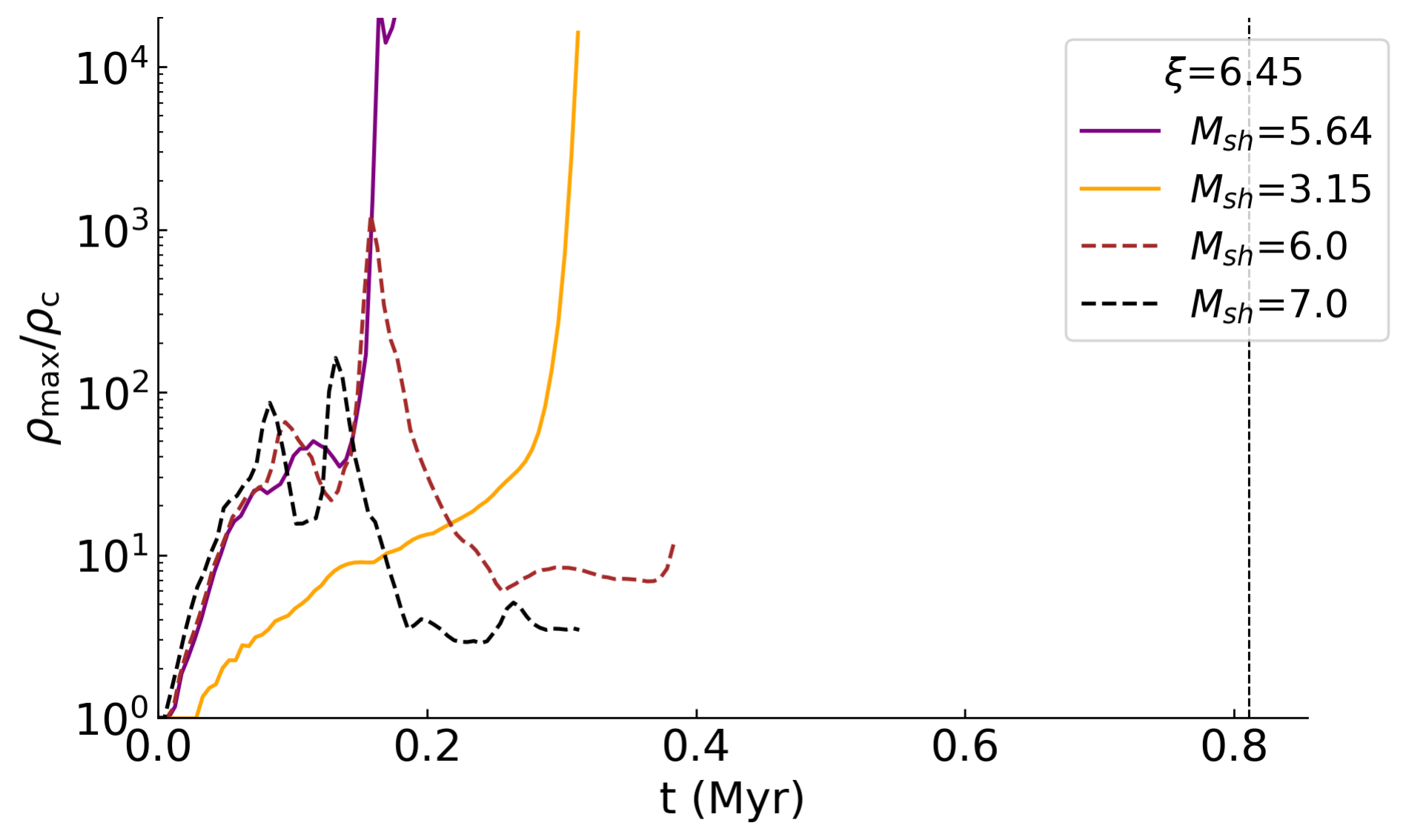}
          \hspace{2.0cm} (e) $\xi$=6.45
        \end{center}
      \end{minipage}
      
    \end{tabular}
      
    \begin{tabular}{c}
      \begin{minipage}{0.5\hsize}
        \begin{center}
          \includegraphics[clip, width=6.7cm]{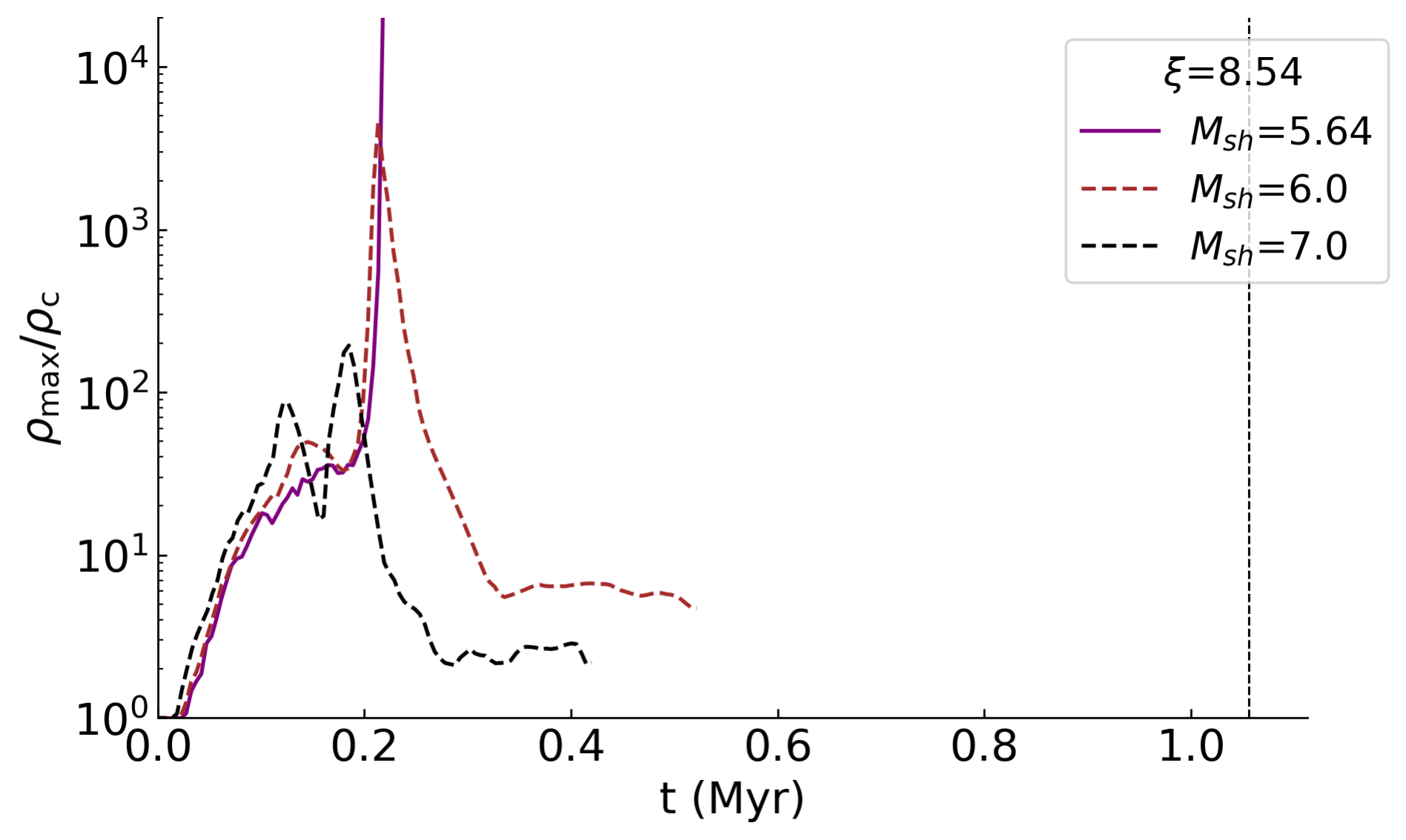}
          \hspace{2.0cm} (f) $\xi$=8.54
        \end{center}
      \end{minipage}

    \begin{minipage}{0.5\hsize}
        \begin{center}
          \includegraphics[clip, width=6.7cm]{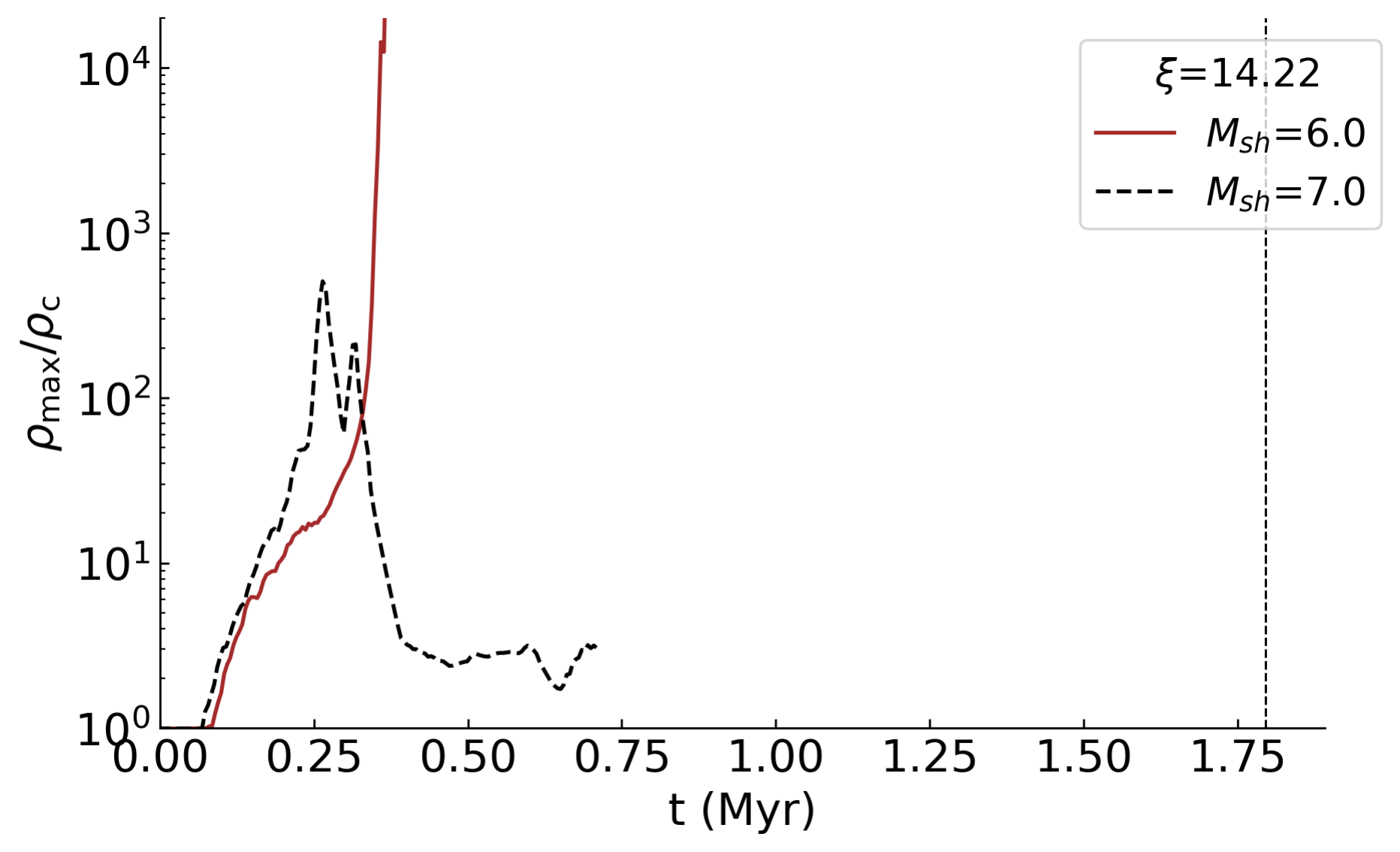}
          \hspace{2.0cm} (g) $\xi$=14.22
        \end{center}
      \end{minipage}
      
    \end{tabular}
    \vskip5pt  
    \caption{Ratio of the maximum density to the initial cloud central density $\rho_{\rm max}/\rho_{\rm c}$ as functions of time at different shock \edit1{Mach numbers}. If a sink particle is introduced during \edit1{the} shock-cloud evolution, the transition is represented by a solid line\edit1{;} otherwise, \edit1{it is} represented by \edit1{a} dashed line. \edit1{The vertical} dashed line \edit1{indicates} the free fall time $t_{\rm ff}$.}
    \label{fig:maximum_density}

\end{figure*}

In this appendix, we show results of cases with different dimensionless radii. 
Figure \ref{fig:maximum_density} shows \edit1{the} evolution of density ratio $\rho_{\rm max}/\rho_{\rm c}$ at different dimensionless radius. \edit1{For} $\xi$=3.22, \edit1{the} density evolution depends on $M_{\rm sh}$. 
For triggered-collapse cases, \edit1{the} maximum density increases monotonically or after rebounding (e.g., $\xi$=4.05 and $M_{\rm sh}$=1.20 case), inducing gravitational collapse. \edit1{Moreover}, when $M_{\rm sh}$ is lower or higher, the maximum density increases at the beginning but decreases to lower values after 
rebounding without cloud collapse. For $\xi$=4.05, all of them correspond to no-collapse cases.

\begin{figure*}
  
    \begin{tabular}{c}

      \begin{minipage}{0.5\hsize}
        \begin{center}
          \includegraphics[clip, width=6.7cm]{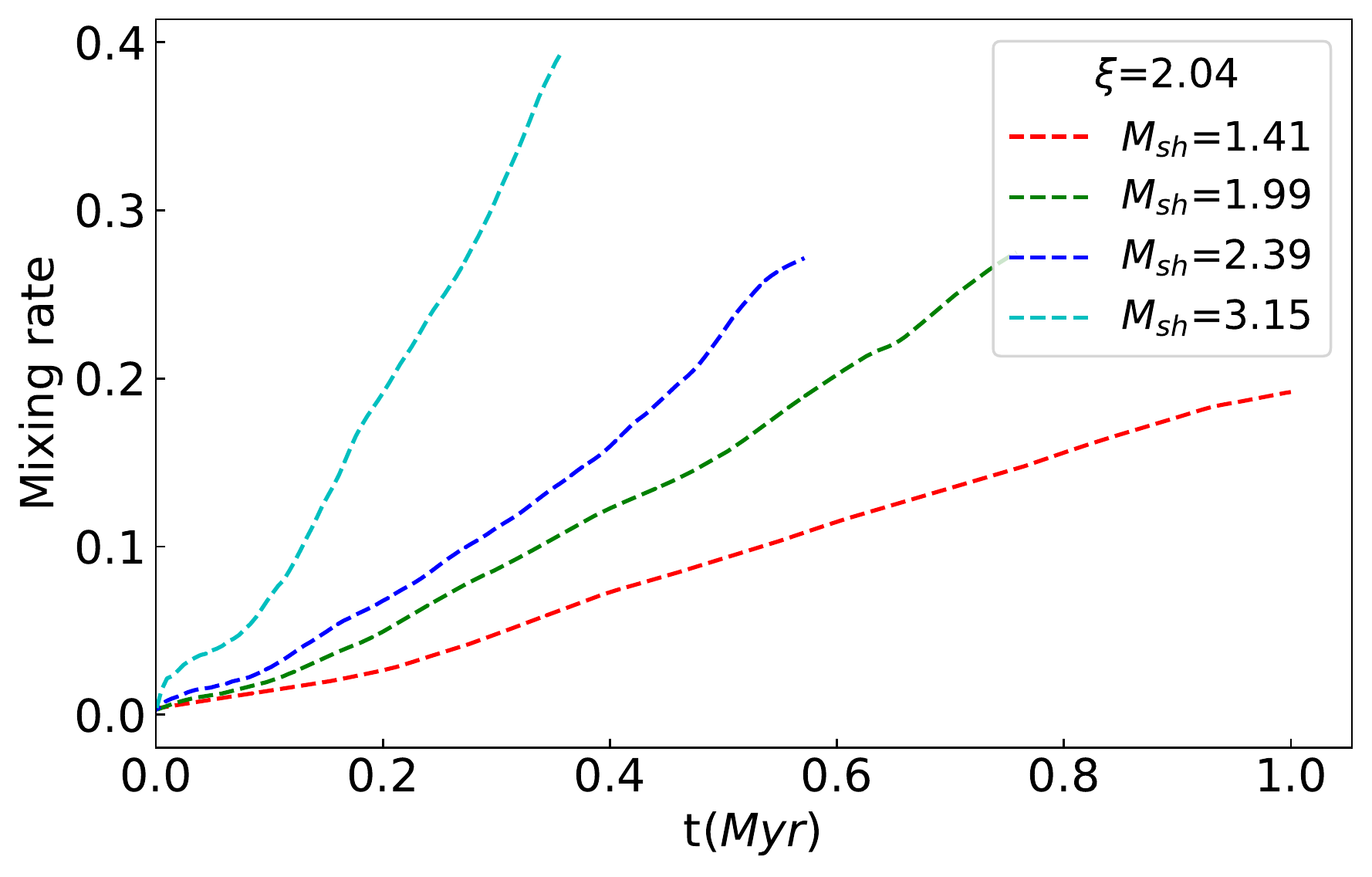}
          \hspace{2.0cm} (a) $\xi$=2.04
        \end{center}
      \end{minipage}

      \begin{minipage}{0.5\hsize}
        \begin{center}
          \includegraphics[clip, width=6.7cm]{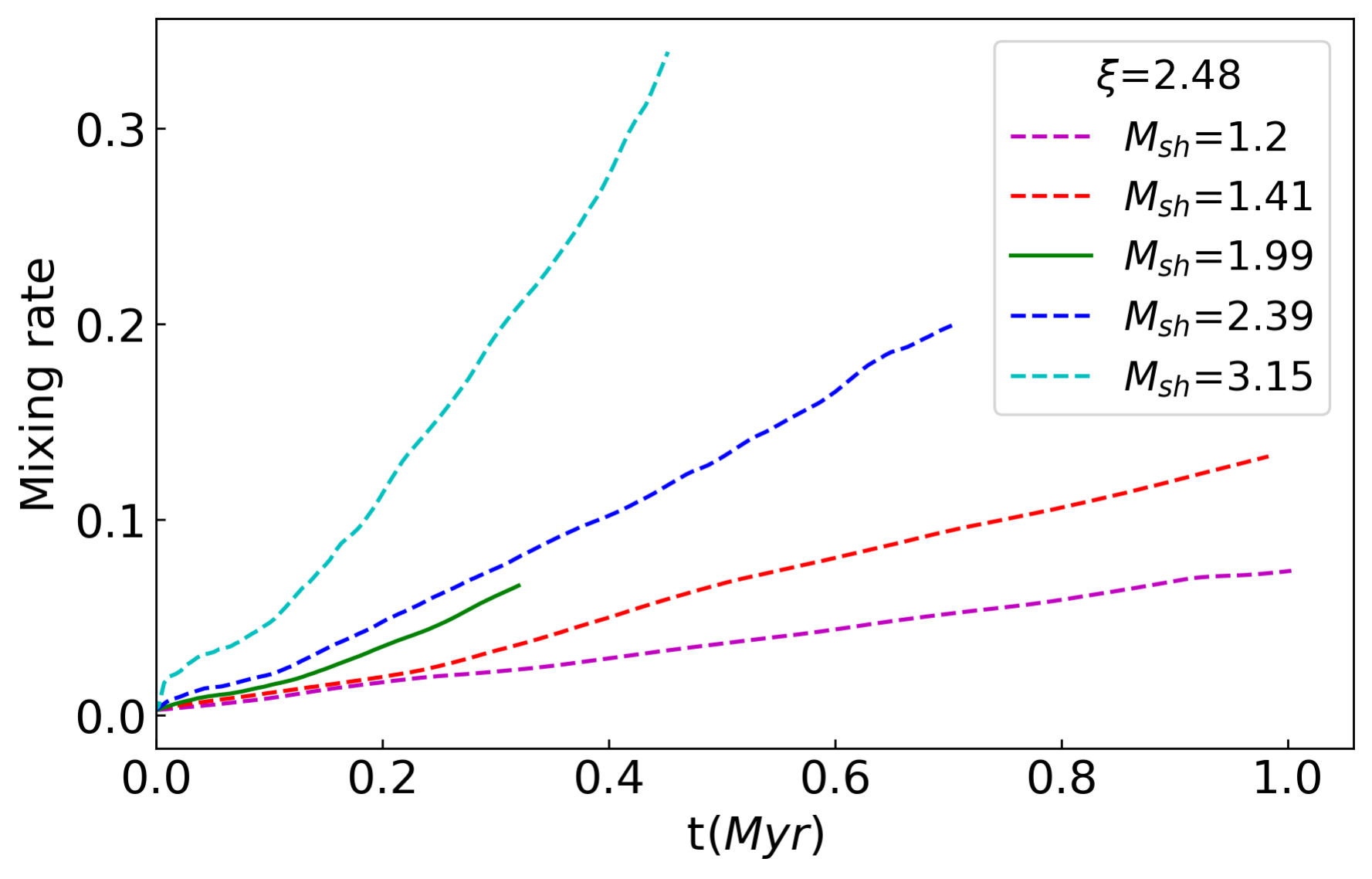}
          \hspace{2.0cm} (b) $\xi$=2.48
        \end{center}
      \end{minipage}
    
    \end{tabular}

    \begin{tabular}{c}

      \begin{minipage}{0.5\hsize}
        \begin{center}
          \includegraphics[clip, width=6.7cm]{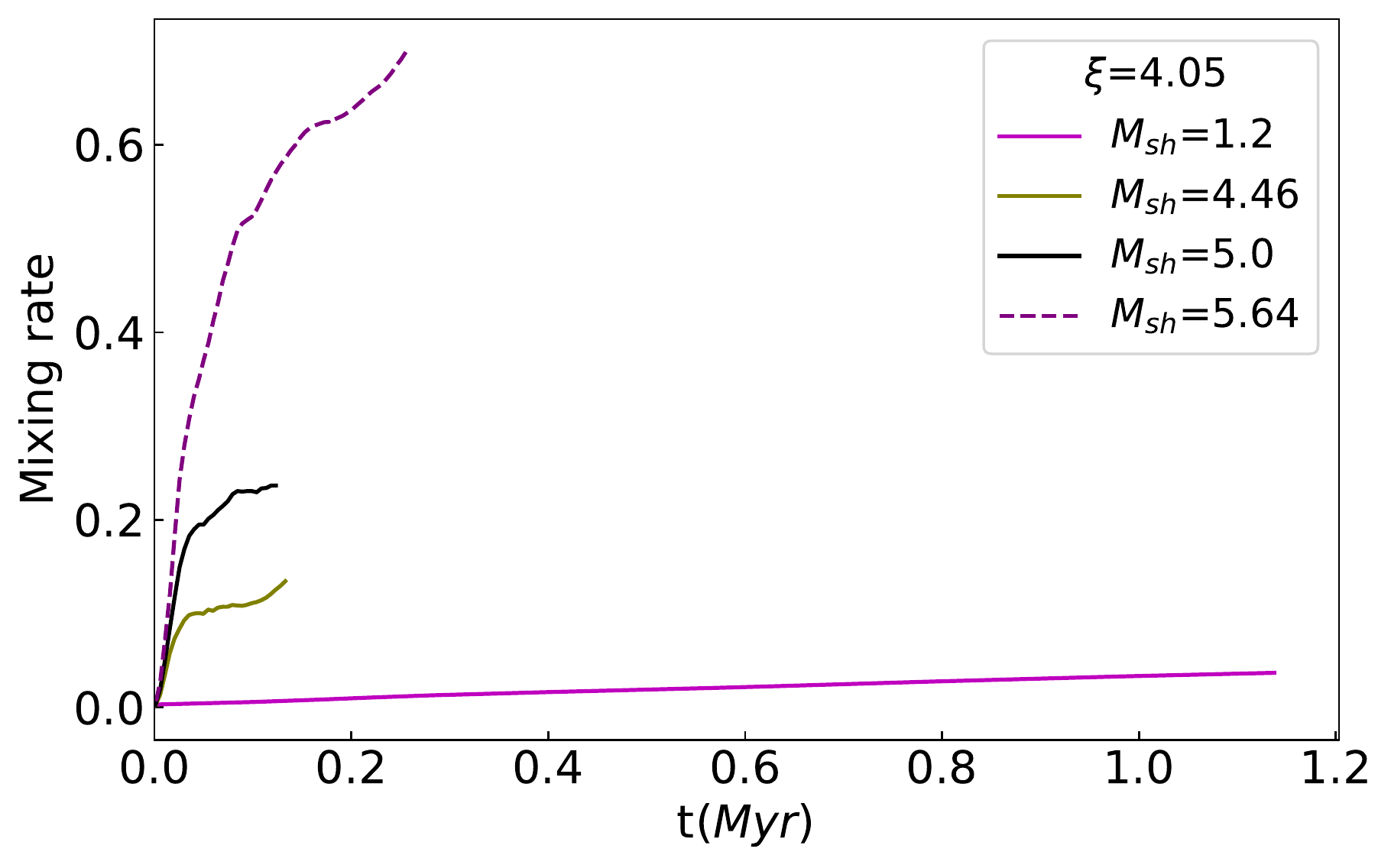}
          \hspace{2.0cm} (d) $\xi$=4.05
        \end{center}
      \end{minipage}

      \begin{minipage}{0.5\hsize}
        \begin{center}
          \includegraphics[clip, width=6.7cm]{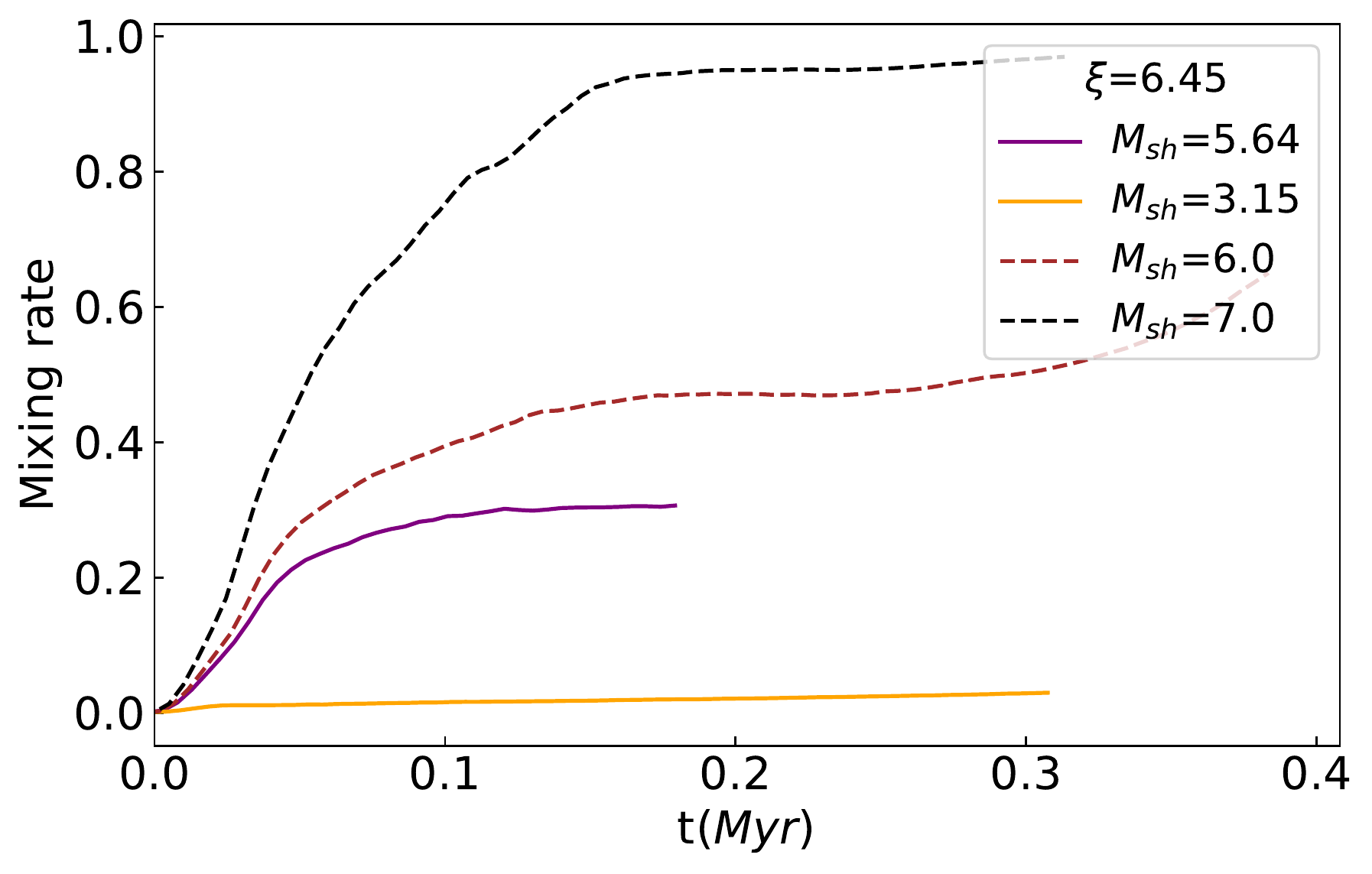}
          \hspace{2.0cm} (e) $\xi$=6.45
        \end{center}
      \end{minipage}
      
    \end{tabular}
     
    \begin{tabular}{c}

      \begin{minipage}{0.5\hsize}
        \begin{center}
          \includegraphics[clip, width=6.7cm]{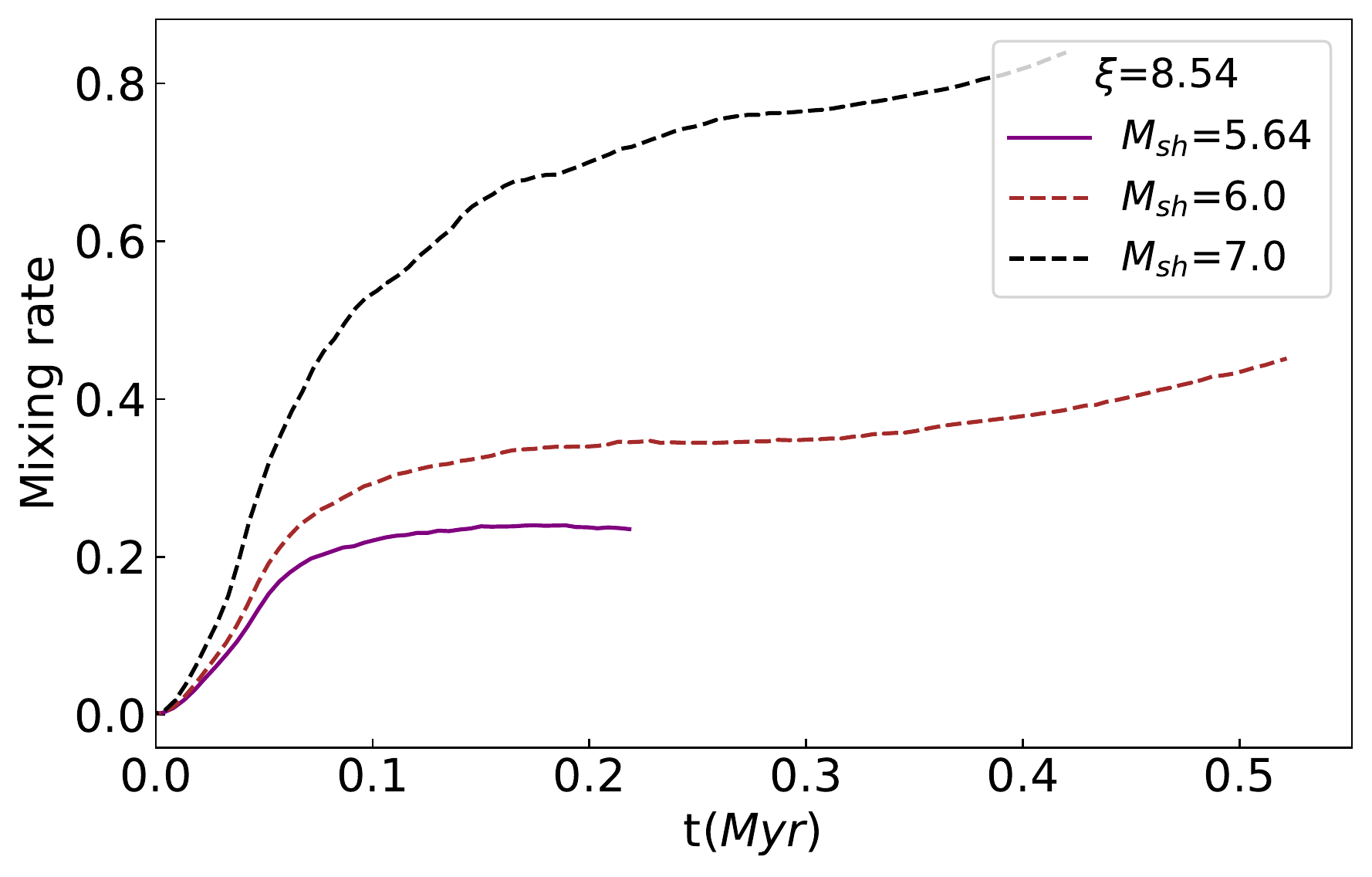}
          \hspace{2.0cm} (f) $\xi$=8.54
        \end{center}
      \end{minipage}

    \begin{minipage}{0.5\hsize}
        \begin{center}
          \includegraphics[clip, width=6.7cm]{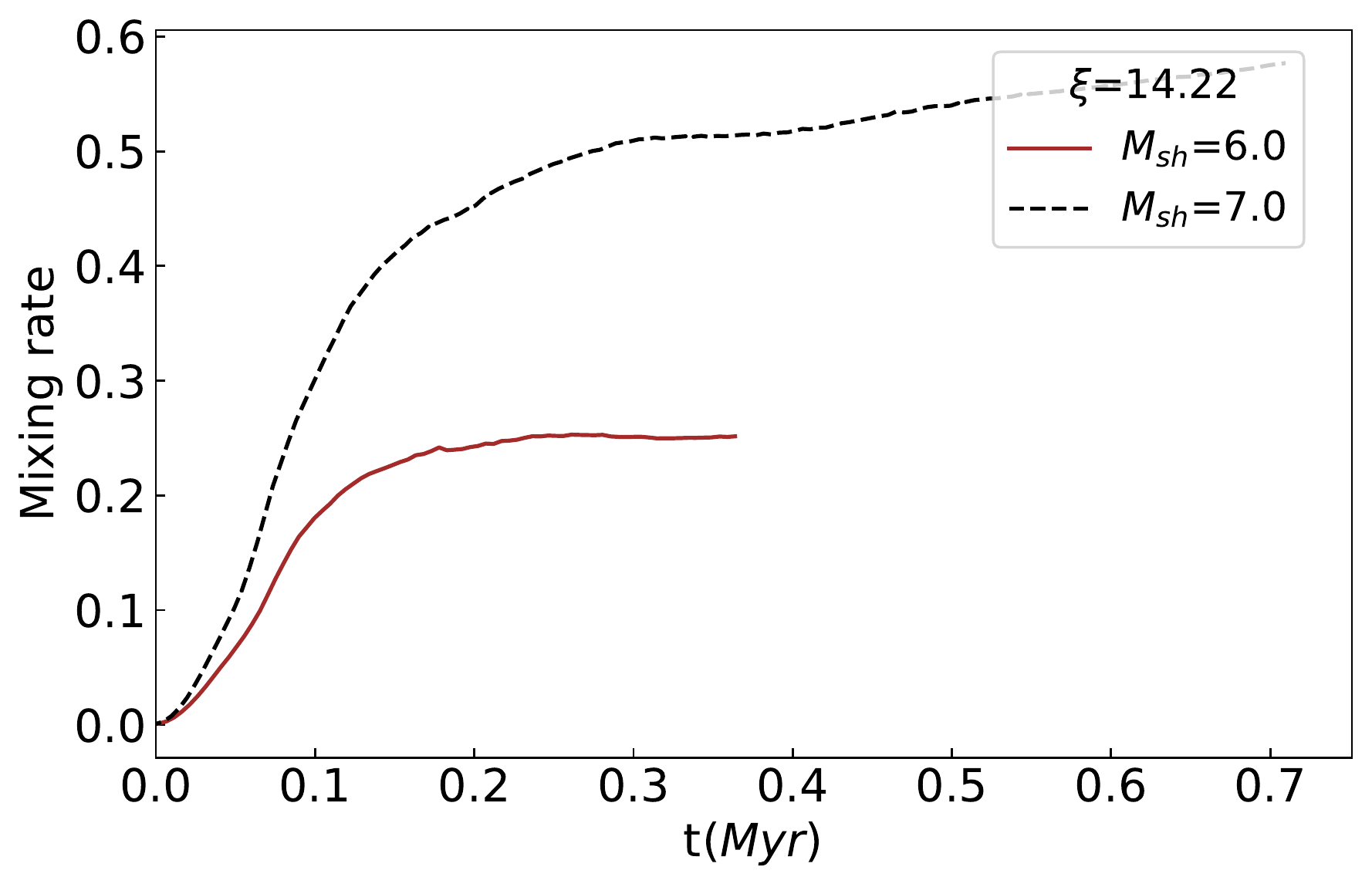}
          \hspace{2.0cm} (g) $\xi$=14.22
        \end{center}
      \end{minipage}
    \end{tabular}
    \vskip5pt  
    \caption{Mixing rate defined in Equation (\ref{eq:mixing rate}) as functions of time after shocks arrive \edit1{at a} cloud at different shock \edit1{Mach numbers}. If a sink particle is introduced during \edit1{the} shock-cloud evolution, the transition is represented by a solid line.
    After the time at which \edit1{a} sink particle is introduced, \edit1{the} transition is not shown. In cases in which a sink particle is not introduced, the transition is represented by a dashed line.}
    \label{fig:mixing_rate}

\end{figure*}

\begin{figure*}
  
    \begin{tabular}{c}

      \begin{minipage}{0.5\hsize}
        \begin{center}
          \includegraphics[clip, width=6.7cm]{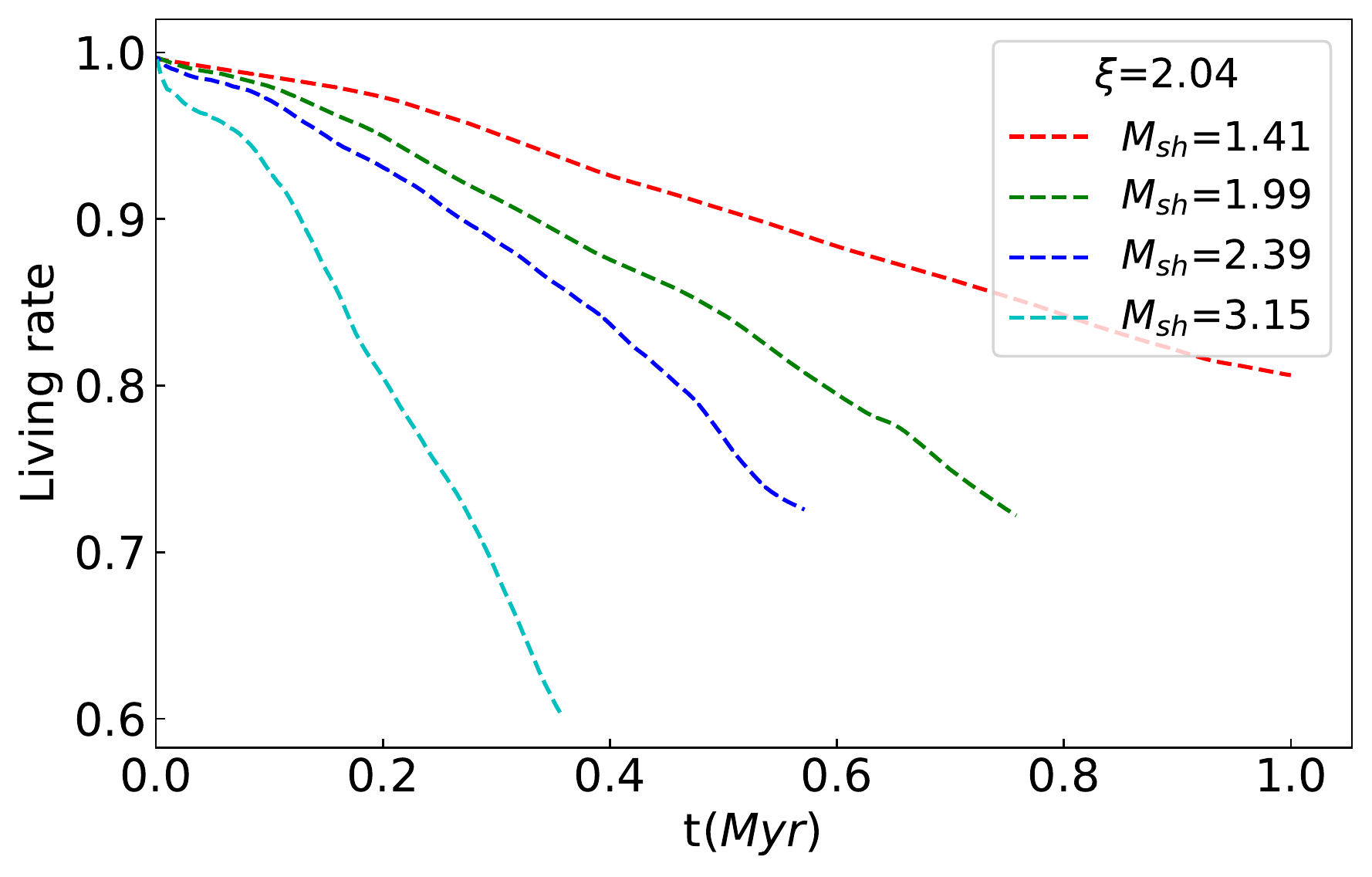}
          \hspace{2.0cm} (a) $\xi$=2.04
        \end{center}
      \end{minipage}

      \begin{minipage}{0.5\hsize}
        \begin{center}
          \includegraphics[clip, width=6.7cm]{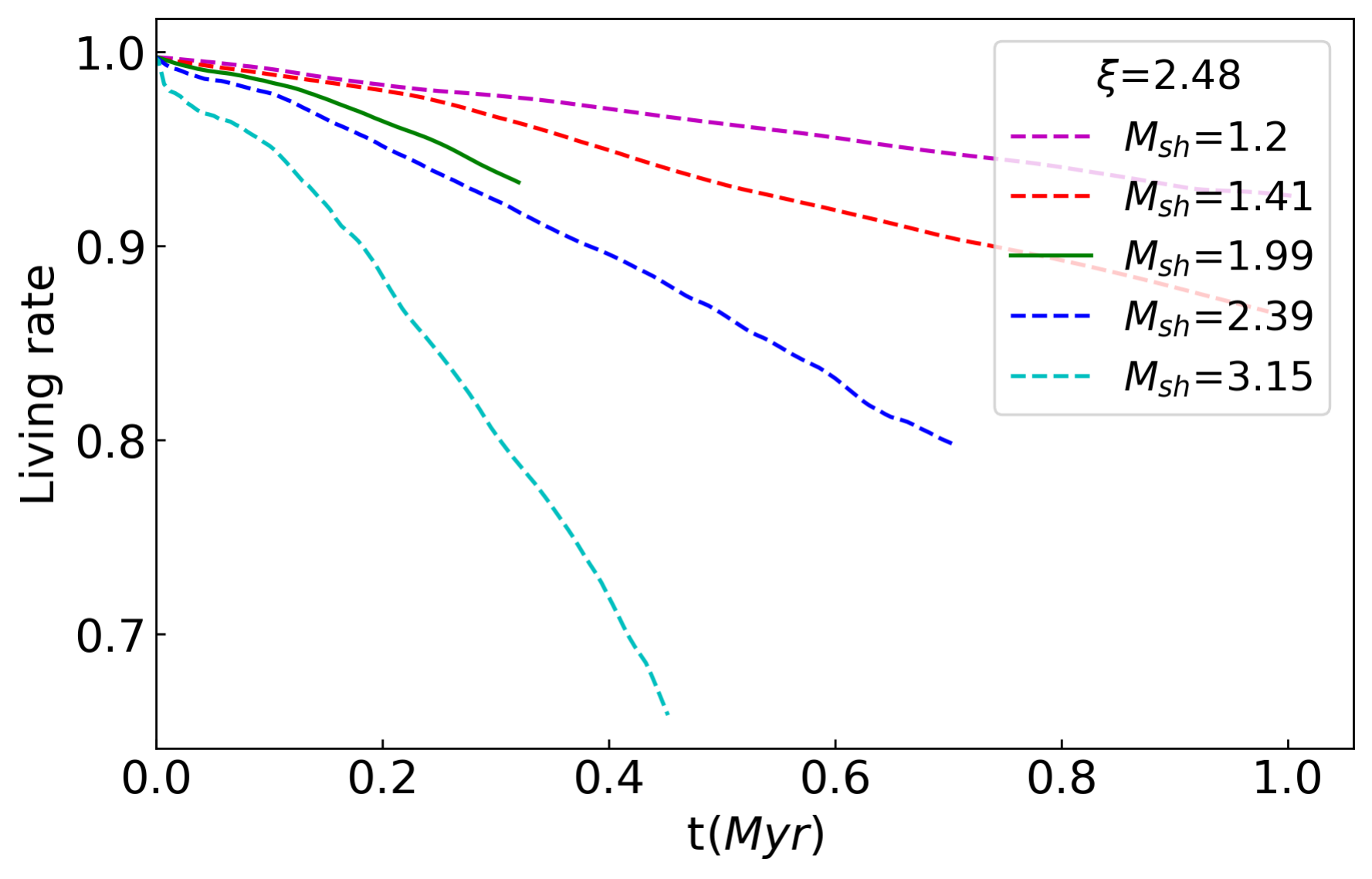}
          \hspace{2.0cm} (b) $\xi$=2.48
        \end{center}
      \end{minipage}
      \end{tabular}
      
      \begin{tabular}{c}

      \begin{minipage}{0.5\hsize}
        \begin{center}
          \includegraphics[clip, width=6.7cm]{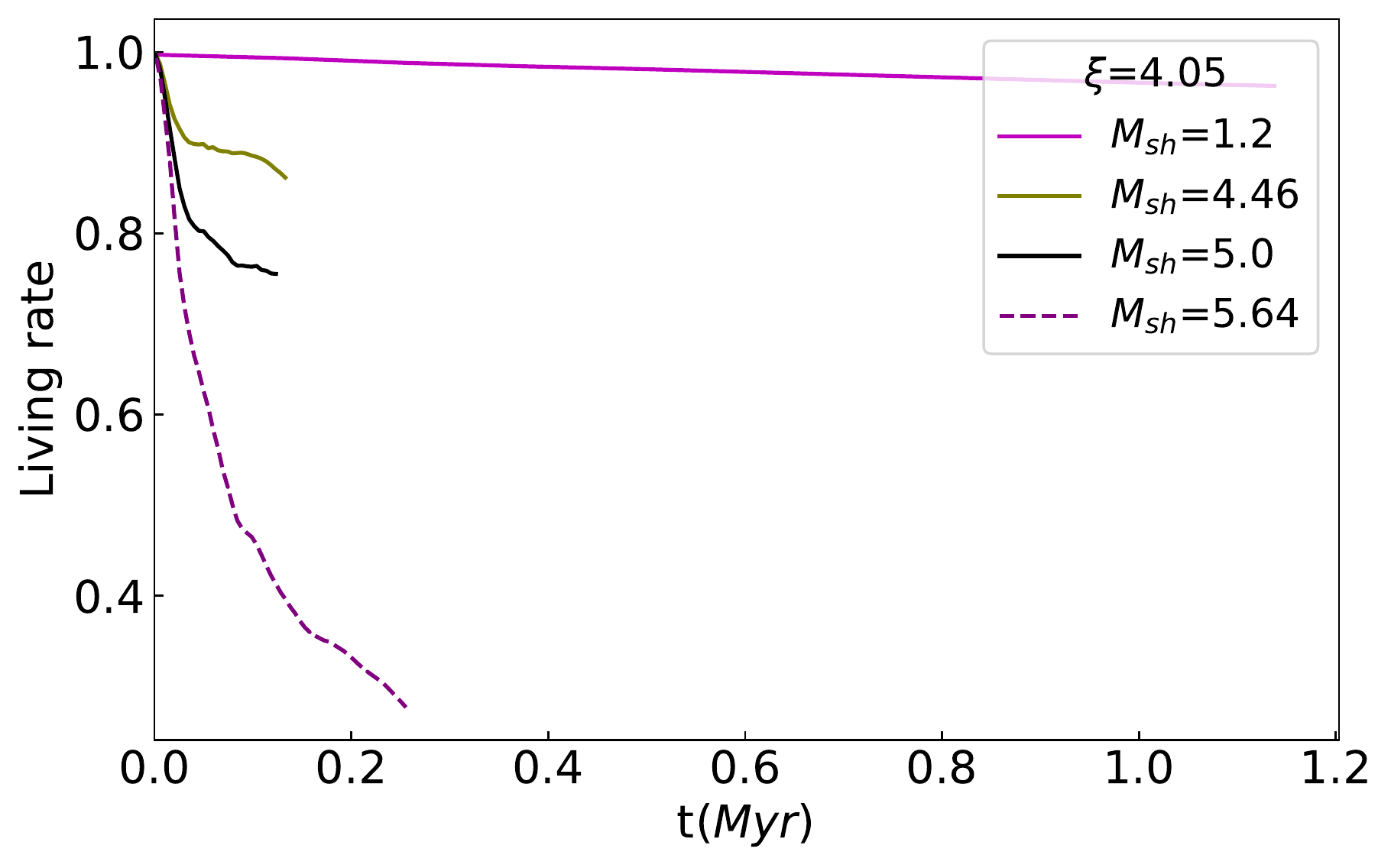}
          \hspace{2.0cm} (d) $\xi$=4.05
        \end{center}
      \end{minipage}


      \begin{minipage}{0.5\hsize}
        \begin{center}
          \includegraphics[clip, width=6.7cm]{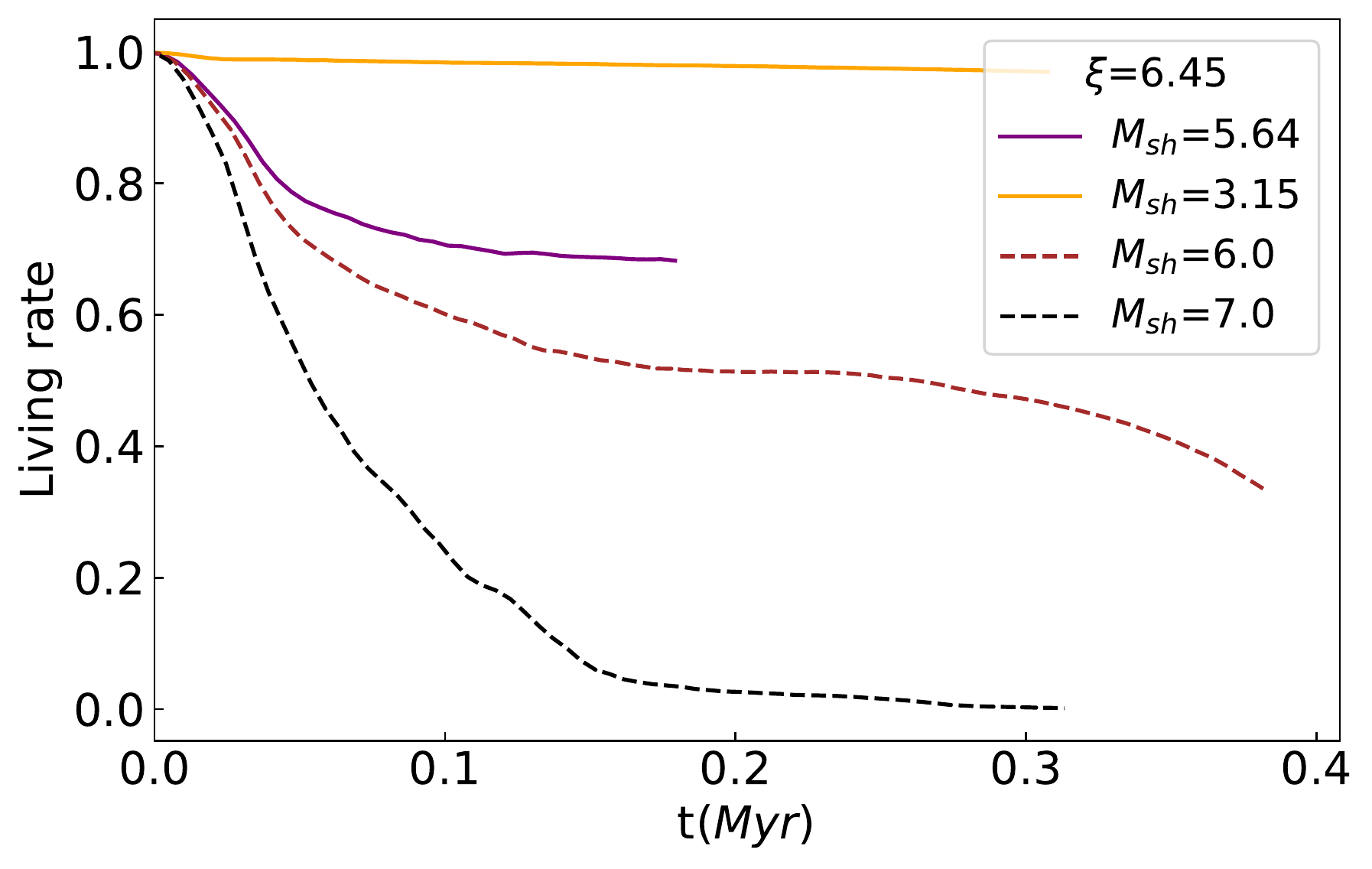}
          \hspace{2.0cm} (e) $\xi$=6.45
        \end{center}
      \end{minipage}
      
      \end{tabular}

      \begin{tabular}{c}
      \begin{minipage}{0.5\hsize}
        \begin{center}
          \includegraphics[clip, width=6.7cm]{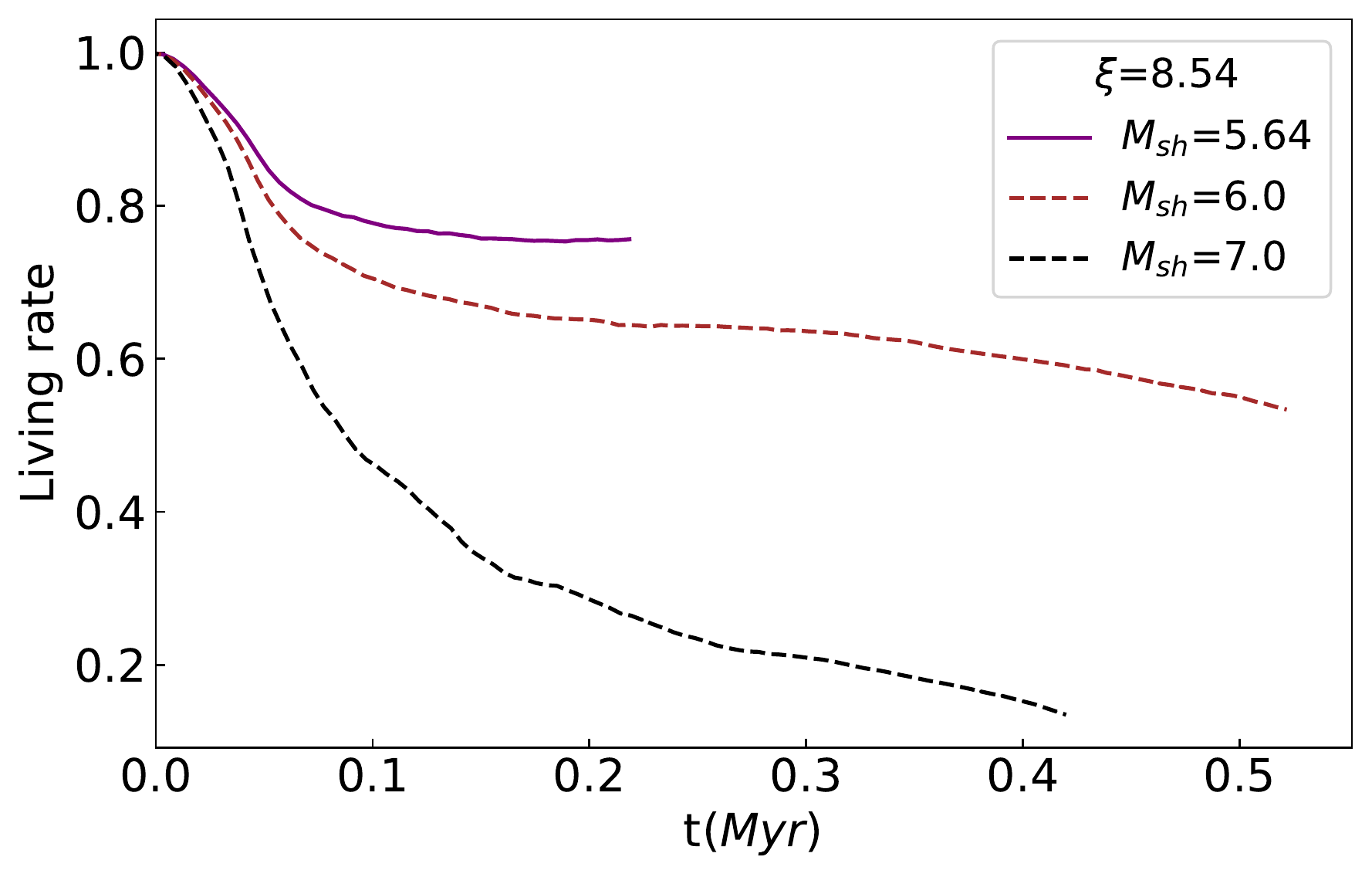}
          \hspace{2.0cm} (f) $\xi$=8.54
        \end{center}
      \end{minipage}

    \begin{minipage}{0.5\hsize}
        \begin{center}
          \includegraphics[clip, width=6.7cm]{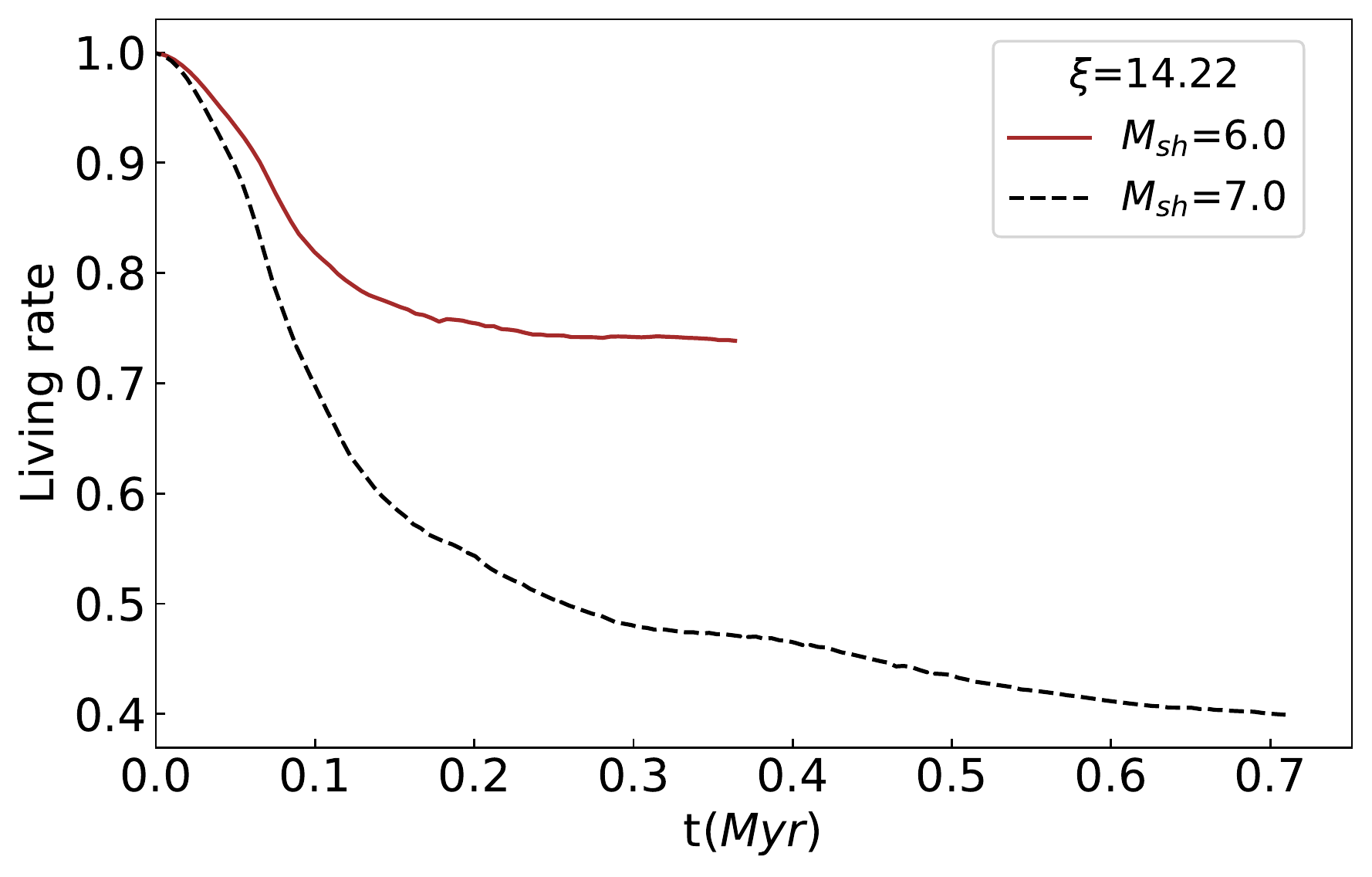}
          \hspace{2.0cm} (g) $\xi$=14.22
        \end{center}
      \end{minipage}
    \end{tabular}
    \vskip5pt  
    \caption{Living rate defined in Equation \ref{eq:living rate} as functions of time after shocks arrive \edit1{at a} cloud at different shock \edit1{Mach numbers}. If a sink particle is introduced during \edit1{the} shock-cloud evolution, the transition is represented by a solid line.
    After the time at which sink particle is introduced, \edit1{the} transition is not shown. In cases in which a sink particle is not introduced, the transition is represented by a dashed line.}
    \label{fig:living_rate}

\end{figure*}

Figure \ref{fig:mixing_rate} and \ref{fig:living_rate} show evolution of \edit1{the} mixing and living \edit1{rates}. \edit1{For} $\xi$=3.22, \edit1{the} larger the Mach number, the shorter the time scale of the mixture with the ambient gas. That is, in all cases, the higher the propagating shock velocity, the faster the destruction of the cloud progress.

\begin{figure*}
  
    \begin{tabular}{c}

      \begin{minipage}{0.5\hsize}
        \begin{center}
          \includegraphics[clip, width=6.7cm]{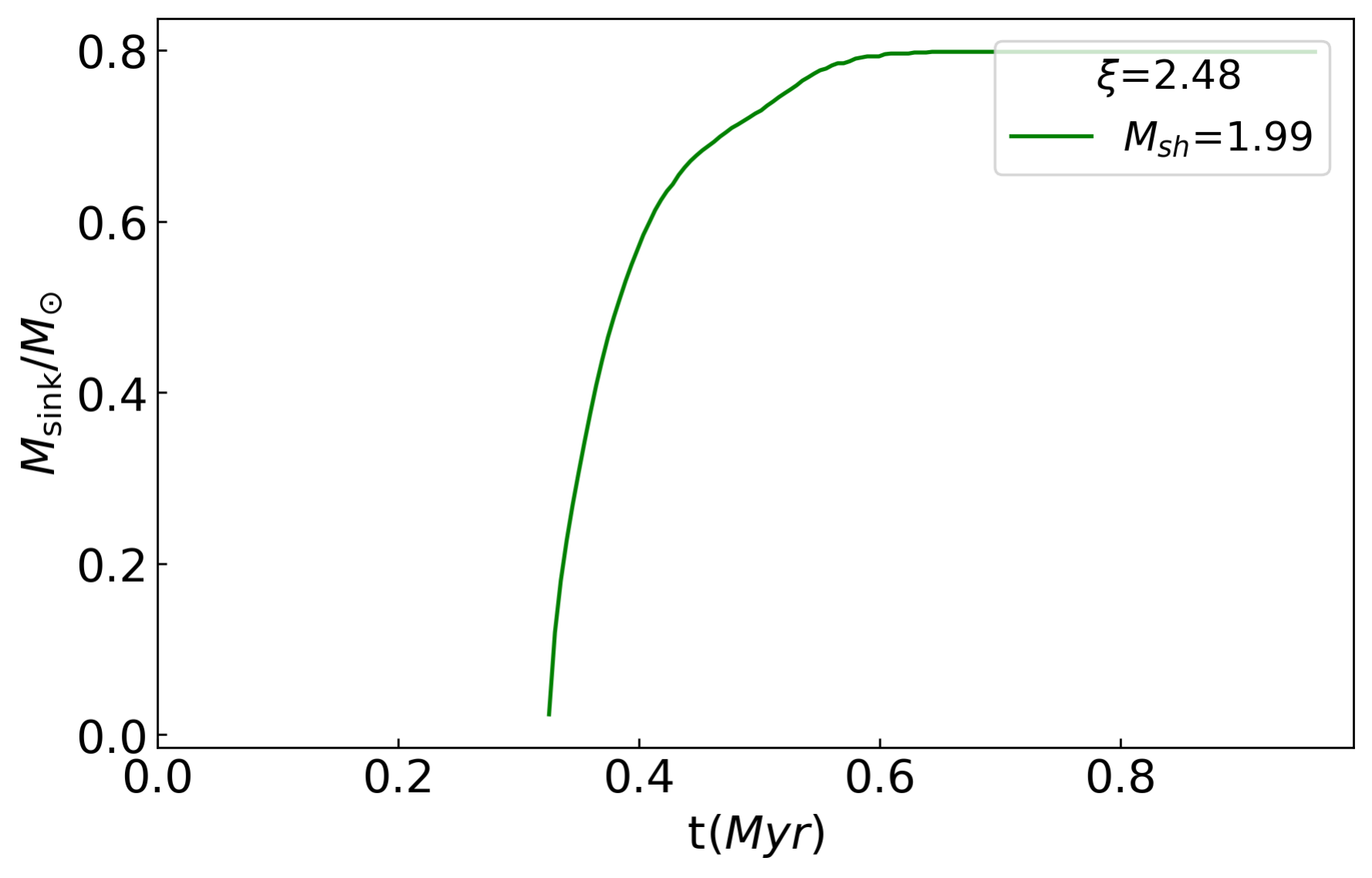}
          \hspace{2.0cm} (a) $\xi$=2.48
        \end{center}
      \end{minipage}


      \begin{minipage}{0.5\hsize}
        \begin{center}
          \includegraphics[clip, width=6.7cm]{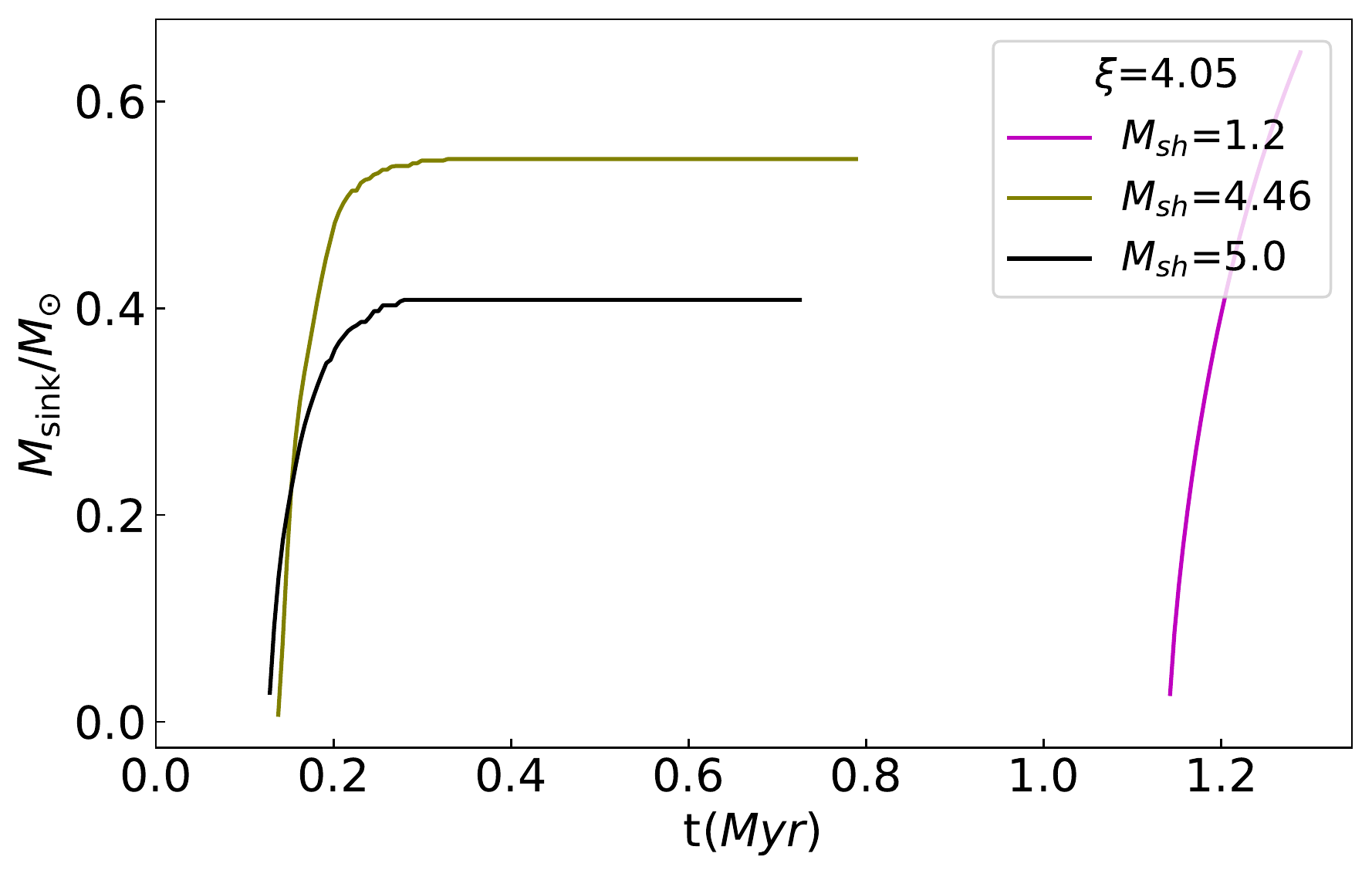}
          \hspace{2.0cm} (c) $\xi$=4.05
        \end{center}
      \end{minipage}
      
    \end{tabular}
      
    \begin{tabular}{c} 
    
      \begin{minipage}{0.5\hsize}
        \begin{center}
          \includegraphics[clip, width=6.7cm]{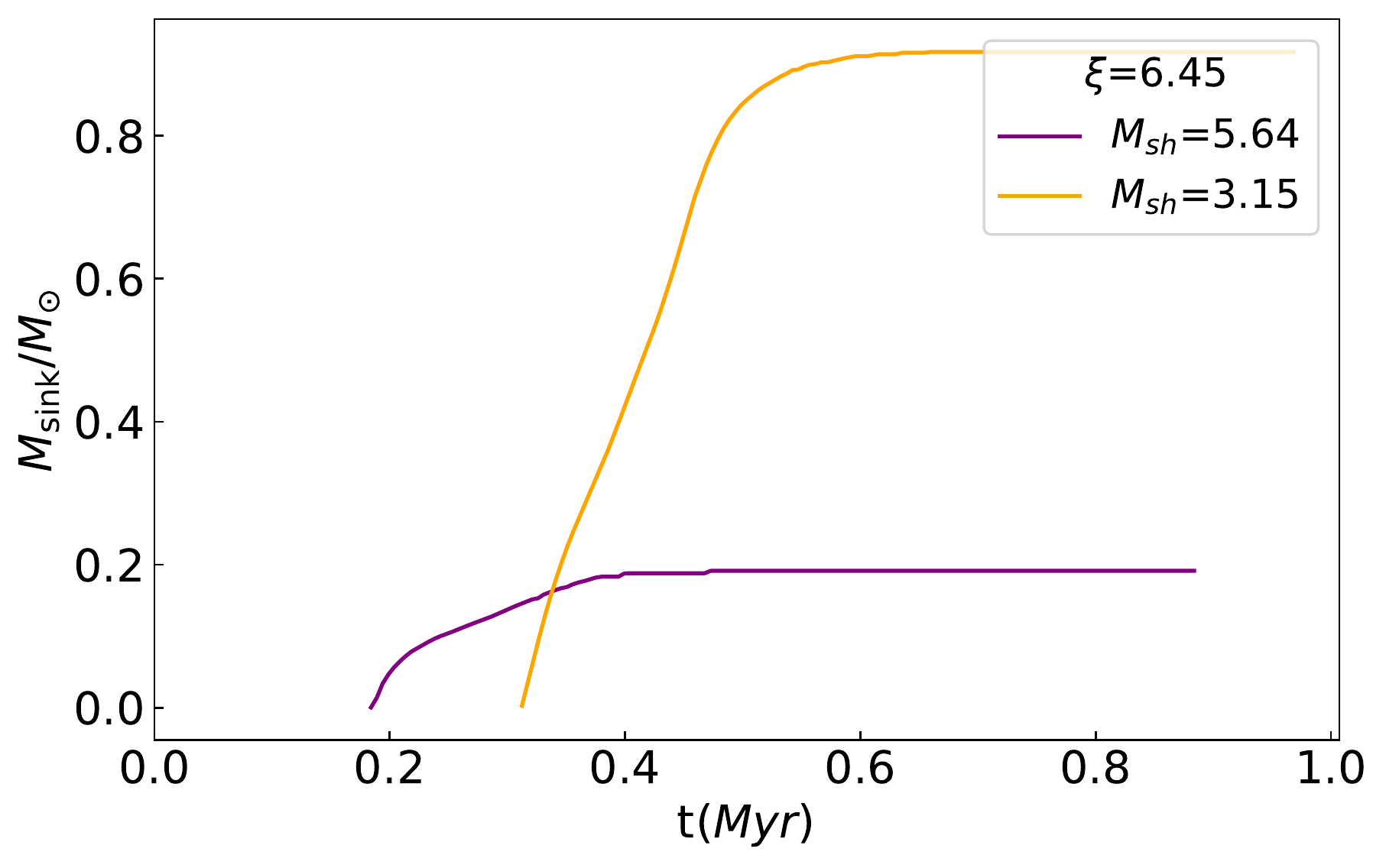}
          \hspace{2.0cm} (d) $\xi$=6.45
        \end{center}
      \end{minipage}
      

      \begin{minipage}{0.5\hsize}
        \begin{center}
          \includegraphics[clip, width=6.7cm]{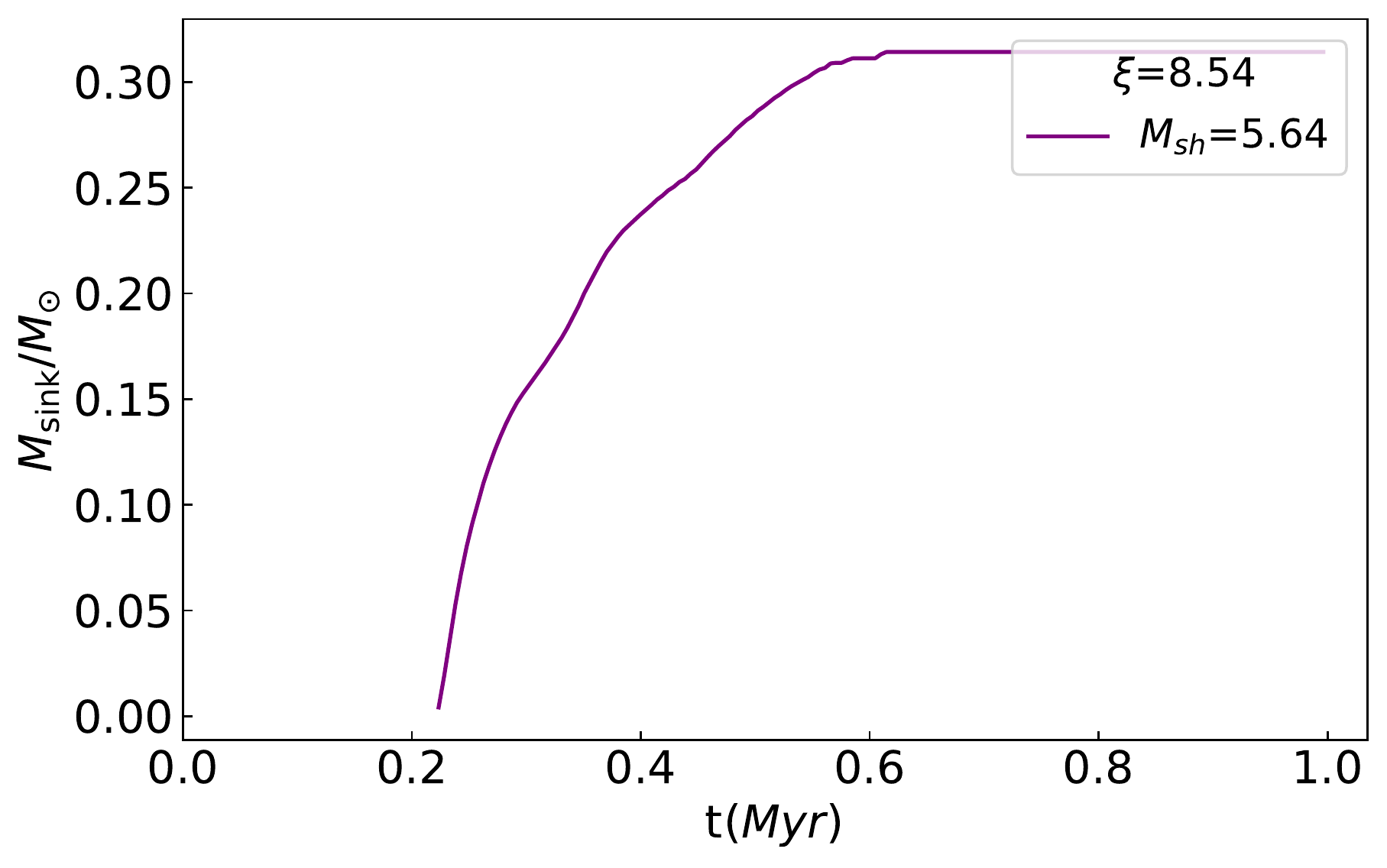}
          \hspace{2.0cm} (e) $\xi$=8.54
        \end{center}
      \end{minipage}
    \end{tabular}
    
    \begin{tabular}{c}

    \begin{minipage}{0.5\hsize}
        \begin{center}
          \includegraphics[clip, width=6.7cm]{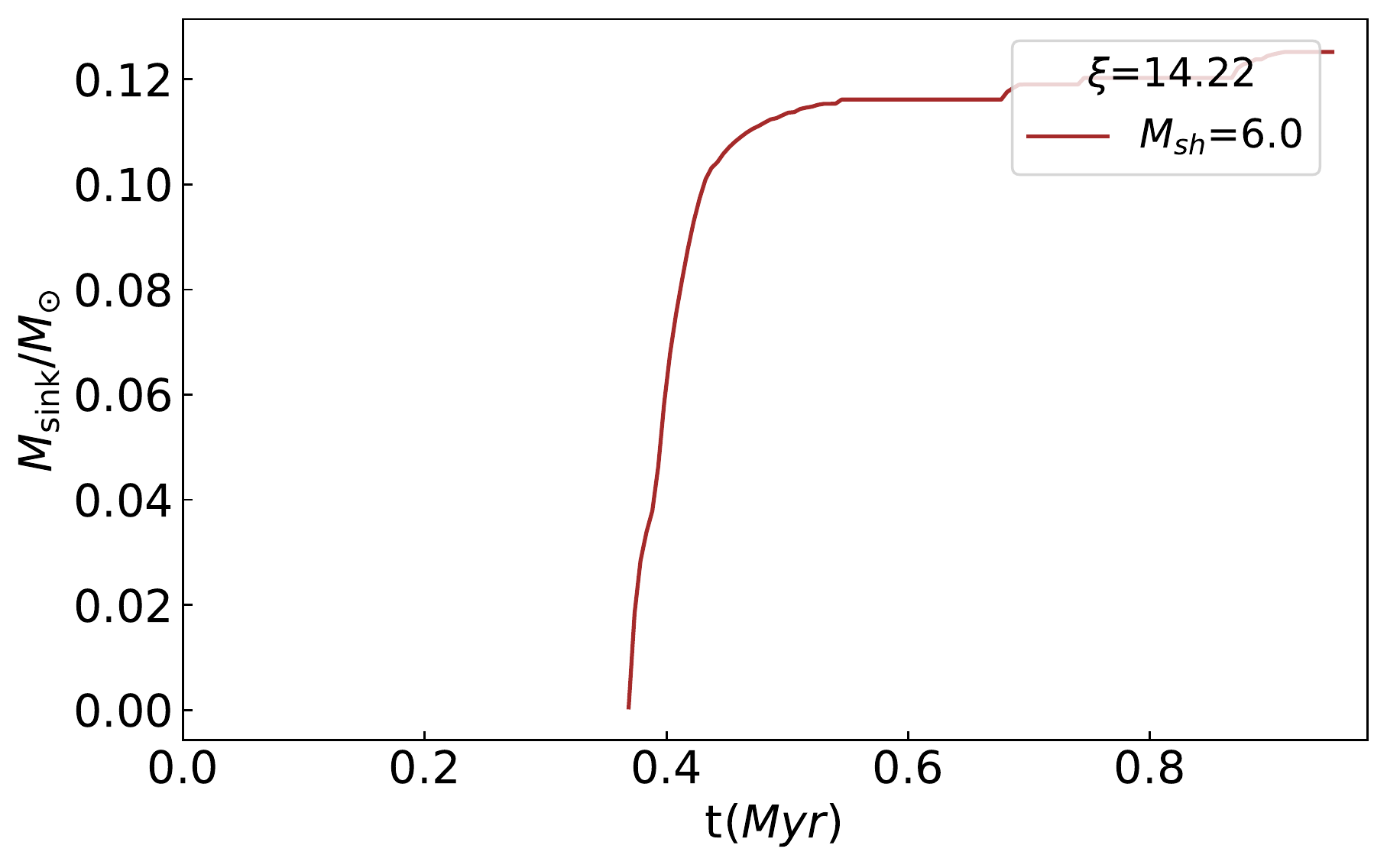}
          \hspace{2.0cm} (f) $\xi$=14.22
        \end{center}
      \end{minipage}
    \end{tabular}
    \vskip5pt  
    \caption{Ratio of sink particle mass \edit1{to} initial cloud mass $M_{\rm sink}/M_{\rm cl}$ as functions of time after shocks arrive at different shock \edit1{Mach numbers}.}
    \label{fig:sink_mass}

\end{figure*}

\begin{figure*}
  
    \begin{tabular}{c}

      \begin{minipage}{0.5\hsize}
        \begin{center}
          \includegraphics[clip, width=6.7cm]{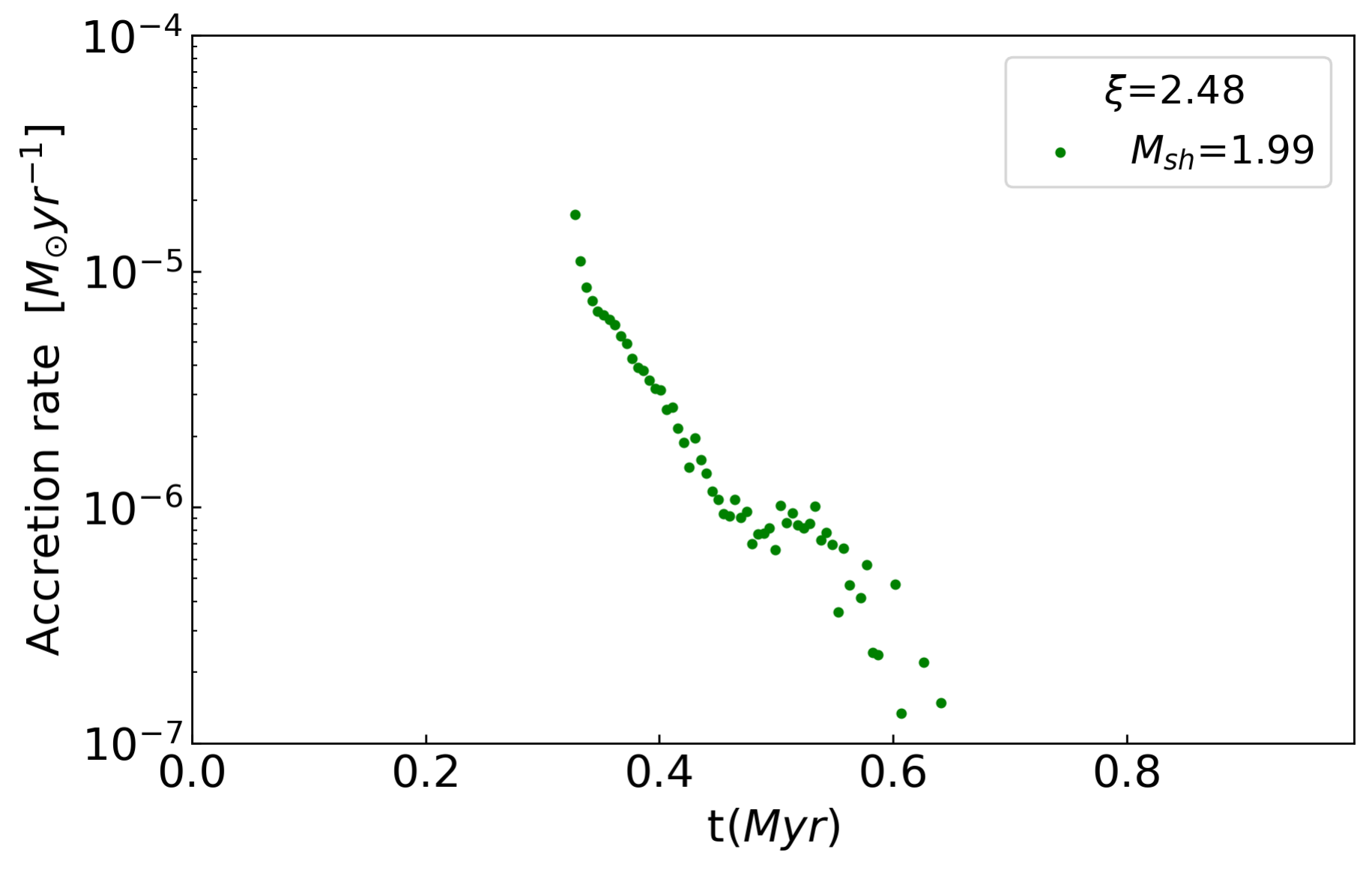}
          \hspace{2.0cm} (a) $\xi$=2.48
        \end{center}
      \end{minipage}


      \begin{minipage}{0.5\hsize}
        \begin{center}
          \includegraphics[clip, width=6.7cm]{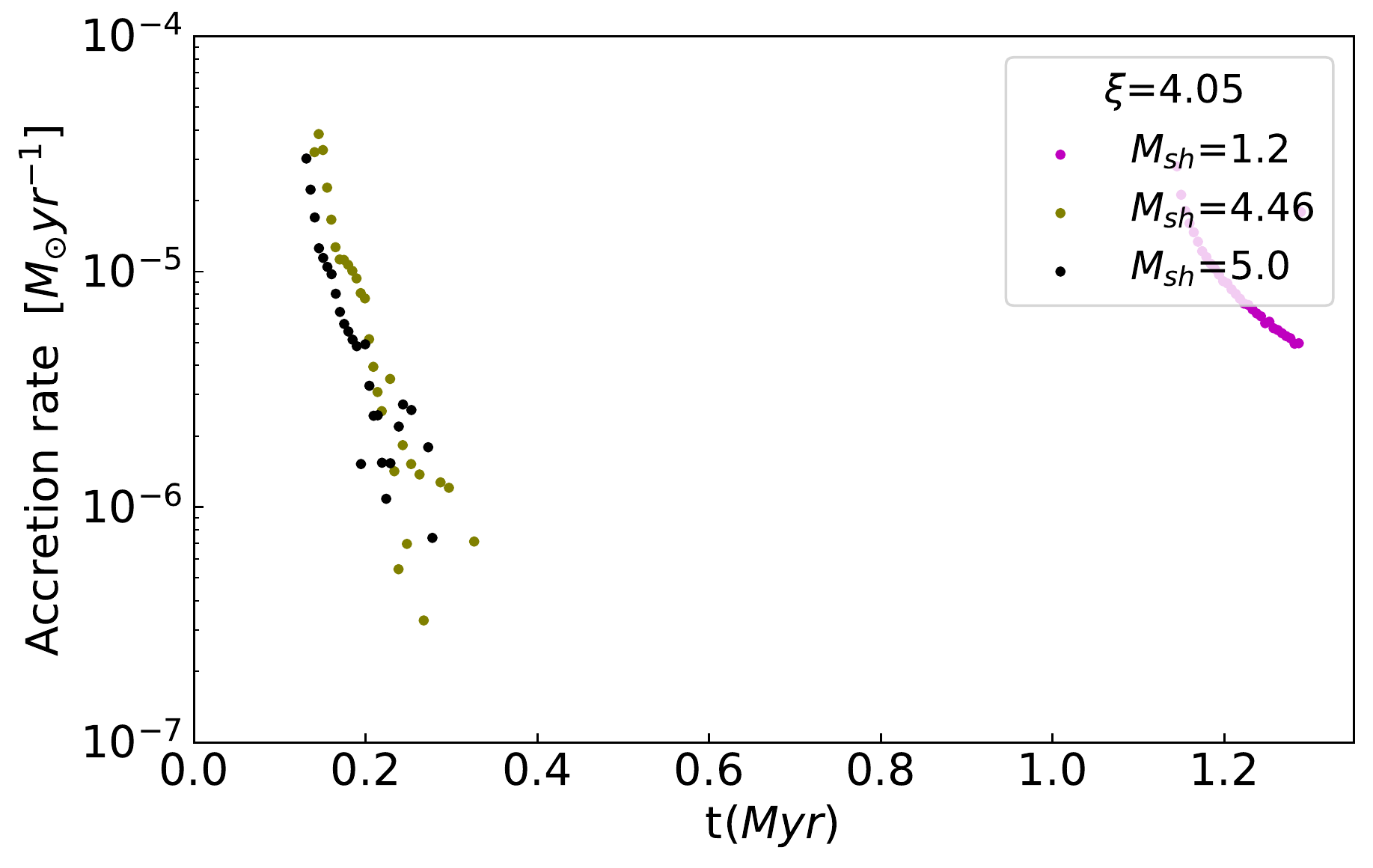}
          \hspace{2.0cm} (c) $\xi$=4.05
        \end{center}
      \end{minipage}
      
    \end{tabular}   
      
    \begin{tabular}{c}  
      \begin{minipage}{0.5\hsize}
        \begin{center}
          \includegraphics[clip, width=6.7cm]{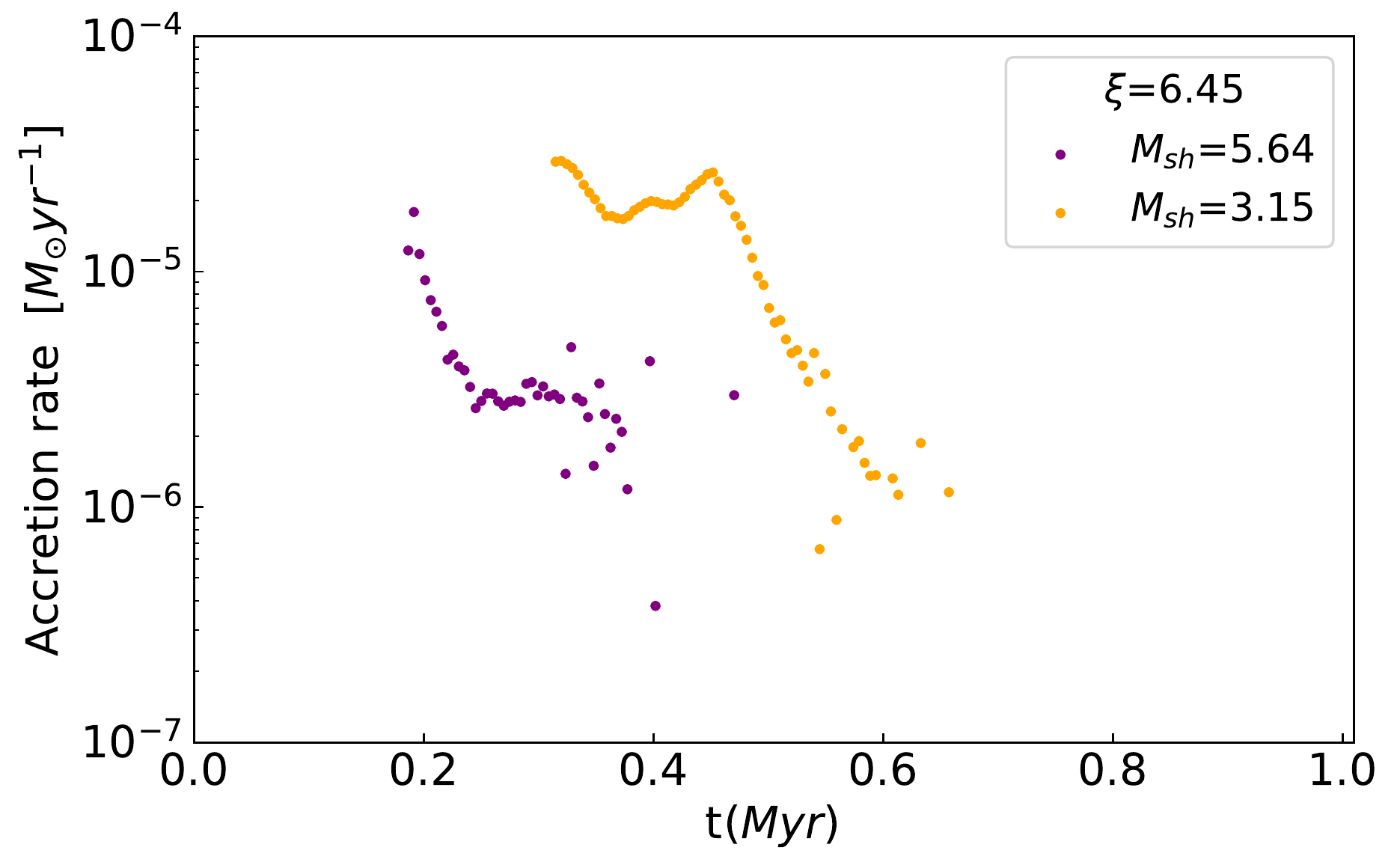}
          \hspace{2.0cm} (d) $\xi$=6.45
        \end{center}
      \end{minipage}

      \begin{minipage}{0.5\hsize}
        \begin{center}
          \includegraphics[clip, width=6.7cm]{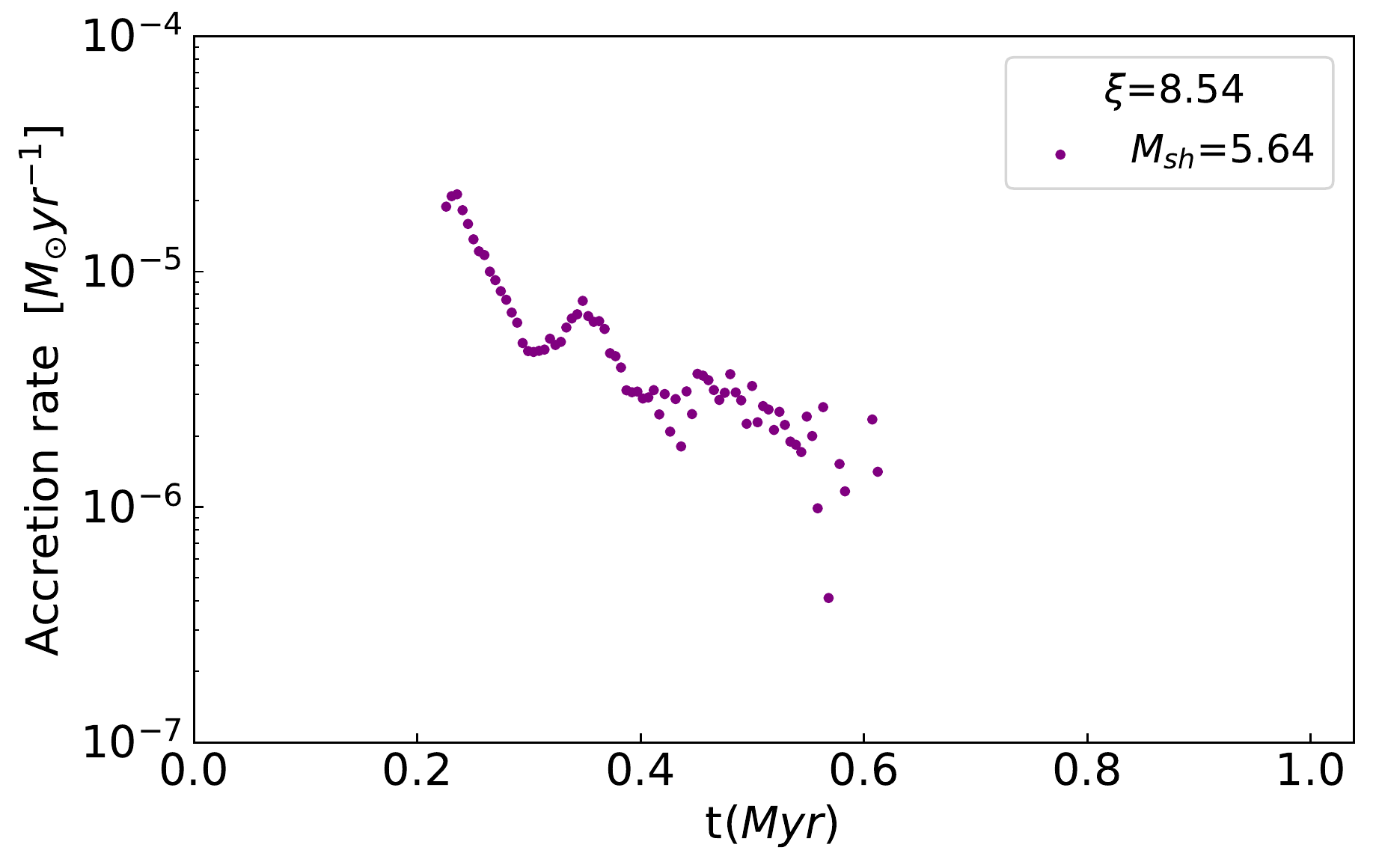}
          \hspace{2.0cm} (e) $\xi$=8.54
        \end{center}
      \end{minipage}
   \end{tabular}

    \begin{tabular}{c}

    \begin{minipage}{0.5\hsize}
        \begin{center}
          \includegraphics[clip, width=6.7cm]{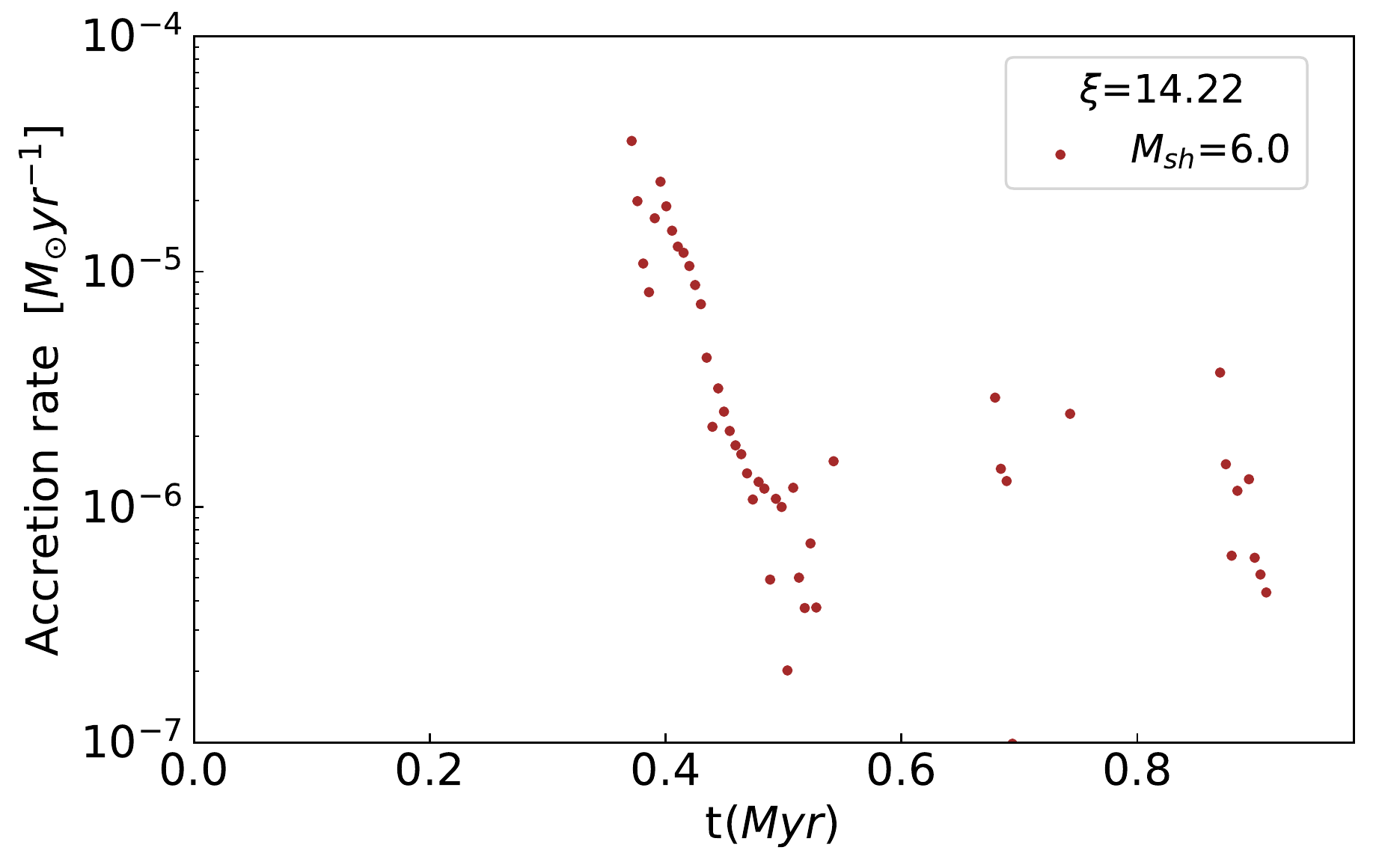}
          \hspace{2.0cm} (f) $\xi$=14.22
        \end{center}
      \end{minipage}
    \end{tabular}
    \vskip5pt  
    \caption{Evolution of \edit1{the} mass accretion rate of sink particles.}
    \label{fig:accretion_rate}

\end{figure*}

Figure \ref{fig:sink_mass} and \ref{fig:accretion_rate} show \edit1{the} evolution of sink particles mass and accretion rates. Figure \ref{fig:formed_result} shows \edit1{the} results of each model.
From $\xi$=4.05 or $\xi$=6.45, we can conclude that the higher the Mach number, the slower the sink particle formation begins and the lower asymptotic sink particles mass. This trend is also the same \edit1{for} $\xi$=3.22.

\section{Moving of sink particles and stripped clouds}

\label{app:Moving}
\begin{figure*}
      \begin{tabular}{c}
      \begin{minipage}{0.5\hsize}
        \begin{center}
          \includegraphics[clip, width=7.5cm]{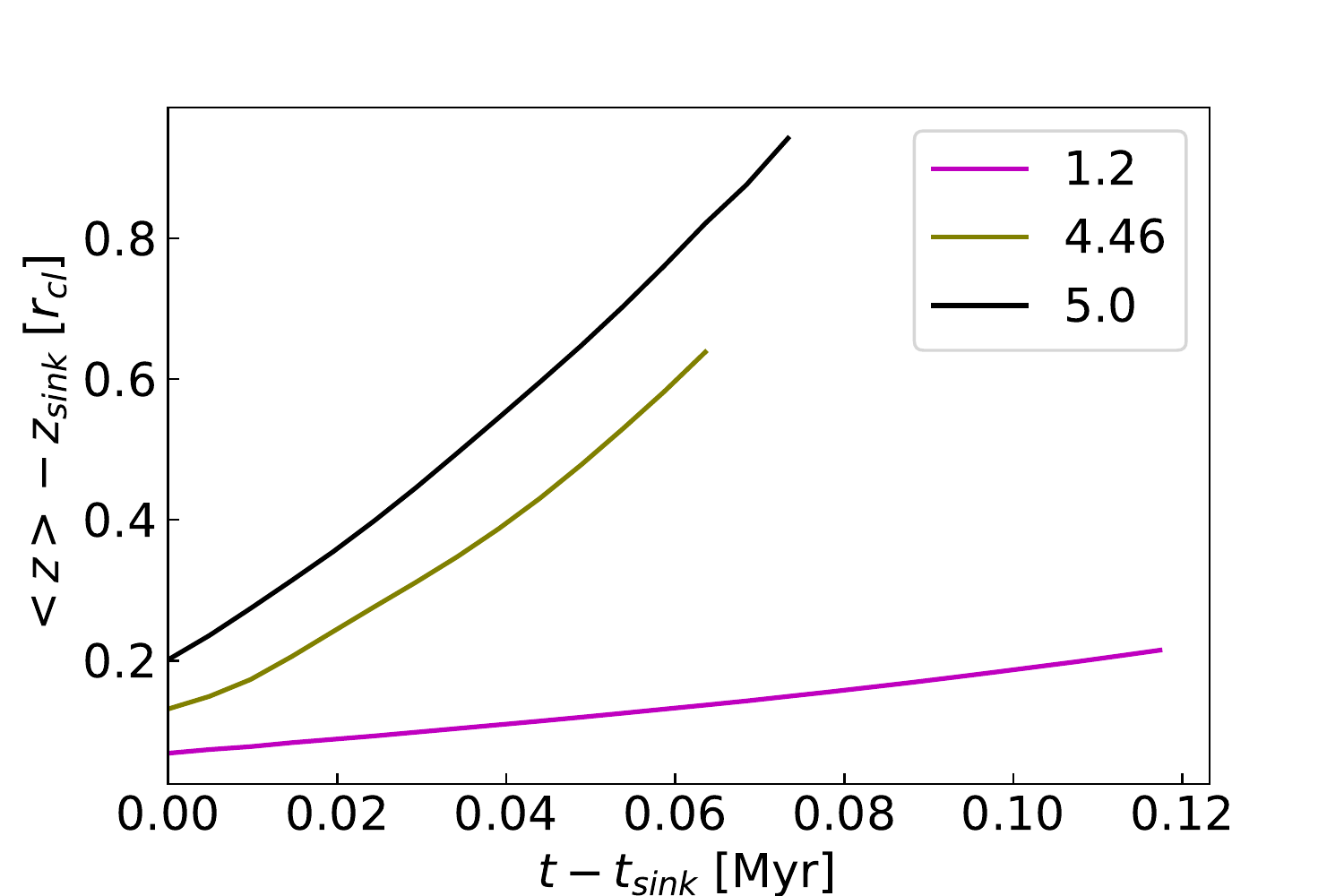}
          \hspace{2.0cm} (a) Relative \edit1{displacement}
        \end{center}
      \end{minipage}

      \begin{minipage}{0.5\hsize}
        \begin{center}
          \includegraphics[clip, width=7.5cm]{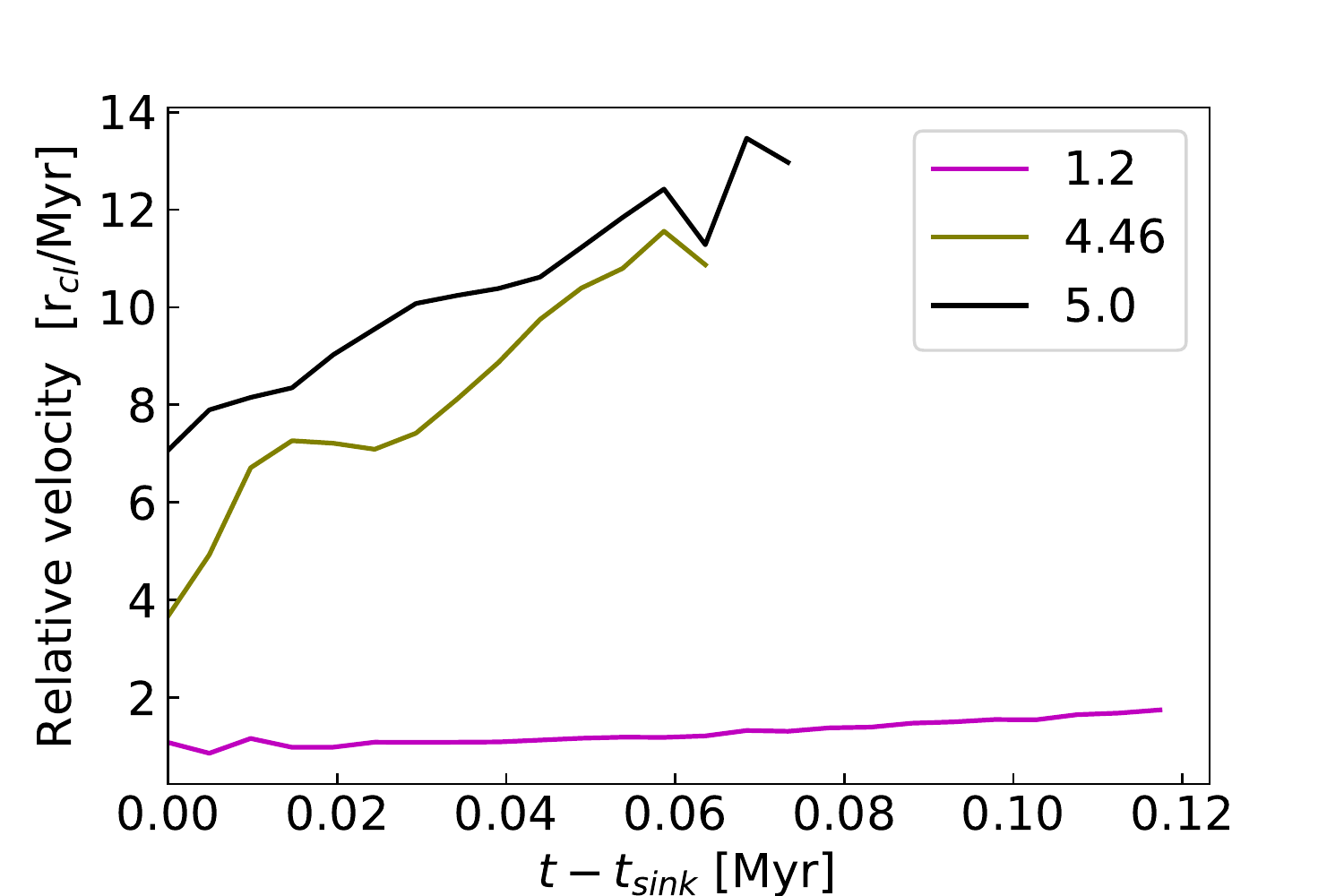}
          \hspace{2.0cm} (b) Relative velocity
        \end{center}
      \end{minipage}

      \end{tabular}
      \vskip5pt  
    \caption{As \edit1{in} Figure \ref{fig:sink_relative} for $\xi=4.05$ and $M_{\rm sh}=1.20, 4.46$\edit1{,} and $5.00$.}
    \label{fig:sink_relative_2}
\end{figure*}

Figure \ref{fig:sink_relative_2} shows the relative \edit1{displacement} and velocity \edit1{between} sink particles and stripped clouds for $\xi=4.05$ and $M_{\rm sh}=$1.20, 4.46\edit1{,} and 5.00 cases.

\clearpage

\bibliography{sample63}{}
\bibliographystyle{aasjournal}



\end{document}